\documentclass{aastex631}

\usepackage{graphicx}
\usepackage{natbib}
\usepackage{epsfig}
\usepackage{multirow}
\usepackage{amsmath} 
\usepackage{threeparttable}
\usepackage{verbatim}
\usepackage{xcolor}

\newcommand{\beq}{\begin{equation}}
\newcommand{\eeq}{\end{equation}}
\newcommand{\red}{\color{black}} 
\newcommand{\Gaia}{{\it Gaia\ }}

\shorttitle{Occultation Predictions}
\shortauthors{French and Souami}

\begin{document}

\title{Earth-based Stellar Occultation Predictions for Jupiter, Saturn, Uranus, Neptune, Titan, and Triton: 2023--2050}
\author[0000-0002-9858-9532]{Richard G. French}
\affil{Department of Astronomy, Wellesley College, Wellesley MA 02481}
\author[0000-0002-0786-7307]{Damya Souami}
\affiliation{LESIA, Observatoire de Paris, Universit\'e PSL, CNRS, Sorbonne Universit\'e, Universit\'e de Paris,\\
 5 place Jules Janssen,\\
  F-92195 Meudon, France}
\affiliation{Departments of Astronomy, and of Earth and Planetary Science, \\ 
501 Campbell Hall \\
University of California, Berkeley, \\
CA 94720, United States of America}
\affiliation{naXys, Department of Mathematics, University of Namur,\\
  Rue de Bruxelles 61,\\
   5000 Namur, Belgium}

\correspondingauthor{Richard G. French}

\email{rfrench@wellesley.edu}
\begin{abstract}
In support of studies of decadal-timescale evolution of outer solar system atmospheres and ring systems, we present detailed Earth-based stellar occultation predictions for Jupiter, Saturn, Uranus, Neptune, Titan, and Triton for 2023-2050, based on the \Gaia DR3 star catalog and near-IR K-band photometry from the 2MASS catalog. We tabulate the number of observable events by year and magnitude interval, reflecting the highly variable frequency of high-SNR events depending on the target's path relative to the star-rich regions of the Milky Way. We identify regions on Earth where each event is potentially observable, and for atmospheric occultations, we determine the latitude of the ingress and egress events. For Saturn, Uranus, and Neptune, we also compute the predicted ring occultation event times. We present representative subsets of the predicted events and highlights particularly promising events. Jupiter occultations with K $\leq$7 occur at a cadence of about one per year, with bright events at higher frequency in 2031 and 2043. Saturn occultations are much rarer, with only two predicted events with K $\leq$5 in 2032 and 2047. Ten Uranus ring occultations are predicted with K$\leq$10 for the period 2023 to 2050. Neptune traverses star-poor regions of the sky until 2068, resulting in only 13 predicted occultations for K$\leq$12 between 2023 and 2050. Titan has several high-SNR events between 2029--2031, whereas Triton is limited to a total of 22 occultations with K$\leq$15 between 2023 and 2050. Details of all predicted events are included in the Supplementary Online Material. \end{abstract}
\keywords{occultations, astrometry, planets: rings, atmospheres}
\parskip 10pt
%
%

\section{Introduction}
Earth-based stellar occultations have proven to be a powerful and versatile tool of solar system discovery and exploration.\footnote{
For somewhat dated but still useful reviews of the stellar occultation technique, including summaries of early observations, see
\cite{Elliot1979} and \cite{Elliot1996}.}
Planetary occultations have revealed the stratospheric thermal structure of Jupiter
\citep{Baum1953,Hubbard1972,Vapillon1973,Veverka1974,Raynaud2004,Hubbard1995,Raynaud2003,Christou2013}, Saturn \citep{Hubbard1997}, Uranus \citep{Sicardy1985}, and Neptune \citep{Roques1994}. They serendipitously led to the discovery of the Uranian rings \citep{Elliot1977,Millis1977}, and with subsequent concerted effort they resulted in the discovery {\red and characterization} of Neptune's ring arcs \citep{Hubbard1986, Sicardy1986,Smith1989,Sicardy1991,DePater2018b,Gaslac2020}, which appear to be evolving over time \citep{Souami2022}. Extensive observing campaigns of the Uranus system yielded the orbital properties of the planet's rings and {\red estimates of} its gravitational field \citep{Nicholson2018,French2023a, French2023b}. Despite the early challenges to accurate predictions for occultations by smaller outer solar system objects, successful {\red airborne and ground-based} occultation observations provided the first convincing detection of the Pluto's tenuous atmosphere \citep{Elliot1989}, {\red determined the properties of its atmospheric waves \citep{Person2008}, and revealed properties of its atmospheric haze from multi-wavelength observations \citep{Person2021}. Multiple occultations by Triton have added to our understanding of its atmosphere \citep{Olkin1997}. The Centaur (2060) Chiron exhibits outgassing behavior \citep{Ruprecht2015} and possibly hosts a ring system \citep{Sickafoose2020}. Over the last decade,} ambitious international observing campaigns have yielded the surprising discoveries of rings around the Centaur (10199) Chariklo \citep{Braga2014,Berard2017, Morgado2021}, dwarf planet (136108) Haumea \citep{Ortiz2017}, and trans-Neptunian object (TNO) (50000) Quaoar \citep{Morgado2023,Pereira2023}, and have begun to provide accurate physical properties of TNOs themselves \citep{Souami2020,Santos2022a,FV2023}.
The availability of highly accurate star positions provided by {\it Gaia} \citep{Gaia2021,Gaia2022} has revolutionized the observing strategy for small target occultations by enabling portable telescopes to be placed along the path of the occultation shadow. {\red Using multiple mobile stations, \cite{Buie2020} measured the size, shape, and astrometric position of TNO (486958) Arrokoth (the flyby target of the New Horizons extended mission) from four stellar occultations.
Recent densely-spaced observations of occultation observations by} by Triton \citep{Marques2022} and Pluto \citep{Young2022a} captured the central flash produced by the refractive focusing by the tenuous atmospheres of these small outer solar system bodies, providing valuable information about their possibly time-variable surface atmospheric pressure, thermal structure, and haze opacity. {\red The shape and duplicity of the {\it Lucy Mission} prime target Polymele were similarly determined from ground station chords separated in the sky plane by only 1.8~km \citep{Buie2022}.}

{\red Meanwhile}, the detailed reconnaissance of the Saturn system by {\sl Cassini} and the ongoing exploration of Jupiter with {\sl Juno}, along with a wealth of {\sl Kepler} and {\sl TESS} observations of exoplanets, have prompted international interest in further exploration of ice giants in our own solar system \citep{Hofstadter2019,Fletcher2020,Blanc2021,Cartwright2021}. Indeed, the most recent National Academies planetary science decadal survey \citep{decadal2022} identifies the Uranus Orbiter and
Probe as 
the highest priority Flagship mission for the decade 2023-2032. 
This invites 
renewed attention to the possibility of future occultation observations of the rings and atmospheres of the giant planets.

{\red In this paper, we identify scientifically useful stellar occultations that can be used to investigate the structure and decadal-timescale evolution of the atmospheres and ring systems Jupiter, Saturn, Uranus, Neptune, Titan, and Triton, complementing results from {\it JWST} and adaptive-optics ground-based imaging. We take advantage of the astrometric accuracy of the {\it Gaia}  DR3 star catalog and recent improvements in the JPL planetary ephemerides} to identify potential occultations for the period 2023-2050. We choose this rather long time period in part to illustrate the extreme time variability of the frequency of high-SNR occultation opportunities, depending on whether the target traverses the dense star fields of the Milky Way (as was the case for Uranus and Neptune in the 1980s), or instead remains for long periods in relatively star-free regions of the sky. For each candidate occultation, we identify regions on Earth where the event is potentially observable, and for atmospheric occultations, we determine the latitude of the ingress and egress events. For Saturn, Uranus, and Neptune, we also compute the predicted ring occultation event times, taking into account the known orbital characteristics of representative rings. We have included Titan and Triton as well, {\red since future occultation observations can provide valuable information about possible seasonal changes in their atmospheres. The uncertainties in {\it Gaia} star positions and proper motions, combined with the estimated current accuracy of the ephemerides of both moons, enable secure identification of potential future occultations worthy of closer attention and refined predictions prior to each event.}

We organize our presentation as follows: Section 2 describes our procedure for identifying candidate occultations, and  Section 3 summarizes the geometrical quantities determined for each potential occultation. The main body of the paper is contained in Section 4, where predicted occultations are summarized for each of the six targets. We include tabulated statistics of the frequency of occultations by stellar magnitude and year, as well as representative detailed figures and tables for a subset of our identified events. Complete figures and tables for all events are included in the Supplementary Online Material (SOM). 
In the final section, we compare the {\it Gaia} and 2MASS star positions, discuss the opportunities for spacecraft occultations, identify some of the uncertainties in the predictions, especially for events far in the future, and highlight the important of continued occultation surveillance of our chosen targets. The Appendix describes the contents of the SOM, including documentation of the machine-readable tables.


\section{Identification of Occultation Candidates}
Previous occultation predictions for the outer planets often required dedicated astrometry and photometry of candidate stars \citep{KM1977, KME1981, Mink1985, Nicholson1988,KM1991,Mink1992}. \cite{Bosh1992} identified stellar occultation candidates for Saturn from the Guide Star Catalog for 1991-1999, and subsequent online predictions for 2000-2009 for Jupiter, Saturn, Uranus, and Neptune by A.~Bosh were posted at \url{http://www2.lowell.edu/users/amanda/occs2000/}. More recently, \cite{Mink1995} made use of the PPM catalog \citep{Roeser1988, Roeser1991, Roeser1993} to produce a statistical overview of stellar occultations between 1950-2050 and summary online tables of predicted events for 2000-2050 with closest-approach distances less than $30''$ for Jupiter\footnote{\url{http://tdc-www.harvard.edu/occultations/jupiter/jupiter.ppm2000.html}}, Saturn\footnote{\url{http://tdc-www.harvard.edu/occultations/saturn/saturn.ppm2000.html}}, Uranus\footnote{\url{http://tdc-www.harvard.edu/occultations/uranus/uranus.ppm2000.html}}, and Neptune\footnote{\url{http://tdc-www.harvard.edu/occultations/neptune/neptune.ppm2000.html}}. \cite{Saunders2022} identified several promising Uranus and Neptune occultations between 2025-2035, including SNR estimates for ground-based and space-based observations.  For the present survey, we expand on these prediction lists by making use of the {\it Gaia} DR3 catalog for stellar positions, proper motions, and G and RP magnitudes \citep{Gaia2022}, for the period 2023-2050. {\red Proper motions in the DR3 catalog are a factor of 2 more accurate than in the DR2 catalog, significantly reducing the prediction uncertainty the smaller targets Titan and Triton towards the end of our prediction period.} We make use as well of the Two Micron All-Sky Survey (2MASS) \citep{Skrutskie2006} for apparent stellar magnitudes in the K band ($\lambda\sim2.2~\mu$m). 

Most high-SNR Earth-based stellar occultations of the outer planets have been observed {\red from large ground-based telescopes} at IR or near-IR wavelengths, taking advantage of the strong methane absorption band near $\lambda=2.2~\mu$m or the weaker methane band near $\lambda=0.89~\mu$m to reduce the observed brightness of the planet relative to the occultation star. With the expectation that future high-SNR outer planet occultations are likely to be observed using IR-sensitive cameras, we used the 2MASS catalog for our initial survey, restricting the K band magnitude to K$\leq$10 or brighter for Jupiter and Saturn, and K$\leq$15 or brighter for Uranus, Neptune, Titan, and Triton. Although scientifically useful observations are possible for stars fainter than K=10 for Jupiter and Saturn, there are many much brighter stars in our prediction list that should provide ample opportunity for repeated observations of Jupiter and Saturn events in the coming decades. For Uranus {\red (in 2023, K=12.76 and V=5.88; in 2050, K=12.36 and V=5.48)} and Neptune {\red (in 2023, K=12.42 and V=7.79; in 2050, K=12.38 and V=7.75)}, past observations have proven scientifically useful for stars as faint as K$\sim$12, but there are relatively few bright events predicted for the coming decade. Given the current interest in planning for a possible ice giant mission in the relatively near future, and with the prospect of larger telescopes and improved instrumentation in the coming years, we have chosen to set a rather faint K band limit for these two planets. We include K, G, and RP magnitudes so that observers can estimate the expected SNR based on their choice of observing wavelength. {\red Titan and Triton (in 2023, K=12.30 and V=13.49; in 2050, K=12.26 and V=13.45) are less affected by the background brightness of their central planets, and scientifically useful occultation observations of these objects can be obtained using moderate-sized telescopes at visual wavelengths. We tabulate the frequency of occultations by these targets as a function of both K and G magnitudes.}

\subsection{Geocentric Predictions}
We employed our well-tested occultation code (RINGFIT) to compute the geometry of potential occultations, adopting a solar system barycenter inertial reference frame as described in \cite{French1993} and modified very slightly by \cite{French2017}. We used NASA's Navigation and Ancillary Information Facility planetary ephemerides (kernel files) and SPICE toolkit \citep{Acton1996} to compute the geocentric apparent positions of the six targets from 2023 to 2050. The planetary and satellite ephemerides used in this study are listed in {\bf Table~\ref{tbl:kernels}}.  We used the planetary constants file {\tt pck00010.tpc} to determine the (possibly time-variable) pole direction and the equatorial and polar radii of the targets.

	\begin{table*} [ht]
	\scriptsize
	\begin{center} 
	\caption{Spice kernels}
	\label{tbl:kernels} 
	\begin{threeparttable}
	\centering
	\begin{tabular}{l l }\hline
	File name & Contents\\
	\hline 
{\tt naif0012.tls} & leap-seconds \\
{\tt de440.bsp} & {\red Solar system ephemeris for all but Uranus events}\\
{\tt jup365.bsp} & Jupiter ephemeris\\
{\tt sat440l.bsp}  & Saturn and Titan ephemerides\\
{\red {\tt ura161.bsp}} & {\red Uranus ephemeris for Uranus events}\\
{\red {\tt peph.ura160.bsp}} & {\red Solar system ephemeris for Uranus events}\\
{\tt nep097.bsp} & Neptune and Triton ephemerides\\
{\tt nep101.bsp} & Neptune ephemeris \\
{\tt earth\_720101\_070426.bpc} & ITRF Earth rotation\\
{\tt earth\_200101\_990628\_predict.bpc} &ITRF Earth rotation\\
{\tt pck00010.tpc}  & planetary constants\\
 	 \hline
	\end{tabular}
  \end{threeparttable}
\end{center} 
\end{table*}

Next, using a subset of the 2MASS catalog restricted to the ecliptic region, we identified all stars brighter than our chosen K magnitude limits that were within 4$''$ of each target's path, based on the 2MASS positions at the catalog epoch of 1996.0. We chose this rather large impact parameter to account for the design specification of $0.5''$ positional accuracy of the 2MASS catalog relative to ICRS, the angular extent of the planetary targets and ring systems, the Earth's angular size as viewed from the target, estimated ephemeris uncertainties, and the effect of potentially large proper motions between the catalog epoch and the times of the predicted events.

With this initial list of candidate stars, we used the {\tt astroquery.vizier} Python interface to query the online VizieR 
 catalog \citep{Ochsenbein2000} to identify the {\it Gaia} DR3 stars that provided the closest matches to the 2MASS star positions. Given the depth of the {\it Gaia} survey, there often remained some ambiguity of the match between a given 2MASS star and its closest {\it Gaia} counterpart. We limited the {\it Gaia} candidates to those with magnitude G$\leq$15 for the Jupiter and Saturn candidate searches, and  G$\leq$19  for the {\red fainter} targets. {\red To find a matching 2MASS catalog entry, we} applied proper motion corrections for the interval between the {\it Gaia} DR3 catalog epoch (J2016) and the 2MASS epoch (J1996), and retained events for which the closest candidate {\it Gaia} star was within {\red an angular separation} $_{\_}r$=1~arcsec of the 2MASS catalog position at epoch. We included a magnitude-dependent correction for the proper motion derived from the {\it Gaia} EDR3 catalog \citep{Cantat2021}, but equally applicable to the DR3 proper motions as well (personal communication T. Contat-Gaudin). We also took account of propagated position error estimates, using the prescription of \cite{Butkevich2014}. Our final identification of the match between the 2MASS and {\it Gaia} catalog entries was based on the proximity of the 2MASS and {\it Gaia} positions and on their relative G and K magnitudes. In the detailed description of our predictions below, we note instances where the {\it Gaia} positions themselves are potentially of reduced accuracy, due to uncertainties in proper motion or in the original catalog astrometry. In Section \ref{section:discussion}, we discuss the distribution of position offsets between the two star catalogs.

As an independent check of our selection algorithm and occultation geometry calculations, we compared our results with geocentric predictions using the open source 
Stellar Occultation Reduction and Analysis package (SORA) \citep{Gomes2022} and the kernel files listed in Table~\ref{tbl:kernels}. Both approaches returned virtually identical sets of candidate occultations. In nearly all cases, the calculated closest approach times of the occultation chord agreed to within a few seconds, and the closest approach sky plane separation of the occultation chord and the target center agreed to within a few km. This level of agreement is quite sufficient for our present purposes. Other occultation prediction software is available from the International Occultation Timing Association (IOTA)\footnote{\url https://occultations.org/observing/software/}, and the geometry of occultation circumstances can also be computed using the NASA/JPL Horizons software.\footnote{\url https://ssd.jpl.nasa.gov/horizons/} Finally, \cite{Yuan2017} have developed an analytic geometry approach to predicting 
ground-based stellar occultations by ellipsoidal solar system bodies. All of our cross-checks indicate that our prediction method is robust at the km level in the sky plane for our target objects.

\subsection{Topocentric Predictions}

The next step in our procedure was to evaluate the observability of each occultation from various locations on Earth {\red representing six geographical regions.}
 As a starting point, we selected a set of 13 observatories around the world, including several that have  been used extensively to observe occultations in the past or for previous occultation predictions \citep{Nicholson1988}. 
{\bf Table \ref{tbl:obslocs}} lists the WGS84 locations of the selected observing sites and their corresponding geographical region and region ID. {\red (In what follows, we will refer to individual sites by their letter codes from this table.) In many cases, the listed sites are close to other major telescopes.} Our current effort is complementary to Lucky Star and IOTA\footnote{\url https://occultations.org/observing/occultation-predictions/major-planet-occultation-predictions/} prediction efforts, which focus instead on relatively short-term observing campaigns, {\red most often for targets that lack accurate long-term ephemerides. We augment these efforts by providing longer-term predictions for the four giant planets, including the detailed circumstances of the ring occultations for Saturn, Uranus, and Neptune, and by presenting decadal timescale predictions for the largest moons of Saturn and Neptune.}

\begin{table*} [ht]
\scriptsize
\caption{Observing Sites}
\label{tbl:obslocs} 
\centering
\begin{tabular}{l l r r r l l}\hline
Code & Name & lat. (deg) & E lon. (deg) & alt (m)  & Region ID & Region name\\
\hline
PIC & Pic du Midi & $+42.9365556 $ &   0.1423056 &  2890 & ENA & Europe \& N. Afr. \\
PAL & Palomar Mtn. (5m) & $+33.3562222 $ & 243.1360056 &  1706 & NAM & North Am. \\
PMO & Purple Mtn. Obs. Nanking & $+32.0666667 $ & 118.8208889 &   364 & EAS & East Asia \\
KPNO & Kitt Peak Natl. Obs. & $+31.9583333 $ & 248.4033333 &  2120 & NAM & North Am. \\
MCD & McDonald Obs. 2.7m & $+30.6715833 $ & 255.9784444 &  2075 & NAM & North Am. \\
TEN & Teide Obs./Tenerife & $+28.3005000 $ & 343.4890944 &  2395 & ENA & Europe \& N. Afr. \\
IRTF & Mauna Kea/IRTF & $+19.8262222 $ & 204.5280000 &  4168 & NAM & North Am. \\
KAV & Kavalur Observatory & $+12.5755556 $ &  78.8316667 &   722 & EAS & East Asia \\
RIO & Rio de Janeiro & $-22.8950556 $ & 316.7770833 &    33 & SAM & South Am. \\
ESO & European Southern Obs. (3.6m) & $-29.2609722 $ & 289.2683056 &  2400 & SAM & South Am. \\
AAT & Siding Spring (AAT) & $-31.2770278 $ & 149.0660833 &  1164 & OCN & Oceania \\
SAAO & So. Afr. Astro. Obs. (Sutherland) & $-32.3795278 $ &  20.8107000 &  1768 & SAF & Southern Africa \\
MSO & Mt. Stromlo Observatory & $-35.3200000 $ & 149.0083333 &   770 & OCN & Oceania \\
\hline
\end{tabular}
\end{table*}

\section{Occultation Geometry}
We restricted our set of occultations to those with Sun-Geocenter-Target (SGT) angle $\geq$45$^\circ$, which eliminated daytime occultations from our survey. For each surviving predicted occultation, we computed the apparent sky plane chord relative to the target center as observed from each of the 13 sites, and from the geocenter. For the topocentric observers, we also computed the elevation angle of the target object, the Sun, and the moon over the course of the occultation. We judged an occultation to be observable during any interval when the Sun was more than 5$^\circ$ below the horizon and the target object was more than 5$^\circ$ above the horizon. Throughout the paper, we refer to these as our usual altitude constraints on observability.

 We assigned a {\it planet event type} to each candidate occultation, indicating the observability of an ingress or egress occultation by the target limb, as described in {\bf Table \ref{tbl:PEventType}}. Any potential occultation of Jupiter, Titan, or Triton of planet event type {\it X} was eliminated from further consideration.
 
\begin{table*} [ht]
\scriptsize
\caption{\red Planet event types for all targets}
\label{tbl:PEventType} 
\centering
\begin{tabular}{l l }\hline
Type & Description\\
\hline
{\it P} & Observable from Earth, but not  from geocenter or any of 13 topocentric sites\\
{\it Pg} & Observable from Earth geocenter but not from any of 13 topocentric sites\\
{\it Pt} & Observable from at least one topocentric sites but not from Earth geocenter\\
{\it Pgt} & Observable from Earth geocenter and at least one topocentric site\\
{\it X} & No planet limb observable from Earth\\
\hline
\end{tabular}
\end{table*}

For Saturn, Uranus, and Neptune events, we also assessed the visibility of possible ring occultations. {\bf Table \ref{tbl:rings}} lists the rings,  semimajor axes or boundaries $a$, and widths $W$ we assumed for our predictions \citep{French2017,DePater2018a,French2023b}. We do not make detailed predictions for possible Jupiter ring occultations, but we do include the main Jupiter ring in our sky plane figures to show locations of the occultation chords relative to the ring system.

\begin{table*} [ht]
\scriptsize
\caption{Planetary {\red ring characteritics}}
\label{tbl:rings} 
\centering
\begin{tabular}{l l r r}\hline
Planet & Ring & $a$ (km)  & $W$ (km)\\
\hline
Jupiter & main ring & 129000& \\
Saturn & C inner edge & 74490& \\  
& B inner edge & 91983& \\
& A inner edge& 122052& \\
& Encke Div inner edge& 133424&\\
& Encke Div outer edge& 133745&\\
& A outer edge& 136774& \\
& F& 140461& \\
Uranus & 6 & 41837.6&1.5\\
  & 5 & 42235.3&2.3 \\
  & 4 & 42571.5&2.3 \\
  & $\alpha$ & 44719.0&8.5 \\
  & $\beta$ & 45661.4 &9.5 \\
  & $\eta$ & 47176.3& 1.6\\
   & $\gamma$ & 47626.8&2.2 \\
   & $\delta$ & 48300.6& 4.6\\
   & $\lambda$ & 50024.9&2.3 \\
   & $\epsilon$ & 51149.6& 58.1\\
Neptune & Galle & 42000&2000 \\
& LeVerrier & 53200& 100\\
& Lassell & 55200& 4000\\
& Arago & 57200& \\
& Galatea & 61953& \\
& Adams & 62933& 15\\
\hline
\end{tabular}
\end{table*}

For Saturn, Uranus and Neptune, the planetary event type is appended with with a {\it ring event type} as defined as in {\bf Table \ref{tbl:REventType}}, depending on the observability of at least one ingress or egress ring occultation event. Any occultation with a combined planetary and ring event type of {\it XX} was eliminated from further consideration.
\begin{table*} [ht]
\scriptsize
\caption{\red Ring event types for Saturn, Uranus, and Neptune}
\label{tbl:REventType} 
\centering
\begin{tabular}{l l }\hline
Type & Description\\
\hline
{\it R} & Observable from Earth, but not  from geocenter or any of 13 topocentric sites\\
{\it Rg} & Observable from Earth geocenter but not from any of 13 topocentric sites\\
{\it Rt} & Observable from at least one topocentric sites but not from Earth geocenter\\
{\it Rgt} & Observable from Earth geocenter and at least one topocentric site\\
{\it X} & No ring occultation observable from Earth\\
\hline
\end{tabular}
\end{table*}

\section{Example predictions}
\begin{figure}
\gridline{\rotatefig{90}{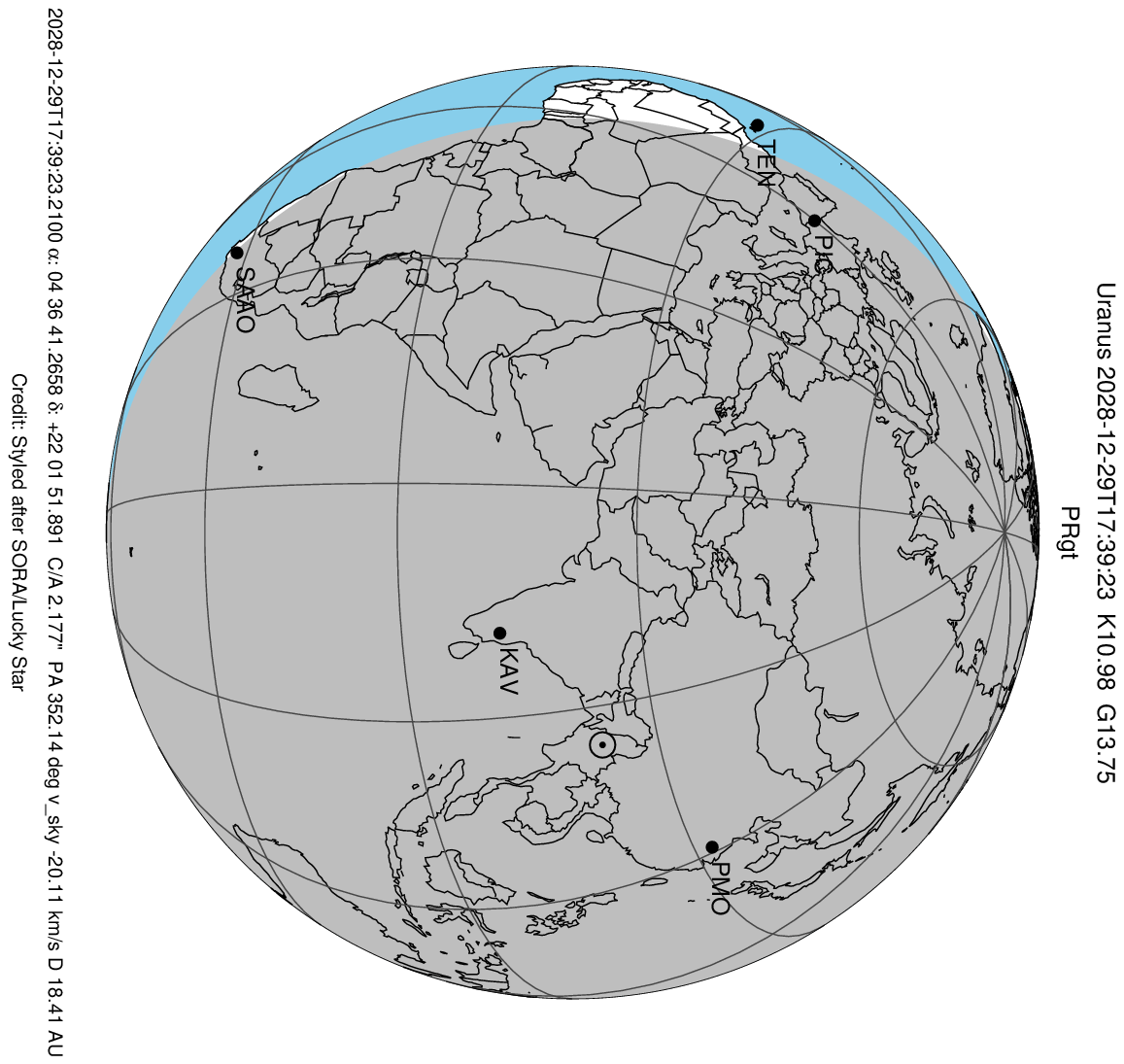}{0.6\textwidth}{}}
\gridline{\rotatefig{90}{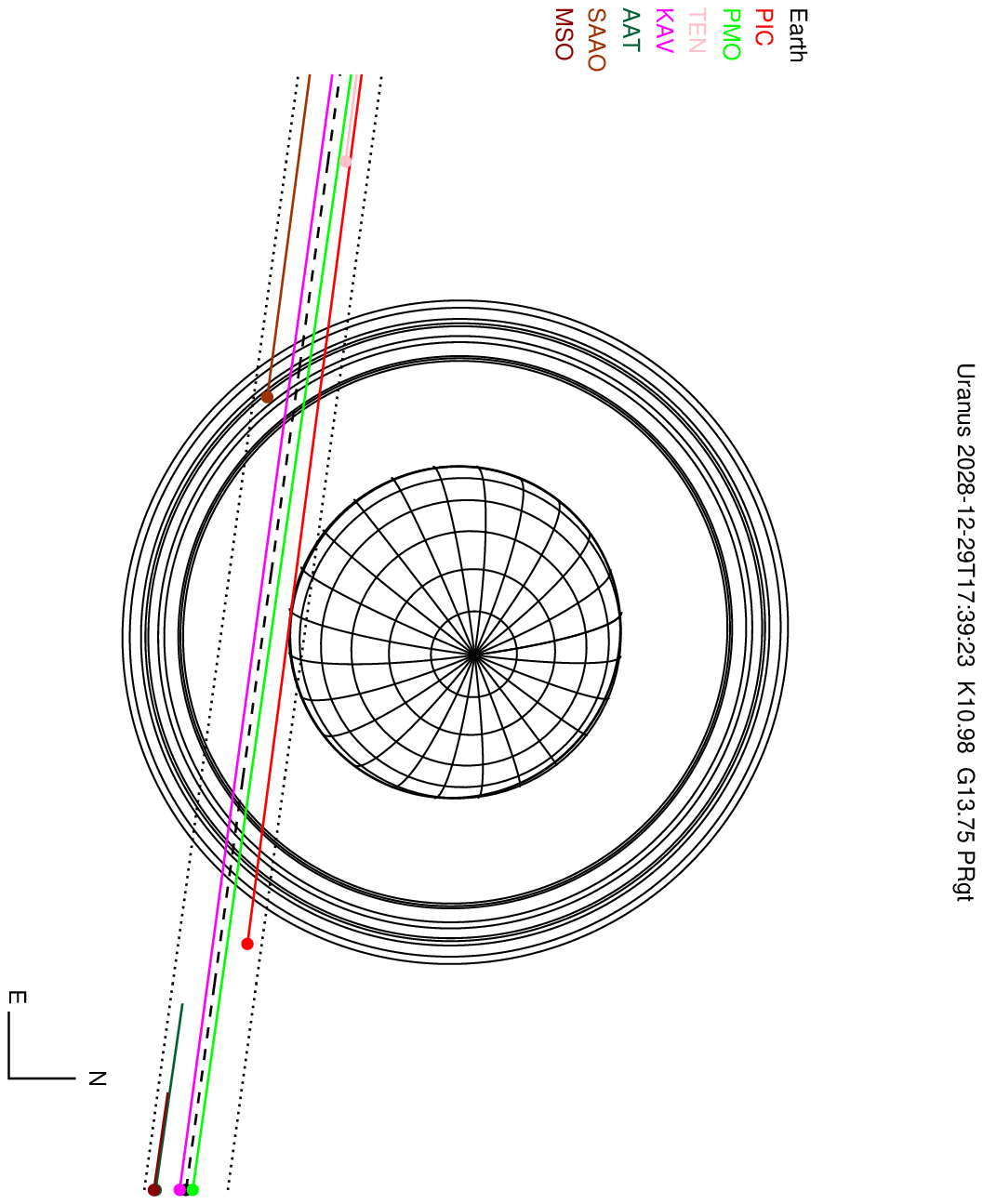}{0.6\textwidth}{}}
\caption{Top: View of Earth from Uranus at the closest approach time of the 2028-12-29 occultation. Bottom: The view of Uranus and its rings from Earth at the midtime of the occultation, showing the sky-plane chords for the labeled observatories. The geocenter path is shown dashed, and the dotted lines on either side demarcate the Earth's projected diameter.}
\label{fig:examples}
\end{figure}

For each surviving occultation on our prediction list, we produced several figures to provide visual overviews of the occultation circumstances. We illustrate these using a predicted Uranus occultation for  {\red 2028-12-29 (K=10.98, G13.75, event type {\it PtRgt}).}
{\bf Figure \ref{fig:examples} (Top)}
shows a shaded view of Earth and the continents as seen from Uranus at the time of the closest approach (C/A) of the geocentric occultation chord to the target center, including the labeled observing sites and the anti-solar point as an open circle. In this instance, the geocentric C/A occurred at 2028-12-29 17:37:23 UTC. Additional details of the event are shown in the figure labels: the K and G magnitudes, the apparent geocentric coordinates of the target star in J2000 coordinates, the closest approach separation in arcsec, the position angle on the sky of the closest approach point (PA), measured North through East, the apparent sky plane {\red velocity ($v_{\rm sky}$) of the target relative to the star at the C/A time,} and $D$, the target distance from Earth in AU. {\red The gray shading marks the region for which the Sun is below the horizon. (For the small targets Titan and Triton, we show where the occultation shadow falls on Earth, rather than a sky plane Earth view of the target.)}

For Jupiter, Saturn, Uranus, and Neptune occultations, we provide a figure showing the Earth view of the target and its ring system in the sky plane, along with occultation chords for any sites during the interval when the target and Sun meet the altitude requirements enumerated above. (For Titan and Triton occultations, we include the projected path of the shadow across the Earth.) {\red {\bf Figure \ref{fig:examples} (bottom)} shows the Uranus sky plane for the sample event. Each occultation chord is coded by the color of the observing site label. The solid dot on each chord marks the earliest point at which the occultation star is within the window of the figure and meets the Sun and target altitude observability requirements. The geocentric chord (marked as Earth) is shown as a black dashed line, bounded by the dotted lines showing the Earth's diameter. In this example, the SAAO chord intersects the outer rings only during egress because the planet was too low in the sky for earlier observations. Only the northernmost PIC chord has a planet occultation.}
 
\begin{figure}
\centerline{\resizebox{4in}{!}{\includegraphics[angle=90]{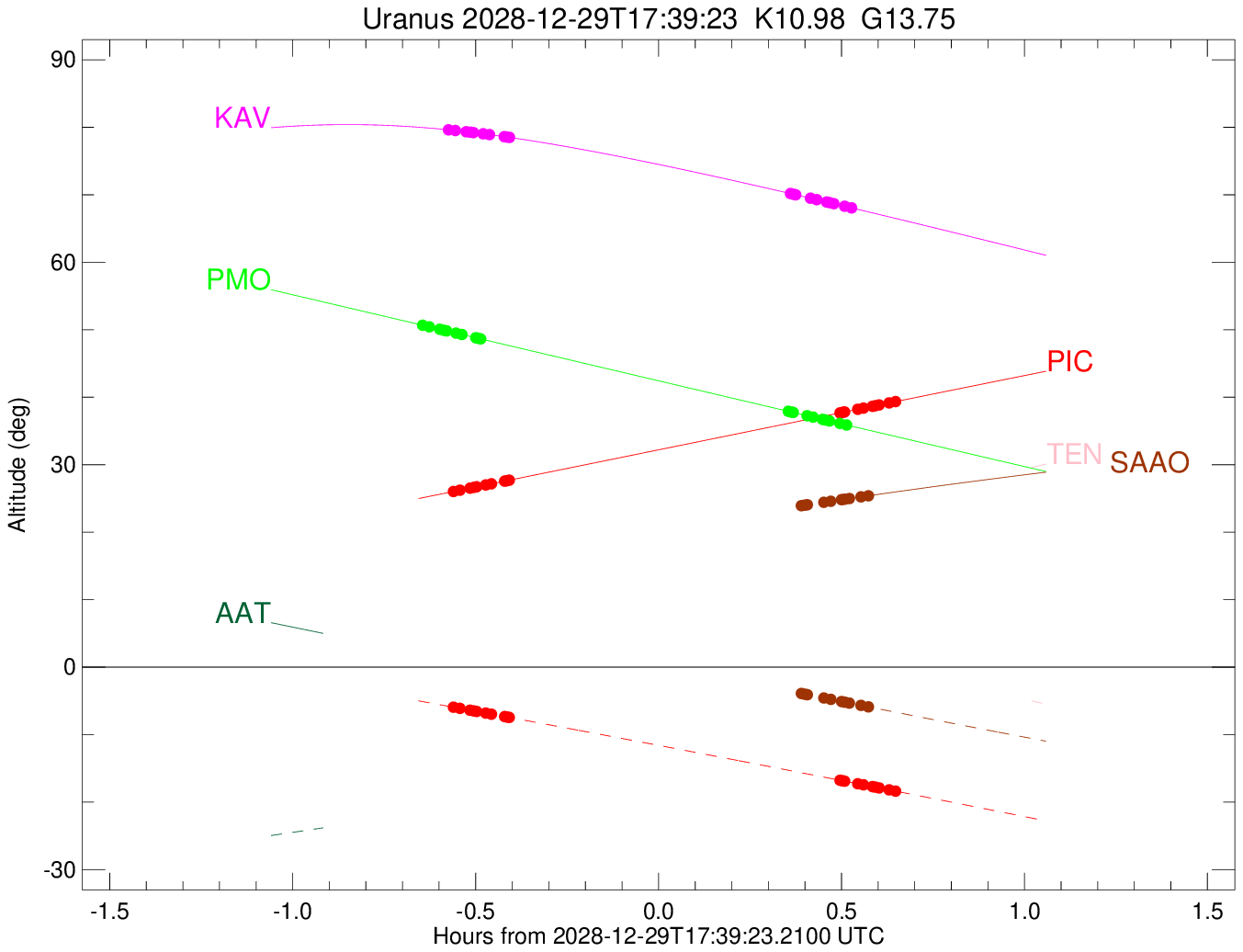}}}
\caption{Altitude of Uranus (solid lines) and the Sun (dashed lines) relative to the the horizon during the 2028-12-29 Uranus occultation, for the labeled observing sites (Table \ref{tbl:obslocs}). See text for details}
\label{fig:altexample}
\end{figure}

Additionally, for each occultation, we include a figure showing the altitude (elevation angle) of the target and Sun above/below the horizon over the course of the event, as viewed from each topocentric site that satisfies the event observability constraints at some point within the plotted vicinity of the closest approach time, as illustrated in {\bf Fig.~\ref{fig:altexample}} for the {\red 2028-12-29} Uranus occultation. For each labeled site, the altitude of Uranus is shown as a solid line and the altitude of theSunis shown as a dashed line. The time interval during which the planet itself occults the star is shown as a thick solid line, and the times of individual ring occultation events are shown as dots. The lines are restricted to the times that meet the simultaneous requirements that the target be at least 5$^\circ$ above the horizon and the Sun be at least 5$^\circ$ below the horizon. The color coding of the observing sites matches that of the corresponding sky plane figure (Fig.~\ref{fig:examples}, bottom panel). {\red For this occultation, KAV and PMO are well situated to observe the ingress and egress ring events at high elevation angle. The northernmost site, PIC, barely misses an atmosphere occultation, shortly after sunset. From SAAO, the planet is low in the sky and the egress ring events are observable shortly after sunset. No ring or atmosphere events are observable from AAT or MSO, where Uranus sinks to below 5$^\circ$ elevation before the ingress ring events, or from TEN, where the Sun is finally $5^\circ$ below the horizon about an hour after the closest approach time, well after the egress ring event times.}

\section{Predicted Occultations}

We now describe the predicted occultations for each of our six targets. Given the large number of events, we include in the main body of the paper a representative subset of the brightest events for most targets, along with summary statistics of the number of occultations by year and stellar magnitude for each target. Complete details for each occultation are included in the summary PDF and text files on the SOM, along with typeset tables of all occultations and machine-readable files for occultations by each target. The organization and contents of the SOM are described in the Appendix.
\begin{deluxetable}{l l l }
\tabletypesize{\scriptsize} 
\tablecaption{Definitions of tabulated results}
\label{tbl:defns}
\tablehead{
\colhead{Quantity\tablenotemark{\scriptsize a}}& 
\colhead{Units} &
\colhead{Description} }
\startdata
Event ID &  & Unique occultation identifier -- see text for details \\
C/A Epoch & UTC & Predicted time of geocentric sky plane chord closest approach to target center\\
ICRS Star Coord& hh mm ss dd mm ss & J2000 apparent geocentric star position at C/A Epoch\\
\ \ \ at C/A Epoch & & \\
$\sigma(\alpha_*)$ & km & Estimated E-W sky plane star position error -- see text for details\\
$\sigma(\delta_*)$ & km & Estimated N-S sky plane star position error -- see text for details\\
$f_0$ & km & Planetary ephemeris offset in the sky plane (east is positive)  \\
$g_0$ & km & Planetary ephemeris offset in the sky plane (north is positive)  \\
C/A & arcsec or mas  or 1000 km& Closest approach separation of geocentric sky plane chord and target center \\
C/A$_p$ & arcsec or 1000 km& Sky plane separation of central planet and occultation star at C/A (Triton/Titan only)\\
PA & deg & Position angle (CCW from north) of C/A point on geocentric sky plane chord \\
\red $v_{\rm sky}$ & km s$^{-1}$ & Geocentric sky plane velocity of target relative to star at C/A \\
Dist & au & Geocentric distance to target at C/A \\
Lat I & deg & planetodetic target latitude at ingress intersection of geocentric sky plane chord\\
& &  and possibly oblate target limb\\
Lat E & deg & planetodetic target latitude at egress intersection of geocentric sky plane chord\\
& &  and possibly oblate target limb\\
K & mag & Apparent K magnitude of occultation star, from 2MASS catalog\\
G & mag & Apparent G  magnitude of occultation star, from {\it Gaia} DR3 catalog\\
G$_*$ & mag & Apparent G magnitude, corrected for sky plane velocity -- see text for details\\
RP & mag & Apparent RP  magnitude of occultation star, from {\it Gaia} DR3 catalog\\
E $\lambda$ & deg & East longitude of sub-Earth point at C/A\\
$\phi$ & deg & Geocentric latitude of sub-Earth point at C/A\\
S-G-T & deg & Sun-Geocenter-Target angle at C/A\\
M-G-T & deg & Moon-Geocenter-Target angle at C/A\\
Event type & & Planet ($P$) and ring ($R$) event characteristics: \\
& & $g$ indicates geocentric event, $t$ indicates topocentric -- see text for details\\
Source ID & & {\it Gaia} DR3 catalog ID \\
DUP & & Source with multiple source identifiers ({\it Gaia} catalog entry)\\
RUWE\tablenotemark{\scriptsize a} & & Renormalized unit weight error ({\it Gaia} catalog entry)\\
2MASS ID\tablenotemark{\scriptsize a} & & 2MASS catalog ID\\
2MASS DUPFLAG\tablenotemark{\scriptsize a} & & if 1, another nearby event with a different STARID shares this 2MASS catalog ID\\
Geocentric Object Pos. & hh mm ss dd mm ss & J2000 apparent geocentric target position at C/A Epoch\\
B & deg & Ring/equatorial plane opening angle as viewed from Earth\\
P & deg & Position angle (CCW from north) of target pole, projected in the sky plane\\
$\dot r$ & km s$^{-1}$ & ring plane radial velocity at geocentric C/A (for ringed targets) \\
$D_*$ & km & Estimated projected diameter of star at target, from SORA -- see text for details\\
$_{\_}r$\tablenotemark{\scriptsize a} & arcsec &angular separation between proper motion-corrected {\it Gaia} position and\\
& & 2MASS star at 2MASS catalog epoch (J1996)\\
{\tt XXX\_N\_TARGETOCCS}\tablenotemark{\scriptsize a} & & Number of observable target occultations by candidate sites in XXX region\\
{\tt XXX\_N\_RINGOCCS}\tablenotemark{\scriptsize a} & & Number of observable ring occultations by candidate sites in XXX region\\
\enddata
\tablenotetext{a} {Contained in machine-readable (MR) tables only.}
\end{deluxetable}

For convenient reference, {\bf Table \ref{tbl:defns}} defines the tabulated results included selectively for each target below, along with additional variables included in the machine-readable prediction files on the SOM. 
The tabulated entries are largely self-explanatory, with the following exceptions:
\begin{itemize}
\item{$\sigma(\alpha_*)$ and $\sigma(\delta_*)$ are estimated star position errors (expressed in km in the sky plane) propagated from the {\it Gaia} DR3 catalog epoch to the time of each occultation, from the method of \cite{Butkevich2014} as implemented in the SORA (using the method {\tt Star.error\_at}), for the appropriate parameters and covariance matrix from the {\it Gaia} EDR3 catalog. These position errors are typically quite small, and result from uncertainties in the proper motion of the star, but not from intrinsic uncertainties in the position of the star at the catalog epoch. The latter are reflected in the RUWE (renormalized unit weight error) values available for each occultation star in the SOM and flagged in the tables below for stars with large positional uncertainties. RUWE values above 1.4 are indicative of less accurate {\it Gaia} positions.
}
\item {$D_*$ is the estimated projected diameter of the star at the target distance. This quantity affects the measured sharpness of transitions in brightness of ring edges and airless target limbs, as well as the scintillation amplitude of planetary atmosphere occultations. The projected star size can be estimated from the color, spectral type, and spectral class of the star, although these estimates can be systematically in error if they do not account for interstellar reddening, which is often significant for occultation stars that lie in the crowded star fields of the Milky Way. For this work, we utilized the {\tt Star.apparent\_diameter} method of the SORA software package \citep{Gomes2022}, which successively interrogates online star catalogs to find the necessary auxiliary apparent magnitudes of the star to estimate the projected star size. In instances where there was ambiguity about whether the candidate catalog stars matched the {\it Gaia} star in question, we omitted the calculated star diameter from our tables.}
\item { $G_* $ is the velocity-corrected apparent G magnitude of the occultation star, according to the prescription 
\beq
G_* = G + 2.5 \log_{10}(v_{\rm{sky}}/20),
\label{eq:Gstar}
\eeq
\noindent where $v_{\rm{sky}}$ is the sky plane velocity in km~s$^{-1}$ \citep{Gomes2022}. This function increases the predicted SNR of slow occultations and downgrades the SNR of rapid occultations.}
\item{{\tt 2MASS DUPFLAG} is set to 1 if another nearby event occurring within one day has the same {\tt 2MASS ID}, indicating an ambiguity in the estimated K magnitude for such events. This occurs primarily when there are crowded star fields, resulting in uncertain matching between the {\it Gaia} star and the corresponding 2MASS star. Such events are flagged in the typeset prediction tables and this variable is included in the machine-readable event tables.}
\item{{\tt XXX\_N\_TARGETOCCS} and {\tt XXX\_N\_RINGOCCS} are contained in the machine-readable tables for each target. They provide an indication of which of the seven global regions listed in Table \ref{tbl:obslocs} are best suited to observe a given occultation. Here, {\tt XXX} is the region code, and the tabulated value indicates how many of the individual sites in Table \ref{tbl:obslocs} are predicted to have an observable target occultation (planet or satellite limb) or an individual ring occultation, subject to the standard altitude constraints. An additional region code -- GEO -- is included to denote geocentric predictions, which are not subject to altitude constraints.}

\end{itemize}

\subsection{Jupiter}\label{sec:Jupiter}

Five stellar occultations by Jupiter have been recorded photometrically in the modern era, beginning with the 1952-11-20 occultation of the V=5.5 star $\sigma$ Arietis \citep{Baum1953}, followed by the 1971-05-13 occultation of $\beta$ Scorpii (V=2.63, 4.92) \citep{Hubbard1972,Vapillon1973,Veverka1974,Raynaud2004}, the 1989-12-13 occultation of SAO 78505 (V=8.7, K=7) \citep{Hubbard1995}, the 1999-10-10 occultation of HIP 9369 (V=7.6) \citep{Raynaud2003}, and  the 2009-08-03/04 occultation of HIP 107302 (45 Cap, V=5.5) \citep{Christou2013}. 

{\red We identified a total of 1844 Jupiter occultations between 2023-2050 for stars brighter than K$\leq$10. {\bf Table \ref{tbl:jupoccpredstats}} lists the number of predicted Jupiter occultations per year.}
The frequency of occultations varies substantially in time, {\red depending on the density of the star fields traversed by Jupiter.} It is quite high for 2031 and 2043, with the interval reflecting Jupiter's orbital period of just under 12 years, as Jupiter crosses a dense region of the Milky Way. The transits of the Milky Way in 2024/25, 2036/37, and 2048/49 show much more modest increases in the frequency of events. Similar variations are seen in the prediction list of \cite{Mink1995}, which includes 57 geocentric Jupiter atmospheric occultations. Our search identified only a handful of Jupiter occultations by stars brighter than K=7 before 2031, and only five for K$\leq$4 between 2023-2050, but there are many events with K$\leq$9. 

\begin{figure}
\centerline{\resizebox{6in}{!}{\includegraphics[angle=0]{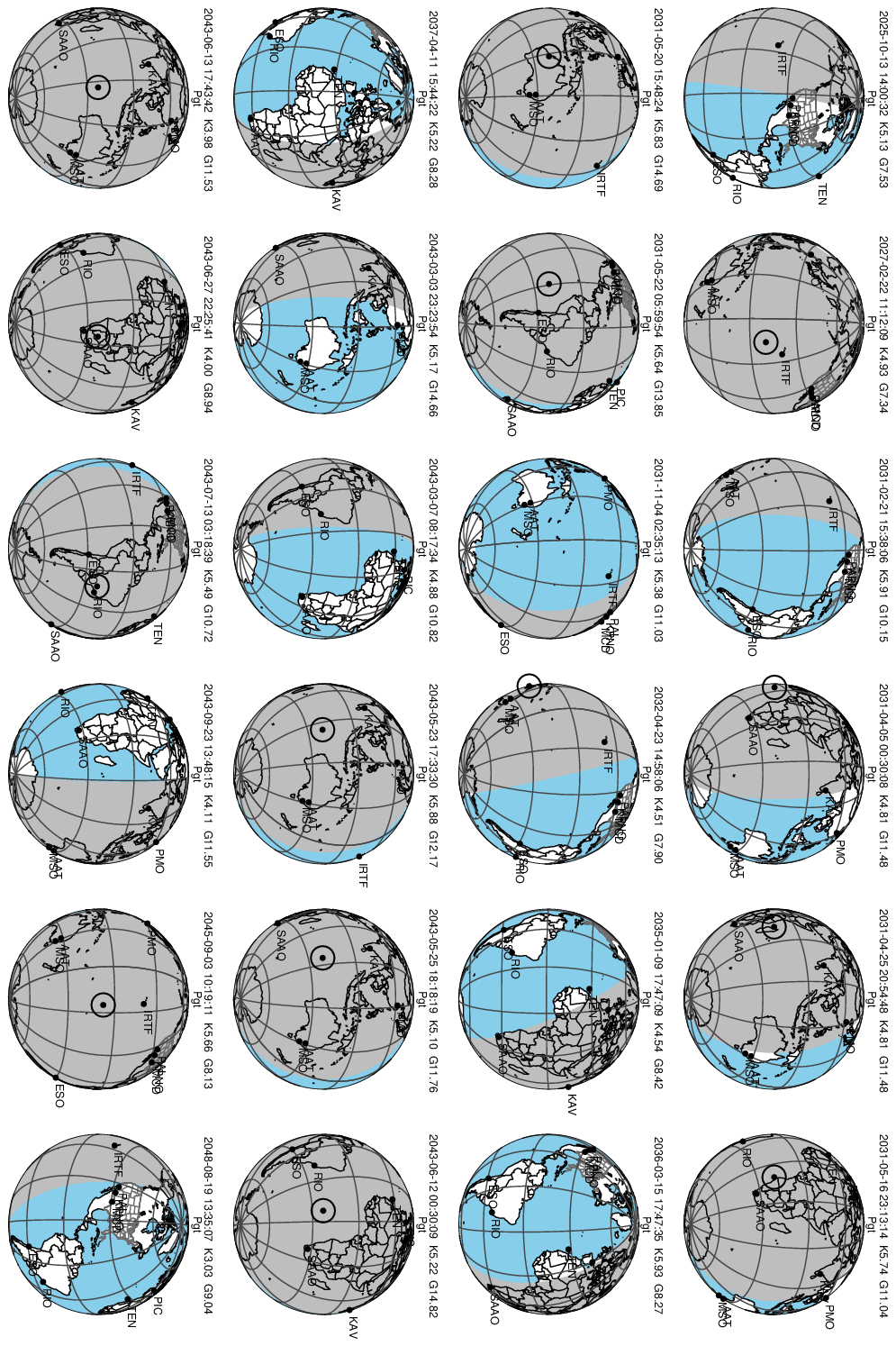}}}

\caption{{\red Gallery of Earth views from Jupiter at mid-occultation for the 24 brightest predicted occultations between 2023-2050 with planetary event type {\it Pgt}. See text for details.}}
\label{fig:JupiterGlobes000}
\end{figure}

\begin{figure}
\centerline{\resizebox{6in}{!}{\includegraphics[angle=0]{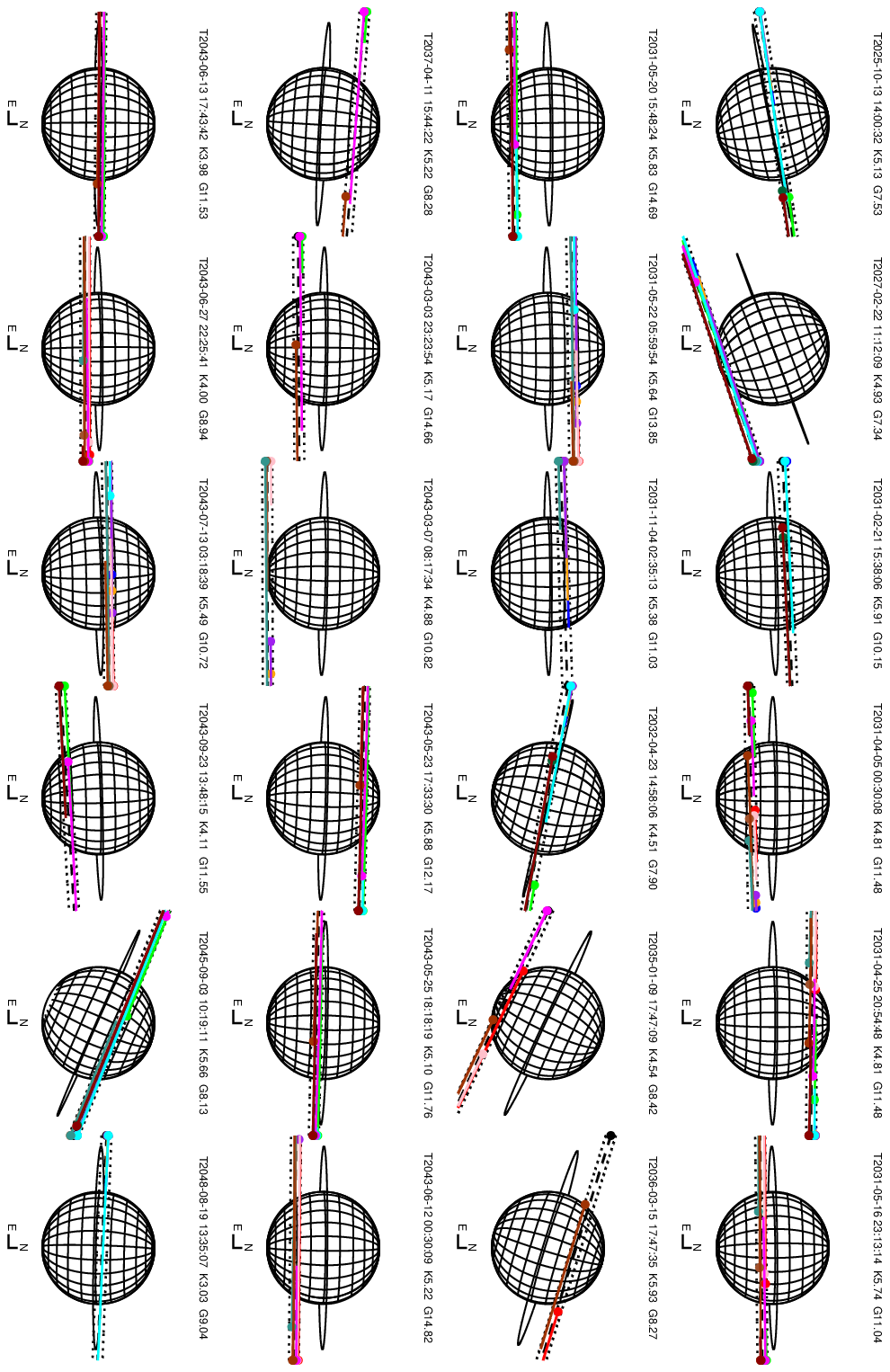}}}
\caption{{\red Gallery of sky plane geocentric views of Jupiter at mid-occultation for the 24 brightest predicted occultations between 2023-2050 with planetary event type {\it Pgt}.See text for details.}}
\label{fig:JupiterSky000}
\end{figure}

{\red To illustrate the range of event geometries and the varying aspect of Jupiter over this time interval, we include here the brightest (K mag) 24 Jupiter occultations between 2023-2050 with planetary event types $\it Pt$ or $\it Pgt$. {\bf Figure~\ref{fig:JupiterGlobes000}} shows a gallery of views of Earth at mid-occultation and {\bf Fig.~\ref{fig:JupiterSky000}} shows the corresponding sky-plane views of Jupiter.}
\begin{deluxetable}{c c c c c c c c }
\tabletypesize{\scriptsize} 
\tablecaption{Jupiter Occultations 2023--2050 }
\label{tbl:jupoccpredstats}
\tablehead{
\colhead{Year}& 
\colhead{K $<$4} &
\colhead{K 4--5} &
\colhead{K 5--6} &
\colhead{K 6--7} &
\colhead{K 7--8} &
\colhead{K 8--9} &
\colhead{K 9--10} \\[-.95em]
}
\startdata
2023 & -- & -- & -- & -- & 1 & 2 & 6 \\
2024 & -- & -- & -- & 2 & 3 & 7 & 15 \\
2025 & -- & -- & 1 & -- & 2 & 7 & 19 \\
2026 & -- & -- & -- & 1 & -- & 4 & 8 \\
2027 & -- & 1 & -- & -- & 1 & -- & 2 \\
2028 & -- & -- & -- & 1 & -- & 1 & 5 \\
2029 & -- & -- & -- & -- & -- & 2 & 2 \\
2030 & -- & -- & -- & 3 & 2 & 3 & 6 \\
2031 & 1 & 2 & 6 & 14 & 90 & 262 & 346 \\
2032 & -- & 1 & -- & 1 & 7 & 10 & 21 \\
2033 & -- & -- & -- & -- & 2 & 3 & 5 \\
2034 & -- & -- & -- & 2 & 1 & 1 & 5 \\
2035 & -- & 1 & -- & 1 & 2 & 3 & 5 \\
2036 & -- & -- & 1 & 1 & 4 & 6 & 26 \\
2037 & -- & -- & 1 & 3 & 2 & 11 & 15 \\
2038 & -- & -- & -- & -- & -- & 3 & 6 \\
2039 & -- & -- & -- & -- & 1 & 1 & 4 \\
2040 & -- & -- & -- & -- & 1 & 2 & 3 \\
2041 & -- & -- & -- & -- & -- & -- & 6 \\
2042 & -- & -- & 1 & -- & -- & 5 & 8 \\
2043 & 1 & 3 & 8 & 23 & 99 & 243 & 332 \\
2044 & -- & -- & -- & 2 & 1 & 5 & 12 \\
2045 & -- & -- & 1 & 2 & -- & -- & 6 \\
2046 & -- & -- & -- & 1 & 2 & 5 & 6 \\
2047 & -- & -- & -- & -- & 1 & -- & 3 \\
2048 & 1 & -- & -- & 2 & 5 & 13 & 26 \\
2049 & -- & -- & -- & 1 & 6 & 8 & 16 \\[0.5em]
Totals & 3 & 8 & 19 & 60 & 233 & 607 & 914 \\

\enddata
\end{deluxetable}

{\bf Table \ref{tbl:jupoccpredK6}} provides detailed information about the {\red 24 events shown in Figs.~\ref{fig:JupiterGlobes000} and \ref{fig:JupiterSky000}.} All of the items in the table are defined in Table \ref{tbl:defns}, with the exception of the Event ID of the form {\tt Tyynnn..}. Here, {\tt T} is a unique target identifier ({\tt J} for Jupiter, {\tt S} for Saturn, {\tt U} for Uranus, {\tt N} for Neptune, {\tt Ti} for Titan, and {\tt Tr} for Triton); {\tt yy} corresponds to the final two digits of the year of the event, and {\tt nnn..} is the chronologically increasing number of predicted occultations by a given target in the specified year. The complete table of predictions is contained in the SOM, with sequential Event IDs without gaps within any given year, although gaps in the numbering occur in the abbreviated tables included in the main body of the paper. 


\begin{longrotatetable}
\movetabledown=6mm
\begin{deluxetable}{c c c D D D D D D D D D D D}
\tablecolumns{14}
\tabletypesize{\tiny}
\tablecaption{Geocentric Jupiter Occultation Predictions 2023--2050}
\label{tbl:jupoccpredK6}
\tablehead{
\\[-2.75em]
\colhead{Event ID} & 
\colhead{C/A Epoch} &
\colhead{ICRS Star Coord at Epoch} & 
\multicolumn2c{$\sigma(\alpha_*)$ (km)} &
\multicolumn2c{C/A  (${'}{'}$)} & 
\multicolumn2c{PA (deg)} & 
\multicolumn2c{$v_{\rm sky}$ (km/s)} &
\multicolumn2c{Dist (au)} &
\multicolumn2c{Lat I (deg)} &
\multicolumn2c{K}& 
\multicolumn2c{G} & 
\multicolumn2c{RP} & 
\multicolumn2c{E $\lambda$ (deg)} & 
\multicolumn2c{S-G-T} \\[-.95em]
\colhead{Event type} & 
\colhead{Source ID} &
\colhead{Geocentric Object Position} &
\multicolumn2c{$\sigma(\delta_*)$ (km)} &
\multicolumn2c{C/A (10$^3$ km)} &  
\multicolumn2c{P (deg)} & 
\multicolumn2c{B (deg)} &
\multicolumn2c{$\dot r$  (km/s)} &
\multicolumn2c{Lat E (deg)} &
\multicolumn2c{$D_*$ (km)} &
\multicolumn2c{G${}_*$} & 
\multicolumn2c{DUP} & 
\multicolumn2c{$\phi$ (deg)} &
\multicolumn2c{M-G-T} 
}
\decimals
\startdata
J250025 & 2025-10-13 14:00:32.12 & 07 41 12.72000  +21 26 47.82082  & 1.0 & 1.093 & 188.48 & 13.98 & 5.144 & 0.4 & 5.132 & 7.526 & 6.792 & 243.2 & 86.7 \\ 
\ \ \ Pgt & 673253445252610816 & 07 41 12.70846  +21 26 46.74017  & 0.7 & 4.076 & 11.37 & 1.51 &  .  & 7.0 & 1.860 & 7.137 & 0 & 21.4 & 7.1 \\ 
J270002* & 2027-02-22 11:12:09.02 & 09 31 38.98066  +15 46 12.26555  & 5.6 & 20.375 & 18.85 & -16.71 & 4.383 & -66.0 & 4.926 & 7.341 & 6.619 & 183.1 & 166.9 \\ 
\ \ \ Pgt & 619023298387097856 & 09 31 39.43671  +15 46 31.54791  & 3.4 & 64.771 & 20.61 & -0.39 &  .  & -69.2 & 1.747 & 7.146 & 1 & 15.7 & 33.4 \\ 
J310050 & 2031-02-21 15:38:06.92 & 17 36 05.07176  $-$22 46 56.99064  & 1.4 & 4.626 & 182.43 & 24.36 & 5.611 & 18.1 & 5.910 & 10.153 & 9.019 & 238.6 & 67.9 \\ 
\ \ \ Pgt & 4116789654396613888 & 17 36 05.05760  $-$22 47 01.61267  & 0.7 & 18.825 & 1.74 & -2.66 &  .  & 16.5 & 2.181 & 10.367 & 0 & -22.8 & 67.8 \\ 
J310215 & 2031-04-05 00:30:08.82 & 17 52 55.94041  $-$22 53 37.68089  & 5.2 & 7.678 & 0.98 & 4.84 & 4.937 & -24.3 & 4.807 & 11.483 & 9.945 & 68.0 & 106.2 \\ 
\ \ \ Pgt & 4070200249968817792 & 17 52 55.94993  $-$22 53 30.00383  & 3.1 & 27.493 & -0.08 & -2.61 &  .  & -26.6 & 6.508 & 9.943 & 1 & -22.9 & 107.3 \\ 
J310218 & 2031-04-25 20:54:48.13 & 17 52 55.94041  $-$22 53 37.68078  & 4.9 & 14.726 & 179.85 & -4.56 & 4.635 & 47.6 & 4.807 & 11.483 & 9.945 & 101.3 & 126.7 \\ 
\ \ \ Pgt & 4070200249968817792 & 17 52 55.94324  $-$22 53 52.40699  & 2.9 & 49.504 & -0.08 & -2.61 &  .  & 47.4 & 6.110 & 9.878 & 1 & -22.9 & 173.5 \\ 
J310286 & 2031-05-16 23:13:14.27 & 17 47 10.16210  $-$22 53 33.02782  & 2.4 & 4.520 & 0.58 & -12.08 & 4.403 & -13.3 & 5.739 & 11.042 & 9.679 & 44.5 & 148.4 \\ 
\ \ \ Pgt & 4068933410786852736 & 17 47 10.16540  $-$22 53 28.50815  & 1.4 & 14.432 & 0.55 & -2.61 &  .  & -13.2 & 2.412 & 10.494 & 0 & -22.9 & 98.0 \\ 
J310312 & 2031-05-20 15:48:24.05 & 17 45 38.86109  $-$22 53 28.30089  & 6.7 & 14.034 & 0.71 & -13.08 & 4.372 & -42.1 & 5.827 & 14.685 & 13.940 & 151.7 & 152.3 \\ 
\ \ \ Pgt & 4068857162169473792 & 17 45 38.87365  $-$22 53 14.26747  & 3.9 & 44.504 & 0.71 & -2.61 &  .  & -42.1 &  .  & 14.224 & 0 & -22.9 & 145.1 \\ 
J310320 & 2031-05-22 05:59:54.18 & 17 44 57.16658  $-$22 52 56.78582  & 5.5 & 10.008 & 180.78 & -13.47 & 4.360 & 29.4 & 5.640 & 13.846 & 12.036 & 297.1 & 154.0 \\ 
\ \ \ Pgt & 4116890534598985728 & 17 44 57.15678  $-$22 53 06.79313  & 3.3 & 31.649 & 0.78 & -2.61 &  .  & 29.4 &  .  & 13.417 & 0 & -22.9 & 164.5 \\ 
J310705 & 2031-11-04 02:35:13.83 & 17 48 17.20920  $-$23 18 07.26401  & 4.3 & 4.775 & 181.72 & 33.83 & 5.867 & 20.1 & 5.384 & 11.027 & 9.670 & 185.6 & 46.3 \\ 
\ \ \ Pgt & 4068768518408378624 & 17 48 17.19876  $-$23 18 12.03675  & 2.6 & 20.318 & 0.42 & -2.20 &  .  & 17.2 & 4.082 & 11.598 & 0 & -23.3 & 170.2 \\ 
J320018* & 2032-04-23 14:58:06.36 & 20 11 35.30687  $-$20 13 00.46270  & 3.6 & 0.327 & 349.41 & 11.96 & 5.019 & 3.1 & 4.506 & 7.896 & 6.941 & 226.6 & 92.9 \\ 
\ \ \ Pgt & 6866344807161369856 & 20 11 35.30260  $-$20 13 00.14155  & 2.5 & 1.189 & -14.24 & -1.13 &  .  & -5.3 & 3.073 & 7.338 & 0 & -20.1 & 110.8 \\ 
J350001 & 2035-01-09 17:47:09.76 & 00 25 26.06476  +01 22 09.00817  & 2.5 & 16.455 & 334.89 & 19.93 & 5.071 & -60.8 & 4.537 & 8.424 & 7.289 & 350.9 & 77.6 \\ 
\ \ \ Pgt & 2547038618688047232 & 00 25 25.59910  +01 22 23.90770  & 1.5 & 60.512 & -25.24 & 2.32 &  .  & -61.1 & 3.543 & 8.420 & 0 & 1.6 & 76.4 \\ 
J360003 & 2036-03-15 17:47:35.37 & 03 01 40.65900  +16 18 54.99790  & 3.3 & 8.494 & 162.82 & 31.32 & 5.557 & 31.6 & 5.933 & 8.265 & 7.547 & 325.1 & 52.4 \\ 
\ \ \ Pgt & 34248069218428800 & 03 01 40.83331  +16 18 46.88312  & 3.0 & 34.231 & -16.98 & 2.89 &  .  & 32.1 & 1.367 & 8.752 & 0 & 16.5 & 170.0 \\ 
J370020 & 2037-04-11 15:44:22.59 & 05 21 58.59041  +22 57 13.24662  & 2.6 & 10.207 & 175.46 & 27.71 & 5.566 & 37.4 & 5.223 & 8.283 & 7.371 & 4.7 & 59.7 \\ 
\ \ \ Pgt & 3414783812488249088 & 05 21 58.64892  +22 57 03.07188  & 1.7 & 41.201 & -3.25 & 2.35 &  .  & 40.1 & 2.262 & 8.637 & 0 & 23.0 & 104.0 \\ 
J430151 & 2043-03-03 23:23:54.17 & 17 58 42.60177  $-$22 59 14.48648  & 17.1 & 7.795 & 0.31 & 21.91 & 5.508 & -27.8 & 5.166 & 14.657 & 12.899 & 117.7 & 73.0 \\ 
\ \ \ Pgt & 4069447879086908416 & 17 58 42.60478  $-$22 59 06.69112  & 11.4 & 31.140 & -0.70 & -2.51 &  .  & -30.0 &  .  & 14.756 & 0 & -23.0 & 18.5 \\ 
J430157 & 2043-03-07 08:17:34.33 & 18 00 34.94357  $-$22 59 30.34468  & 7.0 & 17.902 & 0.13 & 20.44 & 5.455 & -81.9 & 4.879 & 10.823 & 9.399 & 341.4 & 75.9 \\ 
\ \ \ Pgt & 4069438430158980864 & 18 00 34.94648  $-$22 59 12.44286  & 5.0 & 70.827 & -0.90 & -2.50 &  .  & -83.7 & 5.284 & 10.846 & 0 & -23.0 & 24.5 \\ 
J430238 & 2043-05-23 17:33:30.24 & 18 05 37.49681  $-$23 01 49.83575  & 7.0 & 15.665 & 178.58 & -12.59 & 4.370 & 47.6 & 5.876 & 12.169 & 10.655 & 127.4 & 150.6 \\ 
\ \ \ Pgt & 4069573871938648320 & 18 05 37.52495  $-$23 02 05.49575  & 4.7 & 49.653 & -1.44 & -2.43 &  .  & 47.7 & 3.097 & 11.667 & 0 & -23.0 & 32.4 \\ 
J430246 & 2043-05-25 18:18:19.03 & 18 04 45.87252  $-$23 02 25.38066  & 9.4 & 2.705 & 358.66 & -13.12 & 4.354 & -7.8 & 5.102 & 11.764 & 10.161 & 114.0 & 152.8 \\ 
\ \ \ Pgt & 4069584424693604736 & 18 04 45.86795  $-$23 02 22.67675  & 5.8 & 8.542 & -1.35 & -2.43 &  .  & -7.8 &  .  & 11.306 & 0 & -23.0 & 9.3 \\ 
J430271* & 2043-06-12 00:30:09.82 & 17 56 11.96283  $-$23 04 28.69795  & 19.7 & 11.600 & 359.46 & -16.08 & 4.263 & -33.3 & 5.221 & 14.816 & 13.003 & 1.9 & 171.3 \\ 
\ \ \ Pgt & 4069408331058914560 & 17 56 11.95484  $-$23 04 17.09827  & 12.6 & 35.870 & -0.43 & -2.43 &  .  & -33.6 &  .  & 14.578 & 0 & -23.1 & 127.2 \\ 
J430274 & 2043-06-13 17:43:42.54 & 17 55 15.78203  $-$23 04 23.11536  & 11.1 & 0.806 & 179.54 & -16.20 & 4.259 & 2.4 & 3.985 & 11.534 & 9.923 & 101.6 & 173.1 \\ 
\ \ \ Pgt & 4070137715246205568 & 17 55 15.78250  $-$23 04 23.92118  & 7.1 & 2.489 & -0.33 & -2.43 &  .  & 2.1 & 11.331 & 11.305 & 0 & -23.1 & 104.8 \\ 
J430339 & 2043-06-27 22:25:41.69 & 17 47 25.33650  $-$23 04 45.90459  & 2.9 & 5.388 & 0.16 & -16.11 & 4.256 & -14.9 & 4.005 & 8.936 & 7.611 & 15.1 & 171.5 \\ 
\ \ \ Pgt & 4068789787093980672 & 17 47 25.33758  $-$23 04 40.51642  & 1.7 & 16.633 & 0.52 & -2.42 &  .  & -15.7 & 5.263 & 8.701 & 0 & -23.1 & 71.1 \\ 
J430433 & 2043-07-13 03:18:39.12 & 17 39 41.37292  $-$23 03 52.48115  & 3.1 & 3.900 & 180.58 & -13.77 & 4.318 & 12.1 & 5.488 & 10.722 & 9.429 & 285.0 & 155.3 \\ 
\ \ \ Pgt & 4116554874369464960 & 17 39 41.37006  $-$23 03 56.38046  & 1.8 & 12.212 & 1.35 & -2.40 &  .  & 10.3 & 2.499 & 10.317 & 0 & -23.1 & 75.7 \\ 
J430537 & 2043-09-23 13:48:15.37 & 17 37 39.31108  $-$23 14 07.51003  & 15.9 & 10.435 & 3.46 & 15.65 & 5.223 & -35.1 & 4.111 & 11.549 & 10.549 & 55.7 & 85.0 \\ 
\ \ \ Pgt & 4116713135287242624 & 17 37 39.35681  $-$23 13 57.09370  & 8.8 & 39.534 & 1.57 & -2.22 &  .  & -39.1 &  .  & 11.283 & 0 & -23.3 & 155.2 \\ 
J450004 & 2045-09-03 10:19:11.65 & 22 33 41.12388  $-$10 31 16.26580  & 2.6 & 8.605 & 157.89 & -15.83 & 3.997 & 21.0 & 5.663 & 8.127 & 7.389 & 201.2 & 175.2 \\ 
\ \ \ Pgt & 2602776092914341248 & 22 33 41.34351  $-$10 31 24.23785  & 2.4 & 24.947 & -23.93 & 1.29 &  .  & 25.0 & 1.143 & 7.874 & 0 & -10.3 & 82.5 \\ 
J480014 & 2048-08-19 13:35:07.18 & 05 38 13.82080  +22 49 12.67217  & 5.8 & 1.138 & 177.71 & 26.78 & 5.469 & 3.2 & 3.030 & 9.043 & 7.717 & 272.9 & 61.6 \\ 
\ \ \ Pgt & 3404219910928409856 & 05 38 13.82409  +22 49 11.53468  & 3.2 & 4.516 & -1.51 & 2.64 &  .  & 5.0 & 14.563 & 9.360 & 0 & 22.8 & 174.7 \\ 
\enddata
\end{deluxetable}
\end{longrotatetable}

As noted previously, an indication of the relative accuracy of the {\it Gaia} DR3 star positions is given by the RUWE (renormalized unit-weight error) catalog entry. Values of RUWE $\leq$1.4 indicate a reliable astrometric solution.\footnote{https://doi.org/10.17876/gaia/dr.2/45} The RUWE is included in the machine-readable prediction files on the SOM, but to provide an indication of events with less reliable astrometry, we add a superscript of $^*$ to the Event ID of any occultation with RUWE $>1.4$, $^{**}$ for RUWE$>2$, and $^{***}$ for RUWE$>5$. In Section \ref{section:discussion} below, we evaluate the influence of the value of the RUWE on the offsets between the {\it Gaia} and 2MASS star positions. {\red A superscript {$^{a}$} is appended to the Event ID when a nearby event has the same 2MASS ID, resulting in a corresponding K magnitude ambiguity.}

\begin{figure}
\gridline{\rotatefig{90}{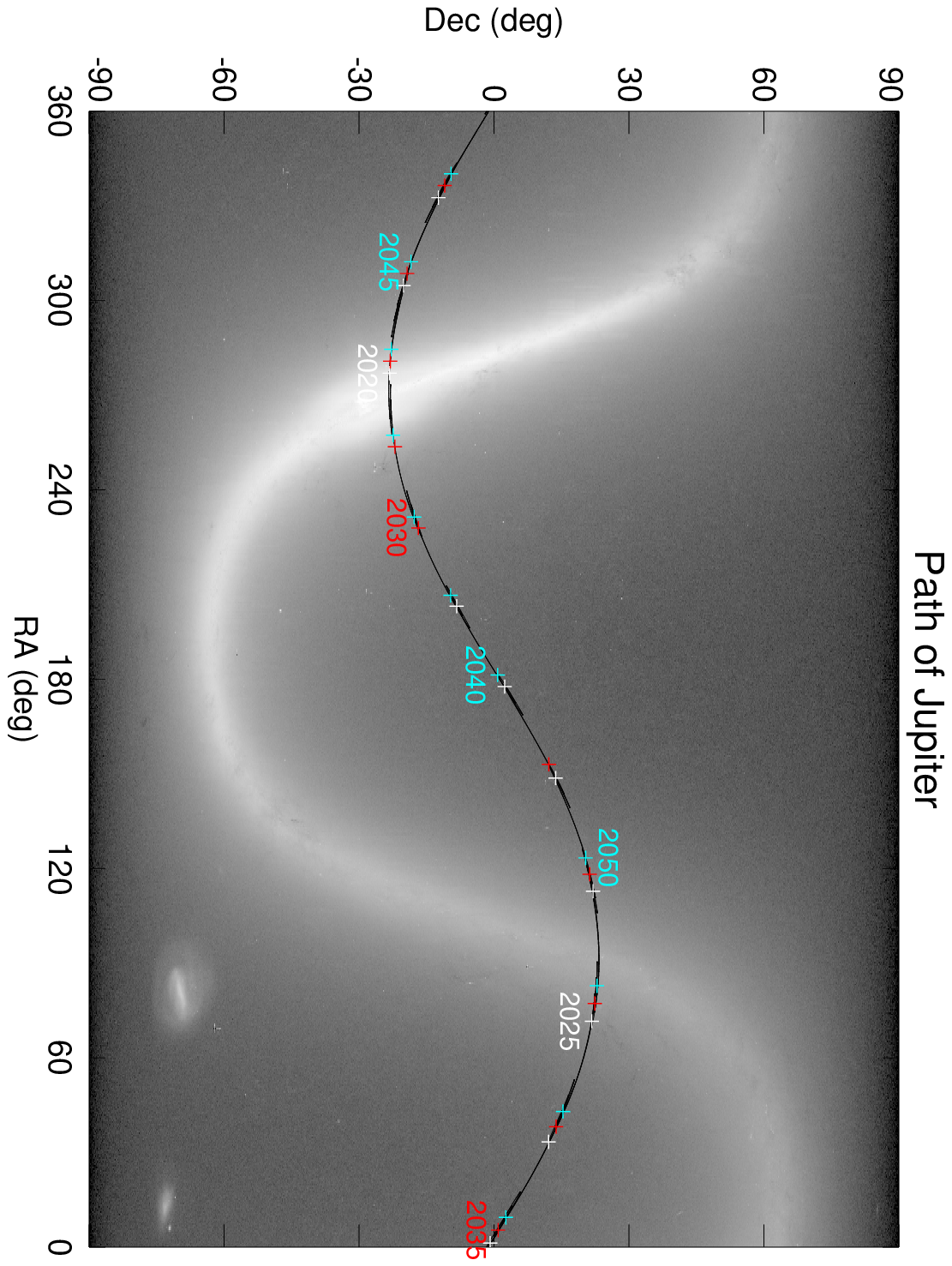}{0.35\textwidth}{}
\rotatefig{90}{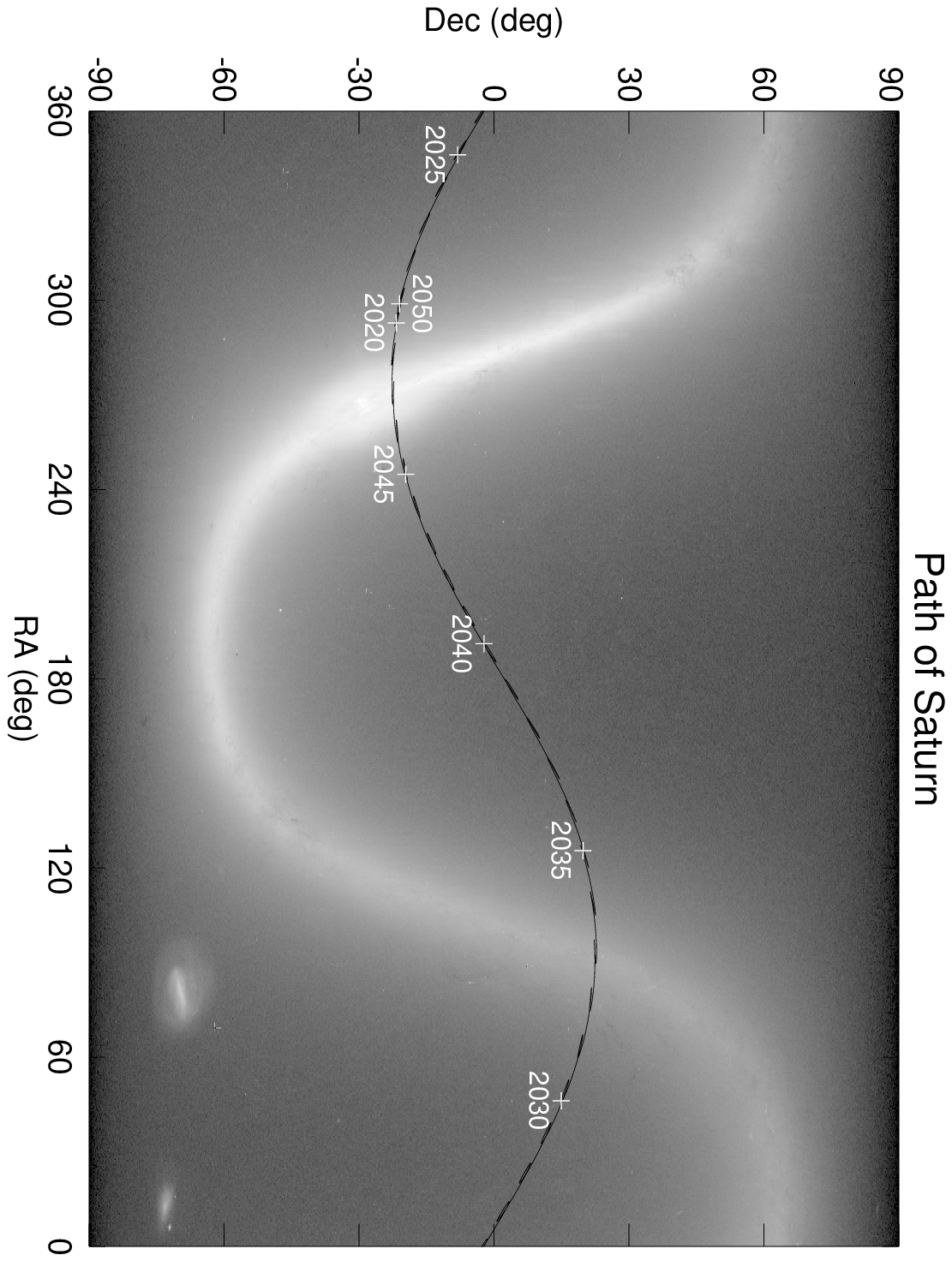}{0.35\textwidth}{}}
\gridline{\rotatefig{90}{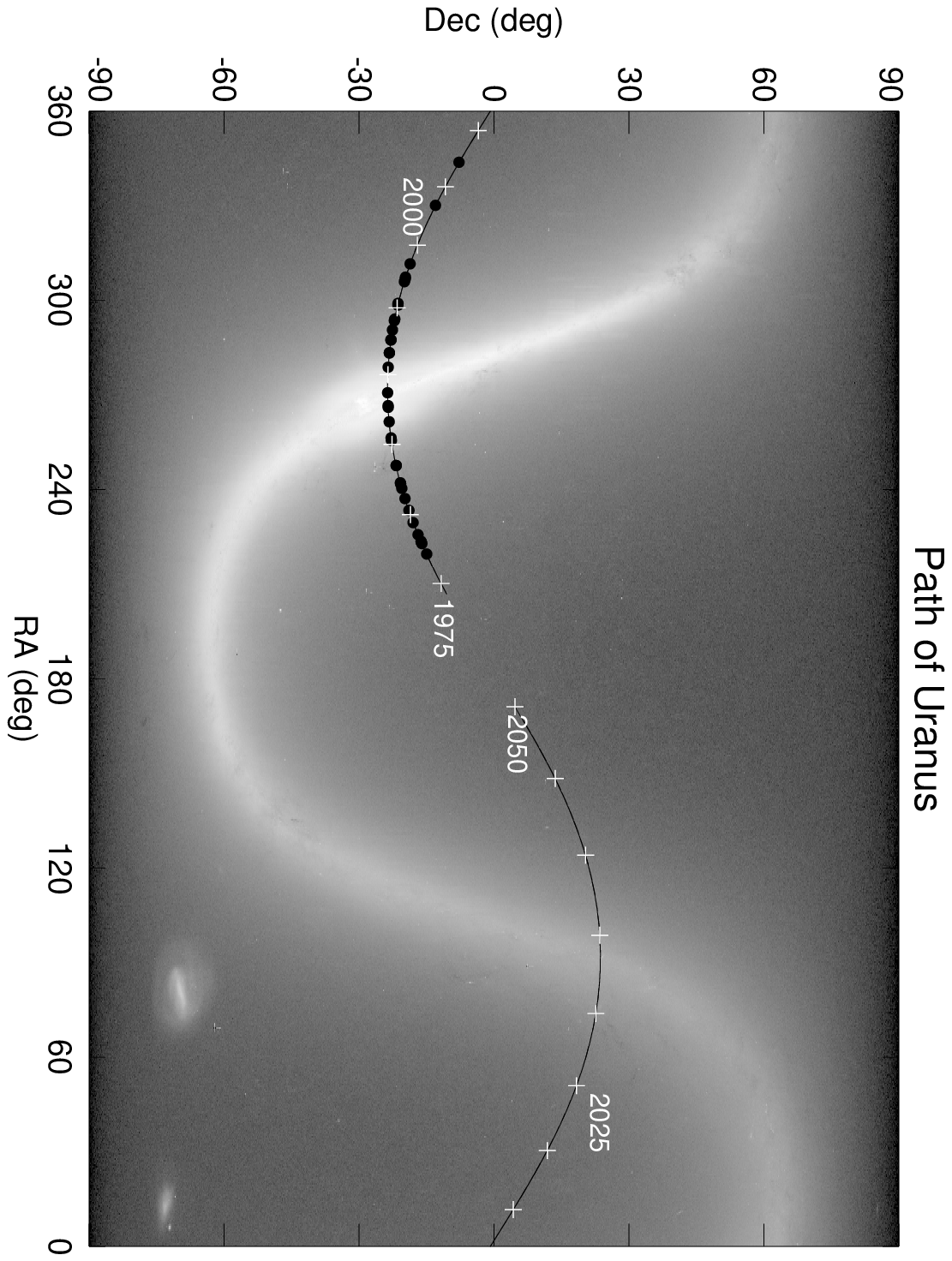}{0.35\textwidth}{}
\rotatefig{90}{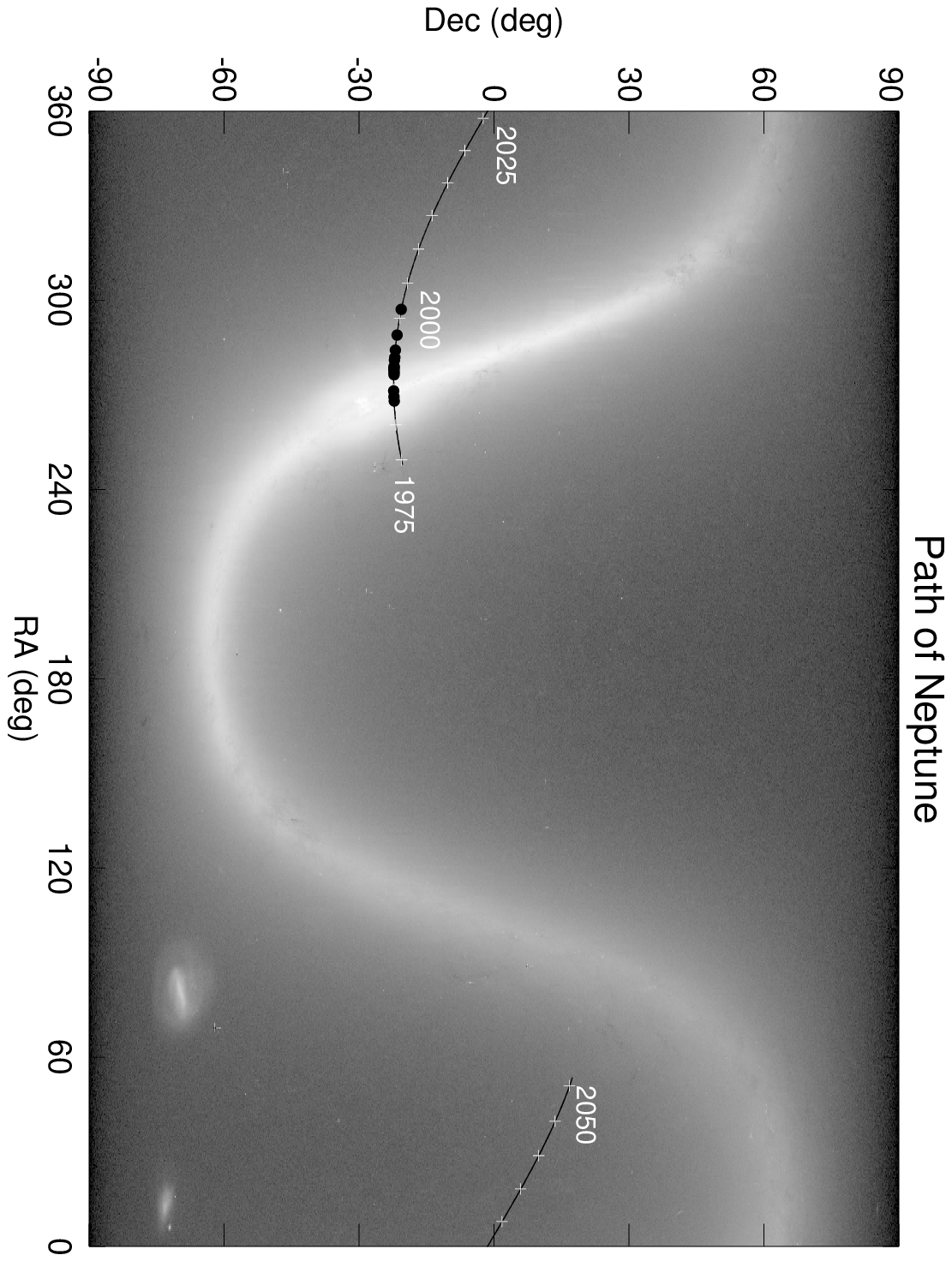}{0.35\textwidth}{}}
\caption{\red Maps of the sky showing the apparent paths of Jupiter, Saturn, Uranus, and Neptune as viewed from Earth. The brightness of the background is proportional to the logarithmic areal density of stars in the 2MASS catalog \citep{Skrutskie2006} with apparent magnitude K$\leq$15. The Milky Way appears as a bright wavy band. Upper left: Jupiter's position is marked at yearly intervals with white symbols for 2020-2029, red symbols for 2030-2039, and cyan symbols for 2040-2050. Upper right: Saturn's position between 2020-2050 is marked with $+$ symbols at five year intervals as labeled. Lower left: The position of Uranus between 1975 and 2050 is marked with $+$ symbols at five year intervals. The filled black dots mark the position of Uranus at times of previous stellar occultation observations between 1977-2007, described in detail in \cite{French2023a}. 
Uranus next crosses the Milky Way in 2026-2027, resulting in a higher frequency of occultations during that period. Lower right: The position of Neptune between 1975 and 2050 is marked with $+$ symbols at five year intervals. The filled black dots mark the position of Neptune at times of previous stellar occultation observations, both published \citep{Roques1994} and under active investigation. Neptune's slow motion across the sky will be far removed from the Milky Way for more than three decades, accounting for the small number of high-SNR Neptune occultations in our prediction list.}
\label{fig:starpaths}
\end{figure}

The path of Jupiter across the sky is shown in {\bf Fig.~\ref{fig:starpaths} (upper left)} from 2020 to 2050 -- an interval of $\sim2.5$ jovian orbital periods -- resulting in multiple intermittent crossings of the dense star fields of the Milky Way, shown as a wavy line. 
A survey of the gallery figures reveals several promising Jupiter occultations in the next few years. The {\red very bright (K=4.93) 2027-02-22} occultation samples Jupiter's south polar region and is visible from sites in Australia, Hawaii and the southwest United States. Figures \ref{fig:JupiterGlobes000} and \ref{fig:JupiterSky000} {\red include} seven occultations in 2031, when Jupiter crosses the Milky Way in the direction of the galactic center. Several of these events have multiple chords that sample a variety of latitudes on Jupiter. Relatively few occultations are predicted for 2033-2042, but there are many bright star events in 2043, when Jupiter's path lies in the direction of the galactic center, with multiple chords and a wide range of sampled latitudes during the atmosphere occultations.

Although we have not evaluated possible Jupiter ring occultations, it is clear from the sky plane figures that such events are rare because the orientation of the ring ansae is very nearly parallel to the sky plane chords, resulting in few predicted ring intersections. This is a consequence of the low inclination of Jupiter's orbit relative to the ecliptic and its pole being nearly normal to its orbital plane 

The results shown here are just a small subset of the predicted events, and we encourage observers to consult the SOM for detailed information about the full set of Jupiter occultations, some of which are only marginally fainter than the restricted range of events in the figures and tables included in the main body of the paper.


\subsection{Saturn}\label{sec:Saturn}

The 1989-07-03 occultation of the bright star 28 Sgr (K=1.48) was predicted by \cite{Taylor1983} and enabled the first detailed post-{\it Voyager} look at the  structure of Saturn's rings \citep{French1993,Nicholson2000} and stratosphere \citep{Hubbard1997}. The event also featured a central flash, when multiple stellar images were detected along the limb of Saturn near mid-occultation \citep{Nicholson1995}, and an occultation by Titan \citep{Sicardy1999}. An occultation of the star GSC 6323-01396 (V = 11.9) by Saturn's rings was observed with the High-Speed Photometer on the Hubble Space Telescope ({\it HST}) on 1991-10-02/03 \citep{Elliot1993}, providing useful information about the pole direction and radius scale of Saturn's ring system.
On 1998-11-14, Saturn and its rings occulted the star GSC 0622-00345 \citep{Harrington2010}, yielding information about gravity waves in Saturn's upper atmosphere. Subsequently, the {\it Cassini} orbital tour provided extensive observations of the Saturn's rings and the atmospheres of both Saturn and Titan, but future Earth-based stellar occultations by Saturn and its rings could still provide valuable information about spatial and temporal variations in the rings and the planet's stratosphere.

Our search resulted in the identification of 290 predicted occultations by Saturn and/or its rings for the period 2023-2050, for a limiting magnitude K$\leq$10. {\bf Table \ref{tbl:satoccpredstats}} lists the number of predicted events per year as a function of K magnitude, {\red excluding years with no predicted events.}
Each entry in the table is of the form {\tt P/R}, where {\tt P} is the number of predicted planetary atmosphere occultations with event types {\tt P, Pg, Pgt}, or {\tt Pt}, and {\tt R} is the number of predicted ring occultations with event types that include {\tt R, Rg, Rgt}, or {\tt Rt}. We predict a total of  267 atmosphere events and 284 ring events for Saturn, considerably fewer than the 1844 predicted Jupiter atmosphere occultations (Table \ref{tbl:jupoccpredstats}). 
In comparison, the prediction list of \cite{Mink1995} based on the PPM catalog includes just 14 geocentric Saturn atmospheric occultations, an indication of their rarity compared to Jupiter events.

%

\begin{deluxetable}{c c c c c c c }
\tabletypesize{\scriptsize} 
\tablecaption{Saturn Planet/Ring Occultations 2023--2050 }
\label{tbl:satoccpredstats}
\tablehead{
\colhead{Year}& 
\colhead{K $<$ 5} &
\colhead{K 5--6} &
\colhead{K 6--7} &
\colhead{K 7--8} &
\colhead{K 8--9} &
\colhead{K 9--10}\\[-.95em]
\colhead{ }& 
\colhead{P/R} &
\colhead{P/R} &
\colhead{P/R} &
\colhead{P/R} &
\colhead{P/R} &
\colhead{P/R}  \\[-1.05em]
}
\startdata
2023 & -- & -- & -- & 1/0 & -- & -- \\
2024 & -- & -- & -- & 1/1 & -- & -- \\
2025 & -- & -- & -- & -- & -- & 1/1 \\
2026 & -- & -- & -- & -- & -- & 3/3 \\
2029 & -- & -- & -- & 1/1 & 2/2 & 2/3 \\
2030 & -- & 1/1 & 1/1 & 1/1 & 2/2 & 3/4 \\
2031 & -- & -- & -- & 1/1 & 2/2 & 3/4 \\
2032 & 1/1 & -- & -- & 4/4 & 4/4 & 8/8 \\
2033 & -- & -- & 1/1 & 2/2 & 1/1 & 6/7 \\
2034 & -- & -- & -- & -- & 2/2 & 6/7 \\
2035 & -- & -- & -- & -- & 1/1 & 0/1 \\
2036 & -- & -- & -- & -- & 1/1 & 4/4 \\
2037 & -- & -- & 1/1 & 2/2 & -- & 0/1 \\
2038 & -- & -- & -- & -- & 1/0 & 0/1 \\
2041 & -- & -- & -- & -- & 1/1 & -- \\
2043 & -- & -- & -- & -- & 1/1 & 1/1 \\
2044 & -- & -- & -- & -- & 1/1 & 6/6 \\
2045 & -- & -- & -- & -- & 1/1 & 4/4 \\
2046 & -- & -- & 4/4 & 3/4 & 16/16 & 21/23 \\
2047 & 1/1 & 4/4 & 5/6 & 13/13 & 29/33 & 40/41 \\
2048 & -- & -- & 0/1 & 2/2 & 9/9 & 23/26 \\
2049 & -- & -- & 1/1 & 2/2 & 5/5 & 5/6 \\[0.5em]
Totals & 2/2 & 5/5 & 13/15 & 33/33 & 79/82 & 136/151 \\

\enddata
\end{deluxetable}

Predicted Saturn occultations for 2023-2050 are not only rare, but are also unevenly spaced in time, with clusters of events occurring when Saturn crosses the denser star fields of the Milky Way. This is illustrated in {\bf Fig.~\ref{fig:starpaths}} (upper right), which shows the apparent path of Saturn as viewed from Earth between 2020 and 2050.  Saturn next traverses the Milky Way in 2033, in the less dense direction opposite to the brighter galactic center. In 2048 (half a Saturn orbital period later) Saturn's path crosses the direction of the galactic center, resulting in abundant bright occultations.

\begin{figure}
\centerline{\resizebox{5.5in}{!}{\includegraphics[angle=0]{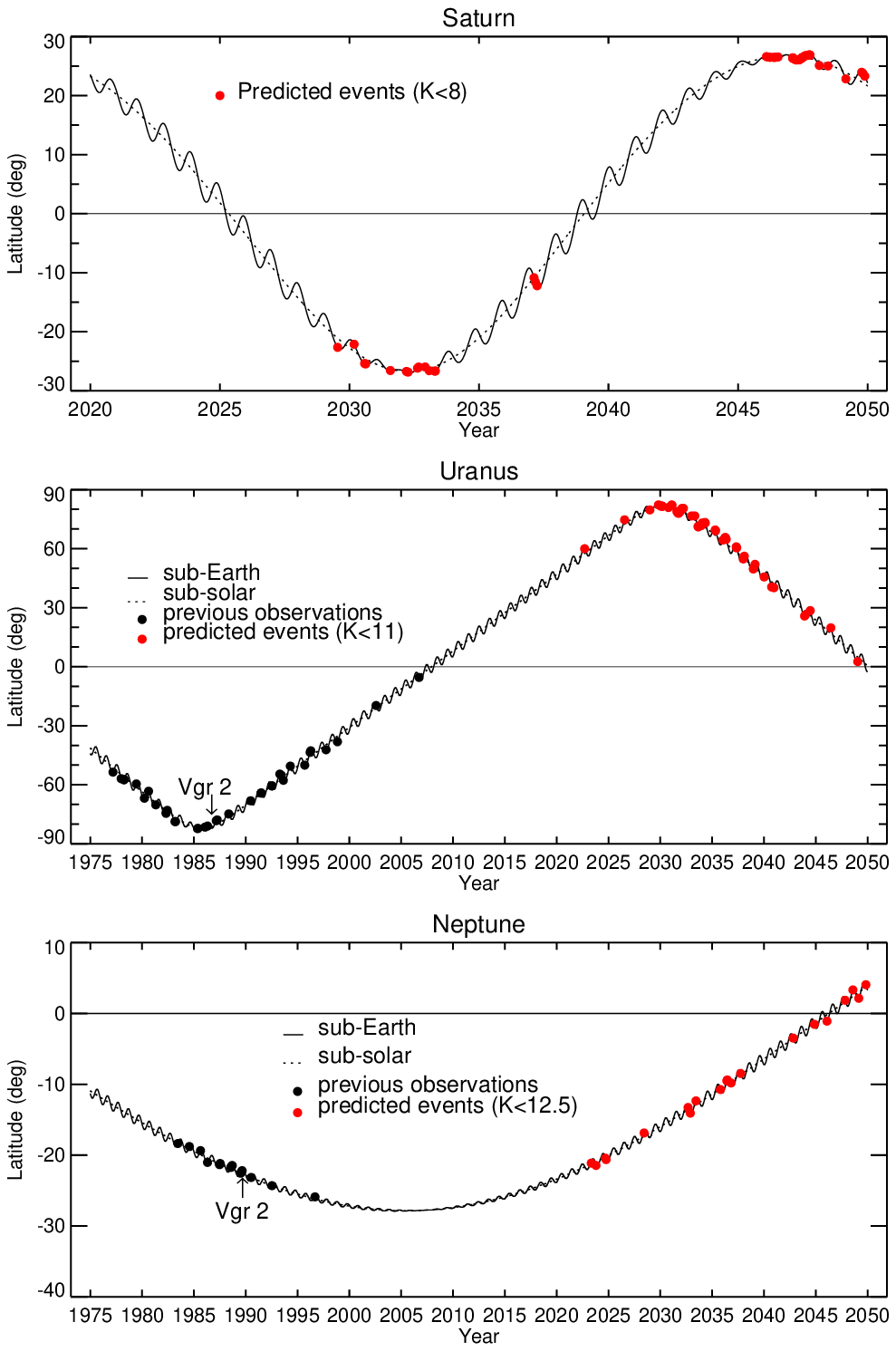}}}
\caption{\red Sub-earth (solid line) and sub-solar (dashed line) latitudes vs time for Saturn, Uranus, and Neptune. These are equivalent to the ring opening angles $B$ and $B'$ defined in the text. For Uranus and Neptune, previous stellar occultation observations are shown as filled black dots. For each planet, selected predicted occultations are shows as red dots, for stars brighter than the cutoff value shown in each panel. The times of the $\it Voyager$ 2 encounters with Uranus and Neptune are marked by arrows.}
\label{fig:BBp}
\end{figure}

The ring opening angle of Saturn's rings varies over time, as shown in {\bf Fig.~\ref{fig:BBp} (top)} for the period 2020-2050. The solid line shows the ring opening angle $B$ as viewed from Earth, with an annual periodic term reflecting the relative inclinations of the orbits of Earth and Saturn. The dashed line shows the ring opening angle $B'$ as viewed from the Sun. The next Saturn ring plane crossings (RPXs) will occur near the years  2025 and  2040. Red dots mark predicted Saturn occultations for stars brighter than K=8. These are most frequent near 2033 and 2048, the times of Milky Way crossings, as noted above, and coincidentally are the times when Saturn's rings are most open as viewed from Earth, enhancing the prospects for high radial resolution of the rings during these occultations.

\begin{figure}
\centerline{\resizebox{6in}{!}{\includegraphics[angle=0]{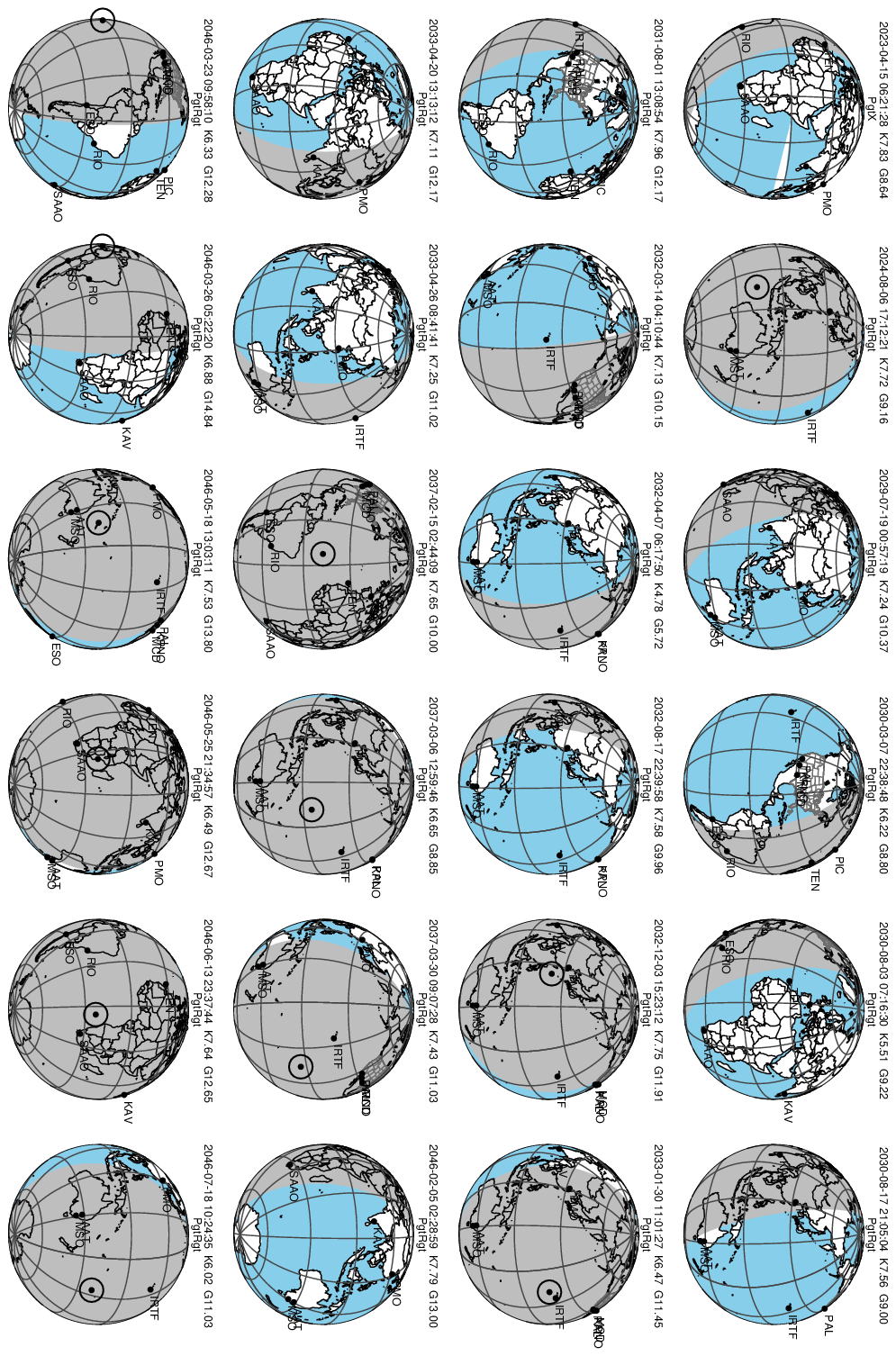}}}
\caption{\red Gallery of Earth views from Saturn at mid-occultation for the 24 brightest predicted events in the K band with composite event type {\it RgtPgt} between 2023-2047.}
\label{fig:SaturnGlobes000}
\end{figure}

\begin{figure}
\centerline{\resizebox{6in}{!}{\includegraphics[angle=0]{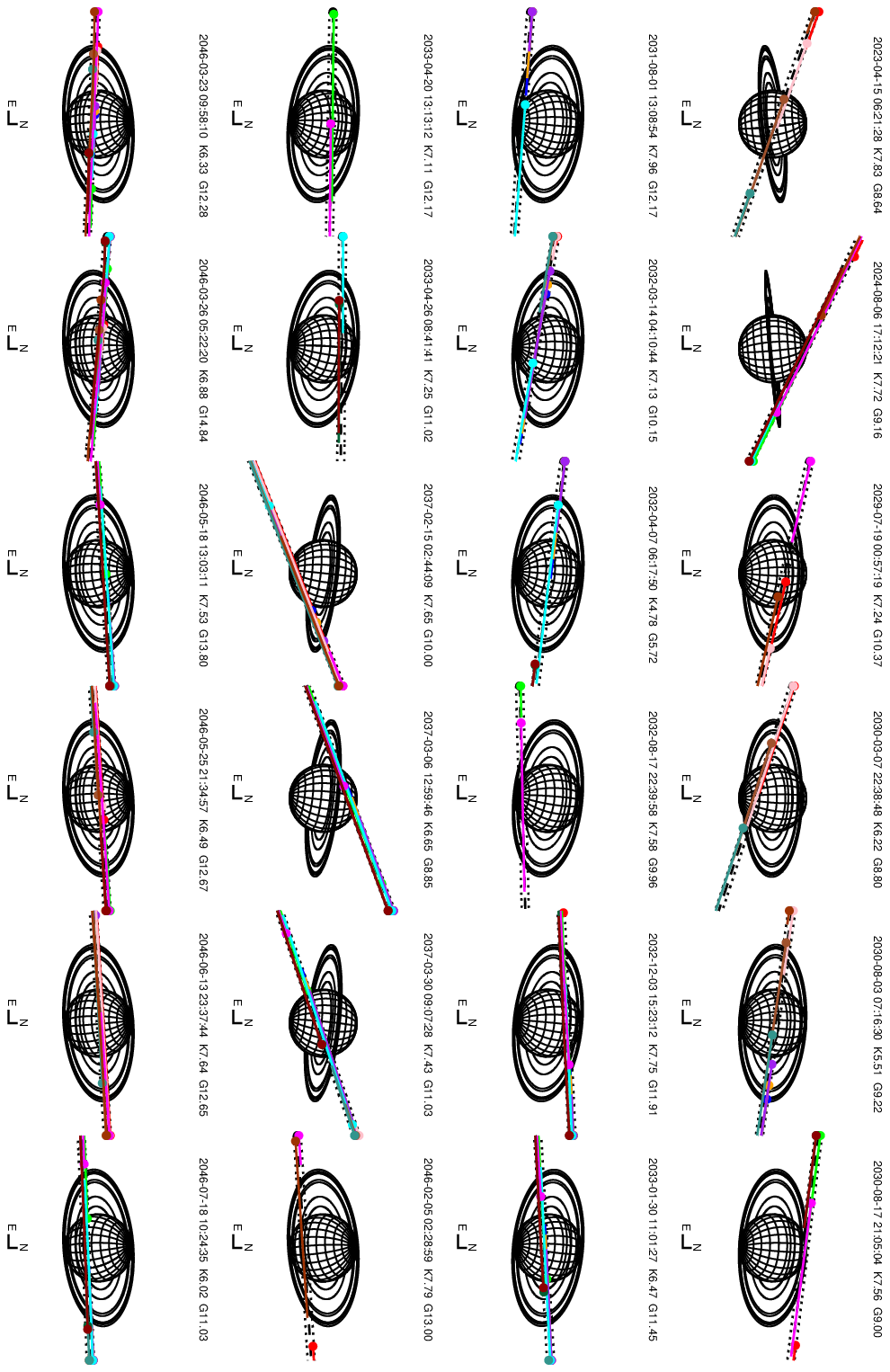}}}
\caption{\red Gallery of sky plane geocentric views of Saturn and its rings at mid-occultation for the 24 brightest predicted events in the K band between 2023-2047 with composite event type {\it RgtPgt}.}
\label{fig:SaturnSky000}
\end{figure}

{\red To illustrate the range of event geometries and the varying aspect of Saturn and its rings, {\bf Fig.~\ref{fig:SaturnGlobes000}} shows a gallery of views of Earth at mid-occultation and {\bf Fig.~\ref{fig:SaturnSky000}} shows the corresponding sky-plane views of Saturn for the 24 brightest predicted events in the K band with composite event type {\it RgtPgt} between 2023-2047. (The end date is chosen to avoid overrepresentation of the large number of predicted events in the period 2047-2050, evident in Table \ref{tbl:satoccpredstats}, owing to Saturn's traversal of the galactic center region of the Milky Way at that time.) {\bf Table \ref{tbl:satoccpred24}} provides detailed information about these events.}

\begin{longrotatetable}
\movetabledown=6mm
\begin{deluxetable}{c c c D D D D D D D D D D D}
\tablecolumns{14}
\tabletypesize{\tiny}
\tablecaption{Geocentric Saturn Occultation Predictions 2023--2047}
\label{tbl:satoccpred24}
\tablehead{
\\[-2.75em]
\colhead{Event ID} & 
\colhead{C/A Epoch} &
\colhead{ICRS Star Coord at Epoch} & 
\multicolumn2c{$\sigma(\alpha_*)$ (km)} &
\multicolumn2c{C/A  (${'}{'}$)} & 
\multicolumn2c{PA (deg)} & 
\multicolumn2c{$v_{\rm sky}$ (km/s)} &
\multicolumn2c{Dist (au)} &
\multicolumn2c{Lat I (deg)} &
\multicolumn2c{K}& 
\multicolumn2c{G} & 
\multicolumn2c{RP} & 
\multicolumn2c{E $\lambda$ (deg)} & 
\multicolumn2c{S-G-T} \\[-.95em]
\colhead{Event type} & 
\colhead{Source ID} &
\colhead{Geocentric Object Position} &
\multicolumn2c{$\sigma(\delta_*)$ (km)} &
\multicolumn2c{C/A (10$^3$ km)} &  
\multicolumn2c{P (deg)} & 
\multicolumn2c{B (deg)} &
\multicolumn2c{$\dot r$  (km/s)} &
\multicolumn2c{Lat E (deg)} &
\multicolumn2c{$D_*$ (km)} &
\multicolumn2c{G${}_*$} & 
\multicolumn2c{DUP} & 
\multicolumn2c{$\phi$ (deg)} &
\multicolumn2c{M-G-T} 
}
\decimals
\startdata
S23001** & 2023-04-15 06:21:28.59 & 22 24 49.19365  $-$11 23 18.66374  & 4.3 & 0.607 & 160.37 & 28.51 & 10.410 & -25.1 & 7.832 & 8.640 & 8.330 & 38.0 & 50.9 \\ 
\ \ \ PgtX & 2602435862784961664 & 22 24 49.20752  $-$11 23 19.23506  & 3.3 & 4.582 & 5.89 & 8.54 &  .  & 34.7 & 0.786 & 9.025 & 0 & -11.3 & 14.5 \\ 
S24001*** & 2024-08-06 17:12:21.22 & 23 18 59.47989  $-$06 43 01.47226  & 7.0 & 8.196 & 153.78 & -15.13 & 8.814 & 87.4 & 7.716 & 9.161 & 8.444 & 136.2 & 146.4 \\ 
\ \ \ PgtRgt & 2631163146681256448 & 23 18 59.72295  $-$06 43 08.82476  & 7.2 & 52.390 & 5.01 & 2.62 &  .  & 34.2 & 0.913 & 8.857 & 0 & -6.6 & 171.7 \\ 
S29002* & 2029-07-19 00:57:19.09 & 03 22 51.43791  +16 17 15.28928  & 2.3 & 3.011 & 167.39 & 22.88 & 9.562 & 10.0 & 7.241 & 10.369 & 9.477 & 99.7 & 63.7 \\ 
\ \ \ PgtRgt & 54522689973491584 & 03 22 51.48358  +16 17 12.35144  & 1.9 & 20.879 & -1.23 & -22.65 & 23.656 & 34.2 & 1.510 & 10.515 & 0 & 16.4 & 158.5 \\ 
S30003 & 2030-03-07 22:38:48.34 & 03 11 19.24980  +15 42 35.64960  & 1.9 & 4.128 & 341.70 & 23.38 & 9.523 & -47.3 & 6.224 & 8.804 & 8.027 & 262.9 & 62.7 \\ 
\ \ \ PgtRgt & 31196722588576128 & 03 11 19.16004  +15 42 39.56916  & 1.6 & 28.512 & -0.88 & -22.12 & 26.008 & -12.2 & 2.163 & 8.974 & 0 & 15.8 & 21.6 \\ 
S30004 & 2030-08-03 07:16:30.77 & 04 19 53.06611  +19 31 03.90032  & 2.4 & 0.761 & 171.77 & 22.68 & 9.482 & -0.3 & 5.508 & 9.224 & 8.203 & 4.4 & 64.1 \\ 
\ \ \ PgtRgt & 48224275053110272 & 04 19 53.07384  +19 31 03.14766  & 1.9 & 5.230 & -2.95 & -25.38 & 23.045 & 11.3 & 3.908 & 9.361 & 0 & 19.6 & 110.3 \\ 
S30005 & 2030-08-17 21:05:04.31 & 04 24 07.02059  +19 39 15.03689  & 2.4 & 8.458 & 173.02 & 16.34 & 9.255 & 62.3 & 7.562 & 9.005 & 8.493 & 143.9 & 77.1 \\ 
\ \ \ PgtRgt & 48279147555452544 & 04 24 07.09344  +19 39 06.64162  & 1.8 & 56.775 & -3.07 & -25.46 & 4.862 & 66.0 & 0.902 & 8.786 & 0 & 19.7 & 48.6 \\ 
S31003 & 2031-08-01 13:08:54.06 & 05 12 56.43379  +21 30 44.94500  & 2.5 & 6.481 & 355.77 & 28.96 & 9.663 & -48.4 & 7.961 & 12.171 & 11.050 & 291.5 & 49.6 \\ 
\ \ \ PgtRgt & 3414247177797508736 & 05 12 56.39959  +21 30 51.40868  & 1.5 & 45.422 & -4.41 & -26.57 & 19.422 & -48.7 & 1.486 & 12.573 & 0 & 21.5 & 152.1 \\ 
S32003 & 2032-03-14 04:10:44.16 & 05 00 57.02045  +21 27 54.29471  & 2.7 & 3.296 & 349.89 & 13.54 & 9.096 & -28.7 & 7.132 & 10.149 & 9.274 & 200.7 & 82.6 \\ 
\ \ \ PgtRgt & 3412182058740339712 & 05 00 56.97903  +21 27 57.53943  & 2.0 & 21.744 & -4.11 & -26.73 & 13.019 & -16.4 & 1.513 & 9.726 & 0 & 21.5 & 49.0 \\ 
S32005 & 2032-04-07 06:17:50.19 & 05 07 55.43570  +21 42 17.13583  & 5.6 & 0.451 & 172.74 & 24.40 & 9.473 & -0.0 & 4.775 & 5.723 & 5.437 & 146.9 & 60.4 \\ 
\ \ \ PgtRgt & 3415147613398691200 & 05 07 55.43982  +21 42 16.68826  & 4.0 & 3.099 & -4.29 & -26.85 & 24.527 & 6.5 & 3.014 & 5.939 & 0 & 21.7 & 98.9 \\ 
S32009 & 2032-08-17 22:39:58.99 & 06 14 57.57298  +22 23 54.85276  & 2.4 & 6.537 & 1.46 & 28.12 & 9.609 & -42.9 & 7.585 & 9.958 & 9.209 & 147.4 & 51.7 \\ 
\ \ \ PgtRgt & 3376976795931337600 & 06 14 57.58501  +22 24 01.38777  & 1.8 & 45.559 & -5.76 & -26.17 & 19.789 & -54.4 & 1.125 & 10.328 & 0 & 22.4 & 171.6 \\ 
S32013 & 2032-12-03 15:23:12.94 & 06 21 14.77166  +22 21 03.85494  & 2.5 & 5.649 & 182.81 & -17.69 & 8.102 & 26.6 & 7.753 & 11.906 & 10.796 & 152.0 & 156.6 \\ 
\ \ \ PgtRgt & 3377160169558092928 & 06 21 14.75173  +22 20 58.21235  & 1.9 & 33.193 & -5.87 & -25.97 & 15.728 & 42.9 & 1.392 & 11.773 & 0 & 22.3 & 166.0 \\ 
S33005 & 2033-01-30 11:01:27.46 & 06 02 28.97310  +22 33 04.32277  & 3.7 & 1.131 & 3.29 & -13.24 & 8.237 & -16.6 & 6.474 & 11.453 & 10.183 & 155.8 & 140.2 \\ 
\ \ \ PgtRgt & 3423718336601355904 & 06 02 28.97781  +22 33 05.45194  & 2.7 & 6.759 & -5.54 & -26.61 & 13.811 & 2.6 & 2.998 & 11.006 & 0 & 22.5 & 146.3 \\ 
S33007 & 2033-04-20 13:13:12.07 & 06 08 27.81207  +22 46 13.59165  & 3.3 & 2.104 & 179.12 & 23.79 & 9.451 & 20.1 & 7.106 & 12.166 & 10.878 & 45.4 & 61.6 \\ 
\ \ \ PgtRgt & 3425270679517659520 & 06 08 27.81442  +22 46 11.48702  & 2.5 & 14.425 & -5.66 & -26.69 & 23.486 & 9.9 &  .  & 12.354 & 0 & 22.8 & 173.8 \\ 
S33010 & 2033-04-26 08:41:41.37 & 06 10 39.39492  +22 46 37.24870  & 2.0 & 4.779 & 179.55 & 26.10 & 9.534 & 39.5 & 7.255 & 11.023 & 10.006 & 108.1 & 56.5 \\ 
\ \ \ PgtRgt & 3425075069527522816 & 06 10 39.39769  +22 46 32.46966  & 1.7 & 33.043 & -5.70 & -26.64 & 22.523 & 29.7 &  .  & 11.312 & 0 & 22.8 & 94.8 \\ 
S37002 & 2037-02-15 02:44:09.31 & 10 07 29.39955  +13 16 06.38810  & 2.1 & 7.429 & 21.68 & -20.15 & 8.248 & -74.9 & 7.654 & 10.003 & 9.294 & 325.9 & 176.7 \\ 
\ \ \ PgtRgt & 3881962210025024896 & 10 07 29.58755  +13 16 13.29056  & 1.6 & 44.440 & -6.13 & -10.85 & 40.042 & -23.3 & 0.905 & 10.011 & 0 & 13.1 & 176.1 \\ 
S37003 & 2037-03-06 12:59:46.54 & 10 01 35.84616  +13 49 33.76525  & 2.4 & 7.240 & 200.56 & -18.69 & 8.292 & 23.0 & 6.648 & 8.853 & 8.166 & 151.3 & 161.8 \\ 
\ \ \ PgtRgt & 615385869108696960 & 10 01 35.67165  +13 49 26.98510  & 1.6 & 43.542 & -6.21 & -11.56 & 34.100 & 73.0 & 1.432 & 8.779 & 0 & 13.6 & 80.2 \\ 
S37004 & 2037-03-30 09:07:28.03 & 09 55 49.78581  +14 19 44.14548  & 2.9 & 1.492 & 18.96 & -12.06 & 8.494 & -38.6 & 7.435 & 11.026 & 9.996 & 184.5 & 136.6 \\ 
\ \ \ PgtRgt & 615301245368208128 & 09 55 49.81917  +14 19 45.55568  & 2.3 & 9.194 & -6.28 & -12.21 & 26.435 & 19.5 & 1.493 & 10.477 & 0 & 14.2 & 28.8 \\ 
S46004 & 2046-02-05 02:28:59.57 & 17 17 34.79616  $-$21 37 16.33741  & 11.7 & 4.411 & 3.56 & 26.40 & 10.548 & -36.4 & 7.791 & 12.996 & 11.607 & 87.4 & 55.7 \\ 
\ \ \ PgtRgt & 4115016039029356416 & 17 17 34.81584  $-$21 37 11.93441  & 7.5 & 33.744 & 4.53 & 26.59 & 22.066 & -34.6 & 2.203 & 13.298 & 0 & -21.7 & 46.3 \\ 
S46020 & 2046-03-23 09:58:10.06 & 17 27 44.73429  $-$21 42 46.33527  & 16.8 & 1.619 & 357.69 & 4.53 & 9.811 & -19.6 & 6.335 & 12.282 & 10.813 & 292.0 & 99.8 \\ 
\ \ \ PgtRgt & 4117826352042608384 & 17 27 44.72967  $-$21 42 44.71744  & 10.8 & 11.518 & 4.77 & 26.49 & 4.758 & -4.4 & 5.115 & 10.670 & 0 & -21.7 & 66.0 \\ 
S46021* & 2046-03-26 05:22:20.36 & 17 27 54.12477  $-$21 42 37.09412  & 44.8 & 0.106 & 175.32 & 3.15 & 9.765 & -9.7 & 6.879 & 14.842 & 13.079 & 358.2 & 102.5 \\ 
\ \ \ PgtRgt & 4117826218935136512 & 17 27 54.12539  $-$21 42 37.19945  & 29.3 & 0.748 & 4.77 & 26.49 & 3.534 & 11.0 &  .  & 12.834 & 0 & -21.7 & 25.7 \\ 
S46029 & 2046-05-18 13:03:11.39 & 17 20 54.76091  $-$21 33 38.52132  & 14.4 & 1.515 & 183.86 & -17.65 & 9.099 & 11.3 & 7.532 & 13.797 & 12.275 & 188.7 & 156.2 \\ 
\ \ \ PgtRgt & 4115052494691697792 & 17 20 54.75360  $-$21 33 40.03263  & 10.7 & 9.996 & 4.60 & 26.47 & 17.434 & 9.6 & 3.182 & 13.661 & 0 & -21.6 & 45.6 \\ 
S46033 & 2046-05-25 21:34:57.15 & 17 18 48.01096  $-$21 31 41.66501  & 11.2 & 0.402 & 183.78 & -18.99 & 9.057 & 3.6 & 6.494 & 12.668 & 11.252 & 53.0 & 163.7 \\ 
\ \ \ PgtRgt & 4115099533179705856 & 17 18 48.00913  $-$21 31 42.06538  & 7.3 & 2.638 & 4.55 & 26.48 & 18.992 & 1.9 & 4.225 & 12.612 & 0 & -21.6 & 53.8 \\ 
S46035 & 2046-06-13 23:37:44.29 & 17 12 50.50336  $-$21 26 19.72822  & 8.6 & 0.525 & 183.61 & -20.12 & 9.021 & 4.5 & 7.641 & 12.649 & 11.303 & 2.0 & 176.2 \\ 
\ \ \ PgtRgt & 4115816792713438208 & 17 12 50.50106  $-$21 26 20.25228  & 5.5 & 3.438 & 4.41 & 26.50 & 20.103 & 2.7 & 1.977 & 12.655 & 0 & -21.5 & 55.9 \\ 
S46043 & 2046-07-18 10:24:35.13 & 17 03 16.69742  $-$21 18 52.85802  & 9.6 & 3.369 & 2.30 & -13.73 & 9.221 & -21.7 & 6.020 & 11.030 & 9.759 & 163.9 & 141.4 \\ 
\ \ \ PgtRgt & 4127635408818049664 & 17 03 16.70716  $-$21 18 49.49172  & 6.3 & 22.531 & 4.17 & 26.54 & 12.823 & -25.5 & 4.183 & 10.622 & 0 & -21.4 & 43.2 \\ 
\enddata
\end{deluxetable}
\end{longrotatetable}

There are only a few occultations with K$\leq 8$ by Saturn between 2023 and 2029, after which there are numerous high-SNR opportunities to observe the atmosphere and the rings unblocked by the planet. The brightest of these is the 2032-04-07 event (K=4.78, G=5.72, V=5.80), with a nearly diametric ring occultation observable from Hawaii. High-SNR atmosphere observations should be possible for this occultation. However, even this bright star is only about 5\% the brightness of the 1989-07-03 occultation star 28 Sgr in the K band, and the icy rings have a high albedo in this wavelength region. The rings are darker in the L band near $\lambda\sim$~3$\mu$m, where future high altitude or spacecraft observations may be possible for this and later occultations.

\subsection{Uranus}\label{sec:Uranus}

The Uranian rings were first detected during the widely observed 1977-03-10 occultation of the bright star SAO 158687 \citep{Bhattacharyya1977, Brahic1977, Chen1978, Elliot1977, Hubbard1977, Millis1977, Morrisby1977, Tomita1977}, and a rich set of subsequent Earth-based occultations revealed that these narrow and sharp-edged rings were eccentric and inclined, precessing under the gravitational influence of the oblate central planet (see \cite{Nicholson2018} and \cite{French2023a} for recent reviews).
Atmospheric occultations during some of these events provided information about the stratospheric temperature profiles (see \cite{Young2001} for results from the 1998-11-06 occultation, and references therein to prior events) and the oblateness of the planet \citep{Baron1989}. 

Our search resulted in the identification of 1173 predicted occultations by Uranus and/or its rings for the period 2023-2050, for a limiting magnitude K$\leq$15. {\bf Table \ref{tbl:uroccpredstats}} lists the number of predicted events per year as a function of K magnitude. The format is the same as for Table \ref{tbl:satoccpredstats}. The distribution of events in time is very non-uniform, resulting from the planet's intermittent traversal of the dense star fields of the Milky Way, separated by long periods in relatively star-free regions of the sky. This is evident in {\bf Fig.~\ref{fig:starpaths} (lower left)}, which shows the apparent path of Uranus in the sky planet as viewed from Earth between 1975 and 2050. Uranus crossed the Milky Way roughly in the direction of the galactic center between 1985-1990, providing abundant opportunities for high-SNR ring and planet occultations. The subsequent reduction in the density of stars along the planet's path, combined with the nearly edge-on aspect of the rings and the decommissioning of high-speed InSb aperture photometers at major observatories, resulted in a virtual absence of Uranus occultation observations in the past two decades. The most recent {\red published observations} were the 2002-11-29 occultation (K=11.4) observed from Palomar Observatory and the 2006-09-20 occultation (V=10.746, K=8.408) observed from the IRTF \citep{French2023a}. {\red The next traversal of the Milky Way will not occur until 2033 in the less dense stellar regions opposite to the galactic center.} Until then, the frequency of high-SNR stellar occultations by Uranus and its rings will be at a typical cadence of one every few years.

\begin{deluxetable}{c c c c c c c c c c}
\tabletypesize{\scriptsize} 
\tablecaption{Uranus Planet/Ring Occultations 2023--2050 }
\label{tbl:uroccpredstats}
\tablehead{
\colhead{Year}& 
\colhead{K$<$7} &
\colhead{K 7--8} &
\colhead{K 8--9} &
\colhead{K 9--10} &
\colhead{K 10--11} &
\colhead{K 11--12} &
\colhead{K 12--13} &
\colhead{K 13--14} &
\colhead{K 14--15} \\[-0.95em]
\colhead{ }& 
\colhead{P/R} &
\colhead{P/R} &
\colhead{P/R} &
\colhead{P/R} &
\colhead{P/R} &
\colhead{P/R} &
\colhead{P/R} &
\colhead{P/R} &
\colhead{P/R} \\[-1.05em]
}
\startdata
2023 & -- & -- & -- & -- & -- & 1/1 & 0/2 & 3/6 & 3/9 \\
2024 & -- & -- & -- & -- & -- & 2/2 & 1/2 & 3/5 & 8/10 \\
2025 & -- & -- & -- & -- & -- & -- & 1/1 & 1/4 & 8/14 \\
2026 & -- & -- & -- & -- & 1/1 & 0/2 & 5/5 & 7/11 & 5/9 \\
2027 & -- & -- & -- & -- & -- & 0/3 & 2/4 & 4/6 & 10/21 \\
2028 & -- & -- & -- & -- & 1/1 & -- & 1/3 & 4/13 & 4/14 \\
2029 & -- & -- & -- & -- & 0/1 & 3/4 & 1/2 & 11/17 & 20/37 \\
2030 & -- & -- & -- & 0/1 & 2/2 & 2/4 & 7/12 & 13/18 & 16/29 \\
2031 & 1/1 & -- & -- & 1/1 & 5/7 & 3/7 & 1/3 & 18/35 & 36/64 \\
2032 & -- & -- & 1/2 & -- & 0/1 & 2/7 & 7/15 & 15/24 & 36/54 \\
2033 & -- & -- & -- & 1/2 & 3/5 & 5/7 & 8/12 & 23/37 & 44/72 \\
2034 & -- & -- & -- & -- & 2/4 & 4/8 & 4/11 & 18/29 & 34/67 \\
2035 & -- & -- & -- & -- & 1/2 & 2/4 & 3/8 & 15/24 & 18/40 \\
2036 & -- & -- & 1/1 & -- & 1/3 & 3/5 & 5/6 & 13/19 & 20/34 \\
2037 & -- & -- & -- & -- & 2/2 & -- & 2/8 & 7/12 & 15/24 \\
2038 & -- & 1/1 & -- & -- & 1/2 & 1/1 & 2/3 & 5/13 & 14/25 \\
2039 & -- & -- & -- & -- & 0/2 & 0/1 & 3/7 & 3/11 & 9/17 \\
2040 & -- & -- & -- & -- & 1/3 & 0/2 & 0/1 & 2/4 & 15/19 \\
2041 & -- & -- & -- & -- & -- & 1/2 & 1/2 & 2/4 & 5/9 \\
2042 & -- & -- & -- & -- & -- & 0/1 & 0/3 & 3/4 & 6/9 \\
2043 & -- & -- & -- & -- & 1/1 & 1/1 & 5/5 & 3/4 & 2/3 \\
2044 & -- & 1/1 & -- & 1/1 & -- & 1/1 & -- & 1/4 & 1/4 \\
2045 & -- & -- & -- & -- & -- & 1/3 & 1/2 & 1/1 & 2/3 \\
2046 & -- & -- & 0/1 & -- & -- & 0/1 & 2/1 & 3/3 & 1/8 \\
2047 & -- & -- & -- & -- & -- & -- & 1/1 & 3/2 & 4/4 \\
2048 & -- & -- & -- & -- & -- & -- & 0/1 & 1/1 & 1/3 \\
2049 & -- & -- & -- & -- & 0/1 & 2/0 & -- & 1/3 & 3/3 \\[0.5em]
Totals & 1/1 & 2/2 & 2/4 & 3/5 & 21/38 & 34/67 & 63/120 & 183/314 & 340/605 \\

\enddata
\end{deluxetable}

The {\red Earth view} of the Uranus ring system varies {\red substantially} over time, as shown in {\bf Fig.~\ref{fig:BBp}} for the period 1975-2050. The solid line shows the sub-Earth latitude, with an annual periodic term reflecting the relative inclinations of the orbits of Earth and Uranus, and the dashed line shows the sub-solar latitude.\footnote{Here, we follow the NASA/JPL Horizons convention of using the IAU definition of the pole direction of Uranus. In our tabulated results, we specify the ring elevation angle $B$, following the convention used at NASA's Planetary Data System Ring-Moon Systems node. This angle is positive on the side of the ring plane defined by positive angular momentum, and negative on the opposite side. The direction of positive angular momentum points toward the IAU-defined north side of the ring plane for Jupiter, Saturn and Neptune, but IAU-defined south side of the ring plane for Uranus, with the result that $B$ is the negative of the IAU-defined sub-earth latitude shown in Fig.~\ref{fig:BBp}.}
 The most recent Uranus ring plane crossings (RPXs) occurred in $\sim$2008 and the next will not occur until $\sim$2050. Black dots mark the times of previously observed occultations, and red dots mark predicted Uranus occultations for stars with K$\leq$11). These are most frequent near 2033, the time of the next Milky Way crossing, as noted above, and coincidentally (just as for Saturn) are the times when the rings are most open as viewed from Earth, enhancing the prospects for high radial resolution of the rings.

The accuracy of our predictions is dependent in part on the accuracy of the Uranus ephemeris. Using Uranus ring occultation observations from 1977 to 2006, \cite{French2023a,French2023b} detected a small but significant quasi-linear drift between the {\tt ura116.bsp} ephemeris and true sky plane positions of Uranus since 1986, the epoch of the {\it Voyager 2} flyby that provided a very accurate instantaneous value for the planet's position. {\red Making use of these ring observations, the Jet Propulsion System (JPL) has released a new Uranus ephemeris ({\tt ura161.bsp}) that removes this drift. 
We have used this revised ephemeris for all of our Uranus predictions.}

\begin{figure}
\centerline{\resizebox{6in}{!}{\includegraphics[angle=0]{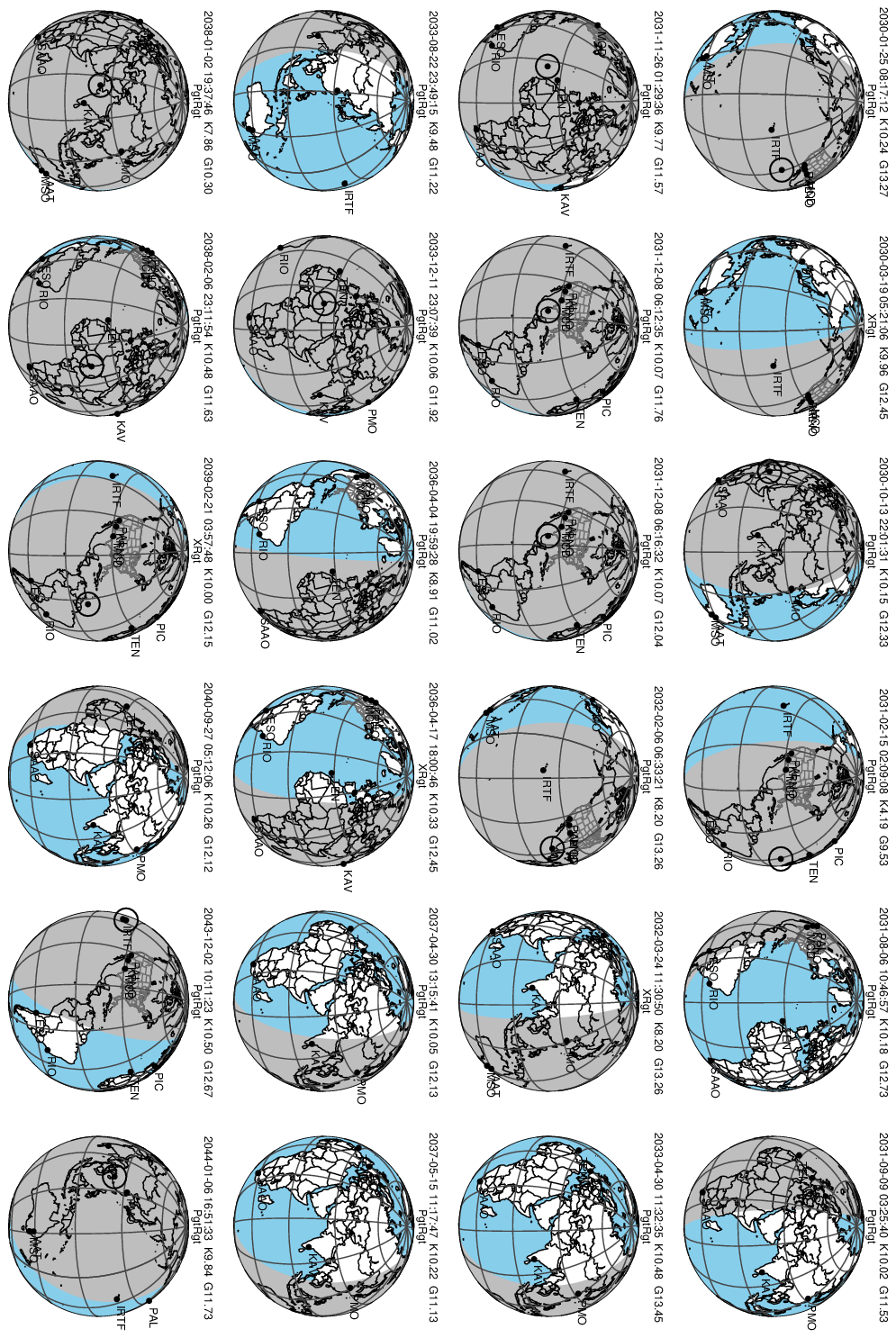}}}\caption{Gallery of Earth views from Uranus at mid-occultation for the 24 brightest predicted events in the K band with ring event type {\it Rgt} for  2023-2050.}
\label{fig:UranusGlobes000}
\end{figure}

\begin{figure}
\centerline{\resizebox{6in}{!}{\includegraphics[angle=0]{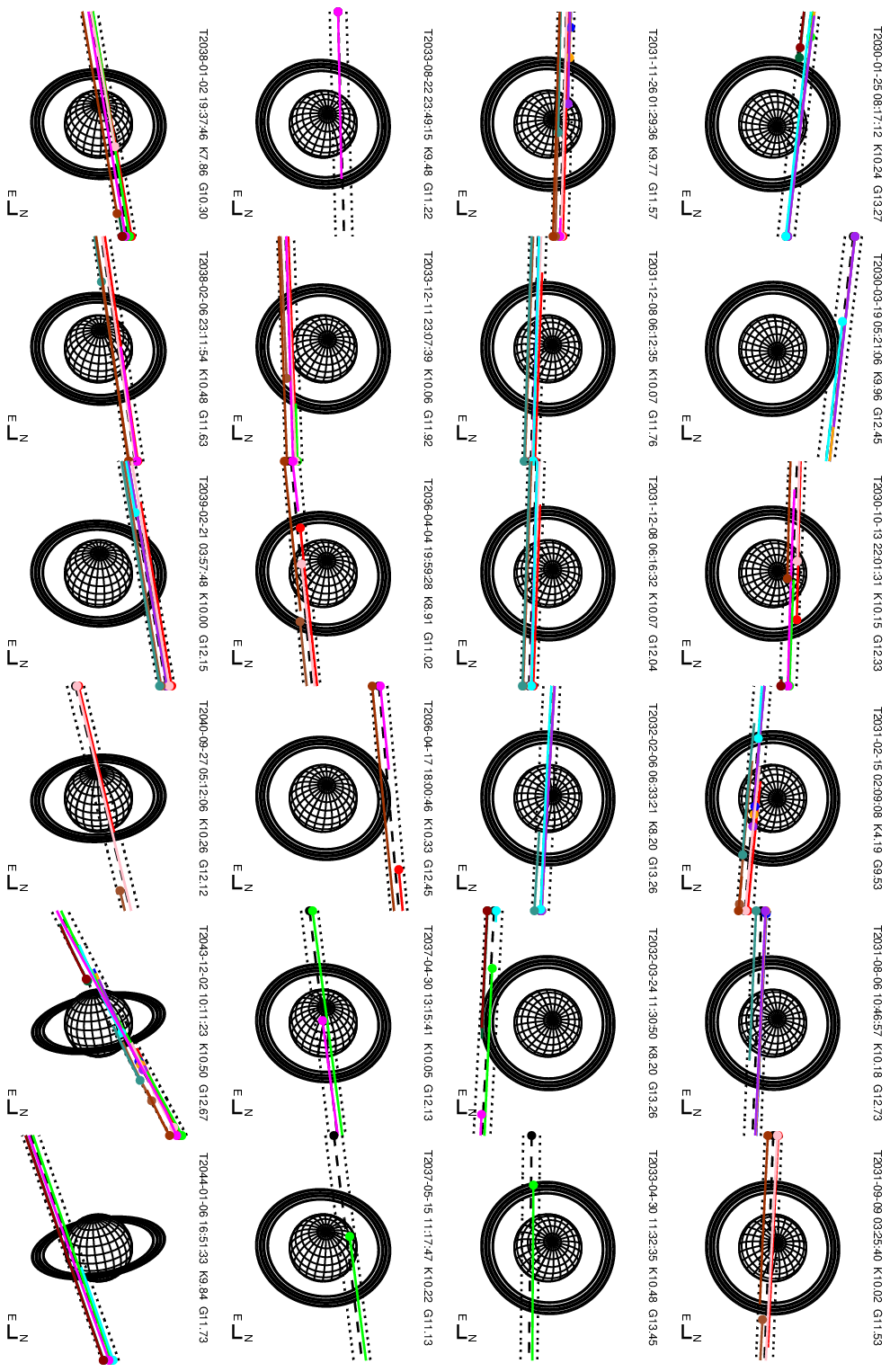}}}
\caption{Gallery of geocentric sky plane views of Uranus and its rings at mid-occultation for for the 24 brightest predicted events in the K band with ring event type {\it Rgt} for  2023-2050.}
\label{fig:UranusSky000}
\end{figure}

To illustrate the range of event geometries and the varying aspect of Uranus and its rings, {\bf Fig.~\ref{fig:UranusGlobes000}} shows galleries of views of Earth at mid-occultation and {\bf Fig.~\ref{fig:UranusSky000}} shows the corresponding sky-plane views of Uranus, for the 24 brightest predicted events in the K band with ring event type {\it Rgt} for 2023-2050. (Note the several of these events have no planet occultations.)
{\bf Table \ref{tbl:uroccpredK6}} provides detailed information this subset of our full prediction list.

The full set of prediction details is available in the SOM. \cite{Saunders2022} identified near-IR 56 Uranus occultations between 2025-01-01 and 2035-12-31 visible from low-Earth orbit, with a solar exclusion angle of $30^\circ$. Our computed geometry matches the subset of these events that satisfy our Sun-Geocenter-Target limit of $\geq 45^\circ$, more appropriate for ground-based observations.

\begin{longrotatetable}
\movetabledown=15mm
\begin{deluxetable}{c c c D D D D D D D D D D D D}
\tablecolumns{16}
\tabletypesize{\tiny}
\tablecaption{Geocentric Uranus Occultation Predictions 2023--2050}
\label{tbl:uroccpredK6}
\tablehead{
\\[-2.75em]
\colhead{Event ID} & 
\colhead{C/A Epoch} &
\colhead{ICRS Star Coord at Epoch} & 
\multicolumn2c{$\sigma(\alpha_*)$ (km)} &
\multicolumn2c{$\sigma(\delta_*)$ (km)} &
\multicolumn2c{C/A  (${'}{'}$)} & 
\multicolumn2c{PA (deg)} & 
\multicolumn2c{$v_{\rm sky}$ (km/s)} &
\multicolumn2c{Dist (au)} &
\multicolumn2c{Lat I (deg)} &
\multicolumn2c{K}& 
\multicolumn2c{G} & 
\multicolumn2c{RP} & 
\multicolumn2c{E $\lambda$ (deg)} & 
\multicolumn2c{S-G-T} \\[-.95em]
\colhead{Event type} & 
\colhead{Source ID} &
\colhead{Geocentric Object Position} &
\multicolumn2c{$f_0$  (km)} &
\multicolumn2c{$g_0$  (km)} &
\multicolumn2c{C/A ($10^3$ km)} &  
\multicolumn2c{P (deg)} & 
\multicolumn2c{B (deg)} &
\multicolumn2c{$\dot r$  (km/s)} &
\multicolumn2c{Lat E (deg)} &
\multicolumn2c{$D_*$ (km)} &
\multicolumn2c{G${}_*$} & 
\multicolumn2c{DUP} & 
\multicolumn2c{$\phi$ (deg)} &
\multicolumn2c{M-G-T} 
}
\decimals
\startdata
U30008 & 2030-01-25 08:17:12.56 & 04 52 12.89939  +22 33 45.77275  & 4.3 & 2.6 & 1.449 & 173.69 & -12.59 & 18.575 & 4.0 & 10.236 & 13.269 & 12.376 & 184.5 & 129.4 \\ 
\ \ \ PgtRgt & 3413198248000757760 & 04 52 12.91090  +22 33 44.33261  &  .  &  .  & 19.520 & -29.32 & 81.59 & 11.662 & 8.5 & 0.703 & 12.767 & 0 & 22.6 & 122.0 \\ 
U30011 & 2030-03-19 05:21:06.97 & 04 52 13.08033  +22 34 11.50958  & 4.1 & 2.8 & 3.582 & 173.06 & 13.45 & 19.419 & . & 9.957 & 12.449 & 11.689 & 176.4 & 76.2 \\ 
\ \ \ XRgt & 3413198449863123072 & 04 52 13.11162  +22 34 07.95411  &  .  &  .  & 50.446 & -29.30 & 81.58 & 0.037 & . & 0.768 & 12.018 & 0 & 22.6 & 96.7 \\ 
U30034* & 2030-10-13 22:01:31.53 & 05 26 40.70040  +23 17 50.97623  & 13.8 & 8.8 & 1.112 & 177.66 & -7.56 & 18.680 & 8.8 & 10.149 & 12.326 & 11.605 & 89.2 & 117.8 \\ 
\ \ \ PgtRgt & 3416213864799115264 & 05 26 40.70373  +23 17 49.86481  &  .  &  .  & 15.068 & 27.62 & 80.90 & 7.273 & 1.0 & 0.630 & 11.270 & 0 & 23.3 & 34.3 \\ 
U31023 & 2031-02-15 02:09:08.50 & 05 09 42.75173  +23 02 04.09133  & 10.6 & 7.2 & 1.057 & 355.51 & -5.18 & 18.741 & -4.0 & 4.194 & 9.526 & 8.105 & 260.7 & 112.7 \\ 
\ \ \ PgtRgt & 3415397339976857728 & 05 09 42.74577  +23 02 05.14505  &  .  &  .  & 14.367 & 0.83 & 82.14 & 5.016 & -5.3 &  .  & 8.059 & 0 & 23.1 & 159.1 \\ 
U31044 & 2031-08-06 10:46:57.05 & 05 40 11.78504  +23 28 48.06683  & 3.8 & 2.1 & 0.830 & 357.74 & 26.32 & 19.776 & -11.1 & 10.176 & 12.730 & 11.951 & 329.0 & 47.9 \\ 
\ \ \ PgtRgt & 3404409812904057216 & 05 40 11.78269  +23 28 48.89603  &  .  &  .  & 11.907 & 42.68 & 78.95 & 25.834 & 3.4 & 0.700 & 13.028 & 0 & 23.5 & 85.7 \\ 
U31055* & 2031-09-09 03:25:40.91 & 05 45 02.14615  +23 31 19.54458  & 18.5 & 11.8 & 0.205 & 357.76 & 11.91 & 19.267 & -10.2 & 10.015 & 11.532 & 10.964 & 47.3 & 79.3 \\ 
\ \ \ PgtRgt & 3427621572816169472 & 05 45 02.14560  +23 31 19.74884  &  .  &  .  & 2.860 & 46.90 & 78.14 & 12.061 & 8.4 & 0.616 & 10.969 & 0 & 23.5 & 6.2 \\ 
U31089 & 2031-11-26 01:29:36.37 & 05 41 02.73872  +23 30 58.00451  & 4.6 & 2.6 & 0.869 & 178.44 & -21.16 & 18.175 & 11.2 & 9.766 & 11.574 & 10.959 & 358.5 & 157.5 \\ 
\ \ \ PgtRgt & 3404413665492320896 & 05 41 02.74047  +23 30 57.13525  &  .  &  .  & 11.459 & 43.36 & 78.79 & 20.814 & -3.7 & 0.677 & 11.635 & 0 & 23.5 & 56.7 \\ 
U31101*\tablenotemark{\tiny a} & 2031-12-08 06:12:35.97 & 05 38 53.52466  +23 30 03.51695  & 53.5 & 29.3 & 0.778 & 358.16 & -23.02 & 18.117 & 3.8 & 10.067 & 11.758 & 10.788 & 275.2 & 170.3 \\ 
\ \ \ PgtRgt & 3404402532936959488 & 05 38 53.52288  +23 30 04.29469  &  .  &  .  & 10.227 & 41.33 & 79.13 & 22.743 & -10.5 & 0.531 & 11.911 & 0 & 23.5 & 93.0 \\ 
U31102\tablenotemark{\tiny a} & 2031-12-08 06:16:32.50 & 05 38 53.49479  +23 30 03.40218  & 299.9 & 182.3 & 0.880 & 358.16 & -23.02 & 18.117 & 3.2 & 10.067 & 12.044 & 10.777 & 274.3 & 170.3 \\ 
\ \ \ PgtRgt & 3404402532934661120 & 05 38 53.49277  +23 30 04.28140  &  .  &  .  & 11.561 & 41.33 & 79.13 & 22.610 & -10.7 & 0.531 & 12.196 & 0 & 23.5 & 93.0 \\ 
U32016 & 2032-02-06 06:33:21.50 & 05 29 26.29907  +23 24 38.41802  & 3.6 & 2.7 & 0.105 & 357.09 & -11.36 & 18.475 & 5.1 & 8.198 & 13.255 & 12.019 & 208.5 & 126.4 \\ 
\ \ \ PgtRgt & 3416228811285405568 & 05 29 26.29869  +23 24 38.52260  &  .  &  .  & 1.403 & 30.93 & 80.49 & 11.427 & -6.0 & 2.858 & 12.641 & 0 & 23.4 & 164.5 \\ 
U32021 & 2032-03-24 11:30:50.46 & 05 29 26.29906  +23 24 38.41771  & 3.8 & 2.9 & 3.250 & 356.93 & 12.00 & 19.233 & . & 8.198 & 13.255 & 12.019 & 87.6 & 79.1 \\ 
\ \ \ XRgt & 3416228811285405568 & 05 29 26.28641  +23 24 41.66337  &  .  &  .  & 45.339 & 30.92 & 80.49 & 5.540 & . & 2.976 & 12.700 & 0 & 23.4 & 72.4 \\ 
U33063 & 2033-04-30 11:32:35.90 & 05 53 12.16049  +23 38 38.54607  & 4.4 & 2.9 & 0.938 & 359.32 & 26.12 & 19.649 & -13.9 & 10.479 & 13.445 & 12.558 & 56.9 & 48.5 \\ 
\ \ \ PgtRgt & 3427825566583885184 & 05 53 12.15968  +23 38 39.48375  &  .  &  .  & 13.364 & 52.70 & 76.63 & 25.673 & 5.1 & 0.598 & 13.735 & 0 & 23.6 & 30.0 \\ 
U33070 & 2033-08-22 23:49:15.47 & 06 20 03.48811  +23 35 31.11235  & 3.8 & 2.7 & 0.957 & 181.53 & 23.43 & 19.536 & -10.4 & 9.480 & 11.220 & 10.622 & 126.6 & 55.1 \\ 
\ \ \ PgtRgt & 3377537095886180096 & 06 20 03.48628  +23 35 30.15520  &  .  &  .  & 13.560 & 66.42 & 71.32 & 23.617 & 19.3 & 0.817 & 11.392 & 0 & 23.6 & 33.7 \\ 
U33117 & 2033-12-11 23:07:39.05 & 06 18 09.99702  +23 38 35.93996  & 3.3 & 2.3 & 1.944 & 2.12 & -22.45 & 18.001 & -6.8 & 10.056 & 11.920 & 11.276 & 27.2 & 165.6 \\ 
\ \ \ PgtRgt & 3425562217600045184 & 06 18 10.00228  +23 38 37.88176  &  .  &  .  & 25.374 & 65.56 & 71.70 & 20.260 & -9.5 & 0.604 & 12.046 & 0 & 23.6 & 56.9 \\ 
U36026 & 2036-04-04 19:59:28.45 & 06 46 18.75718  +23 23 37.50621  & 6.1 & 4.9 & 1.193 & 5.33 & 8.78 & 18.863 & -23.7 & 8.915 & 11.020 & 10.371 & 328.5 & 85.5 \\ 
\ \ \ PgtRgt & 3379627306147808000 & 06 46 18.76526  +23 23 38.69378  &  .  &  .  & 16.326 & 74.62 & 65.71 & 9.058 & 12.2 & 1.166 & 10.126 & 0 & 23.4 & 15.7 \\ 
U36032* & 2036-04-17 18:00:46.68 & 06 47 28.93351  +23 22 14.87368  & 6.8 & 5.0 & 3.590 & 185.10 & 15.08 & 19.076 & . & 10.331 & 12.445 & 11.776 & 345.8 & 73.1 \\ 
\ \ \ XRgt & 3379612733322742784 & 06 47 28.91036  +23 22 11.29723  &  .  &  .  & 49.666 & 74.94 & 65.46 & 3.490 & . & 0.586 & 12.139 & 0 & 23.3 & 171.5 \\ 
U37018 & 2037-04-30 13:15:41.94 & 07 08 05.69920  +22 56 48.96675  & 4.4 & 3.8 & 0.014 & 187.08 & 18.74 & 19.139 & -28.7 & 10.050 & 12.132 & 11.466 & 49.8 & 65.6 \\ 
\ \ \ PgtRgt & 3368245848972447232 & 07 08 05.69911  +22 56 48.95149  &  .  &  .  & 0.201 & 79.84 & 60.92 & 21.238 & 28.8 & 0.647 & 12.061 & 0 & 22.9 & 124.4 \\ 
U37025 & 2037-05-15 11:17:47.17 & 07 10 30.94921  +22 52 40.55814  & 5.1 & 4.7 & 1.328 & 187.14 & 24.91 & 19.352 & -13.2 & 10.224 & 11.134 & 10.790 & 65.2 & 51.7 \\ 
\ \ \ PgtRgt & 3368274504994506112 & 07 10 30.93730  +22 52 39.23937  &  .  &  .  & 18.643 & 80.36 & 60.39 & 26.392 & 26.1 & 0.493 & 11.373 & 0 & 22.8 & 49.3 \\ 
U38002 & 2038-01-02 19:37:46.68 & 07 35 30.42467  +22 07 41.38425  & 6.2 & 5.3 & 0.570 & 189.56 & -22.84 & 17.736 & 36.5 & 7.855 & 10.298 & 9.565 & 77.4 & 169.9 \\ 
\ \ \ PgtRgt & 865677841359417216 & 07 35 30.41789  +22 07 40.82172  &  .  &  .  & 7.330 & 85.08 & 54.75 & 27.384 & -30.7 & 1.806 & 10.442 & 0 & 22.0 & 135.2 \\ 
U38017 & 2038-02-06 23:11:54.78 & 07 29 13.70384  +22 21 36.72236  & 5.1 & 4.2 & 1.137 & 188.56 & -20.02 & 17.817 & 32.5 & 10.480 & 11.628 & 11.215 & 347.7 & 152.8 \\ 
\ \ \ PgtRgt & 866402758821944960 & 07 29 13.69167  +22 21 35.59708  &  .  &  .  & 14.692 & 83.95 & 56.18 & 22.847 & -21.9 & 0.438 & 11.629 & 0 & 22.3 & 122.0 \\ 
U39021 & 2039-02-21 03:57:48.52 & 07 47 22.61794  +21 42 15.37628  & 4.5 & 3.6 & 2.761 & 190.01 & -17.29 & 17.840 & . & 10.003 & 12.151 & 11.482 & 267.0 & 143.0 \\ 
\ \ \ XRgt & 674533650447999360 & 07 47 22.58353  +21 42 12.65574  &  .  &  .  & 35.729 & 87.01 & 52.04 & 15.511 & . & 0.681 & 11.993 & 0 & 21.6 & 167.3 \\ 
U40024 & 2040-09-27 05:12:06.65 & 08 36 22.58698  +19 11 28.83981  & 6.1 & 4.1 & 0.242 & 193.73 & 22.56 & 19.070 & -47.8 & 10.257 & 12.118 & 11.496 & 45.0 & 57.6 \\ 
\ \ \ PgtRgt & 659766114072052608 & 08 36 22.58294  +19 11 28.60272  &  .  &  .  & 3.353 & 93.84 & 40.54 & 34.346 & 50.7 & 0.586 & 12.249 & 0 & 19.0 & 43.2 \\ 
U43014 & 2043-12-02 10:11:23.97 & 09 36 36.40808  +14 59 50.91957  & 6.8 & 7.1 & 0.998 & 207.71 & -2.54 & 18.079 & 62.4 & 10.500 & 12.670 & 11.985 & 280.7 & 108.0 \\ 
\ \ \ PgtRgt & 617976559021588608 & 09 36 36.37606  +14 59 50.03401  &  .  &  .  & 13.083 & 99.88 & 25.75 & 5.511 & -37.7 & 0.530 & 10.428 & 0 & 14.8 & 116.8 \\ 
U44001* & 2044-01-06 16:51:33.19 & 09 33 54.11947  +15 14 29.11231  & 9.3 & 7.0 & 1.565 & 19.50 & -17.60 & 17.593 & 26.5 & 9.838 & 11.732 & 11.121 & 145.2 & 144.5 \\ 
\ \ \ PgtRgt & 618790957835883648 & 09 33 54.15557  +15 14 30.58589  &  .  &  .  & 19.971 & 99.64 & 26.43 & 35.903 & -43.3 &  .  & 11.593 & 0 & 15.0 & 143.5 \\ 
\enddata
\end{deluxetable}
\end{longrotatetable}

\subsection{Neptune}\label{sec:Neptune}

The first modern occultation observation of Neptune occurred on 1968-04-07, when the planet occulted the star BD $-$17 4388 \citep{Kovalevsky1969}. The results were used to estimate the oblateness of Neptune, and this success ushered in a series of subsequent observation campaigns that provided a time and spatial history of the stratospheric temperature of the planet, as reviewed in detail by \cite{Roques1994}. After the occultation discovery of the Uranian rings, concerted efforts were made to detect rings around Neptune, ultimately resulting in the unexpected discovery {\red and subsequent characterization} of incomplete ring arcs 
\citep{Hubbard1986, Sicardy1986,Sicardy1991}. \cite{French1998} combined central flash observations from three atmospheric occultations to derive information about Neptune's stratospheric winds, and \cite{Uckert2014} investigated temperature variations in the upper stratosphere from a 2008-07-23 single-chord stellar occultation of the star USNO-B1.0 0759-0739128 (I=12.60). 

Just as for Uranus, Neptune passed out of the dense star fields of the Milky Way in the mid-1980's, as shown in {\bf Fig.~\ref{fig:starpaths} (lower right)}.  Owing to Neptune's long orbital period (165 years), the next traversal of the Milky Way will not occur until about 2068. Consequently, very few high-SNR opportunities are available for Neptune occultations in the coming decades. Our search resulted in the identification of 88 predicted occultations by Neptune and/or its rings for the period 2023-2050, for a limiting magnitude K$\leq$15. {\bf Table \ref{tbl:nepoccpredstats}} lists the number of predicted events per year as a function of K magnitude, {\red excluding years with no predicted events.} The format is the same as for Table \ref{tbl:uroccpredstats}. 


\begin{deluxetable}{c c c c c c c c }
\tabletypesize{\scriptsize} 
\tablecaption{Neptune Planet/Ring Occultations 2023--2050 }
\label{tbl:nepoccpredstats}
\tablehead{
\colhead{Year}& 
\colhead{K $<$ 9} &
\colhead{K 9--10} &
\colhead{K 10--11} &
\colhead{K 11--12} &
\colhead{K 12--13} &
\colhead{K 13--14} &
\colhead{K 14--15} \\[-0.95em]
\colhead{ }& 
\colhead{P/R} &
\colhead{P/R} &
\colhead{P/R} &
\colhead{P/R} &
\colhead{P/R} &
\colhead{P/R} &
\colhead{P/R} \\[-1.05em]
}
\startdata
2023 & -- & 1/1 & -- & 1/1 & -- & -- & -- \\
2024 & -- & 1/1 & -- & -- & 1/1 & 2/2 & 1/2 \\
2025 & -- & -- & -- & -- & 1/1 & 0/1 & -- \\
2026 & -- & -- & -- & -- & -- & -- & 1/1 \\
2027 & -- & -- & -- & -- & 1/1 & 2/2 & 1/1 \\
2028 & -- & -- & -- & 1/1 & -- & 1/1 & -- \\
2029 & -- & -- & -- & -- & -- & 2/2 & 2/2 \\
2030 & -- & -- & -- & -- & -- & -- & 0/1 \\
2031 & -- & -- & -- & -- & -- & -- & 2/2 \\
2032 & 1/1 & -- & -- & 1/1 & -- & -- & 1/1 \\
2033 & -- & -- & -- & 1/1 & -- & -- & 2/2 \\
2035 & -- & 1/1 & -- & -- & 1/1 & 3/3 & 3/3 \\
2036 & -- & -- & -- & 2/2 & 1/1 & -- & 2/2 \\
2037 & -- & -- & -- & 0/1 & -- & 2/2 & -- \\
2038 & -- & -- & -- & -- & 1/1 & -- & 1/1 \\
2040 & -- & -- & -- & -- & 0/1 & 1/0 & -- \\
2041 & -- & -- & -- & -- & 1/1 & 3/3 & 3/2 \\
2042 & -- & -- & -- & -- & 2/1 & -- & 3/3 \\
2043 & -- & -- & -- & -- & -- & -- & 1/0 \\
2044 & -- & -- & 0/1 & -- & -- & -- & 1/1 \\
2045 & -- & -- & -- & -- & -- & 1/1 & 1/1 \\
2046 & -- & -- & 1/0 & -- & -- & 2/2 & 3/3 \\
2047 & -- & -- & -- & -- & 1/1 & 4/3 & 1/1 \\
2048 & -- & -- & -- & -- & -- & -- & 1/1 \\
2049 & -- & -- & -- & 1/1 & -- & 2/2 & 4/2 \\[0.5em]
Totals & 1/1 & 3/3 & 1/1 & 7/8 & 10/10 & 25/24 & 34/32 \\

\enddata
\end{deluxetable}

The aspect of Neptune's rings varies over time, as shown in {\bf Fig.~\ref{fig:BBp} (bottom)} for the period 1975-2050. The solid line shows the sub-Earth latitude, with an annual periodic term reflecting the relative inclinations of the orbits of Earth and Neptune, and the dashed line shows the sub-solar latitude. Black dots mark the times of previous successful Neptune occultations \citep{Roques1994}, and red dots mark all predicted events from 2023-2050, for K $\leq$12.5. We have not made detailed predictions for individual ring arcs because there is evidence that they are evolving on a decadal timescale, with the leading arcs Libert\'e and Courage having recently faded away \citep{Souami2022}.

\begin{figure}
\centerline{\resizebox{6in}{!}{\includegraphics[angle=0]{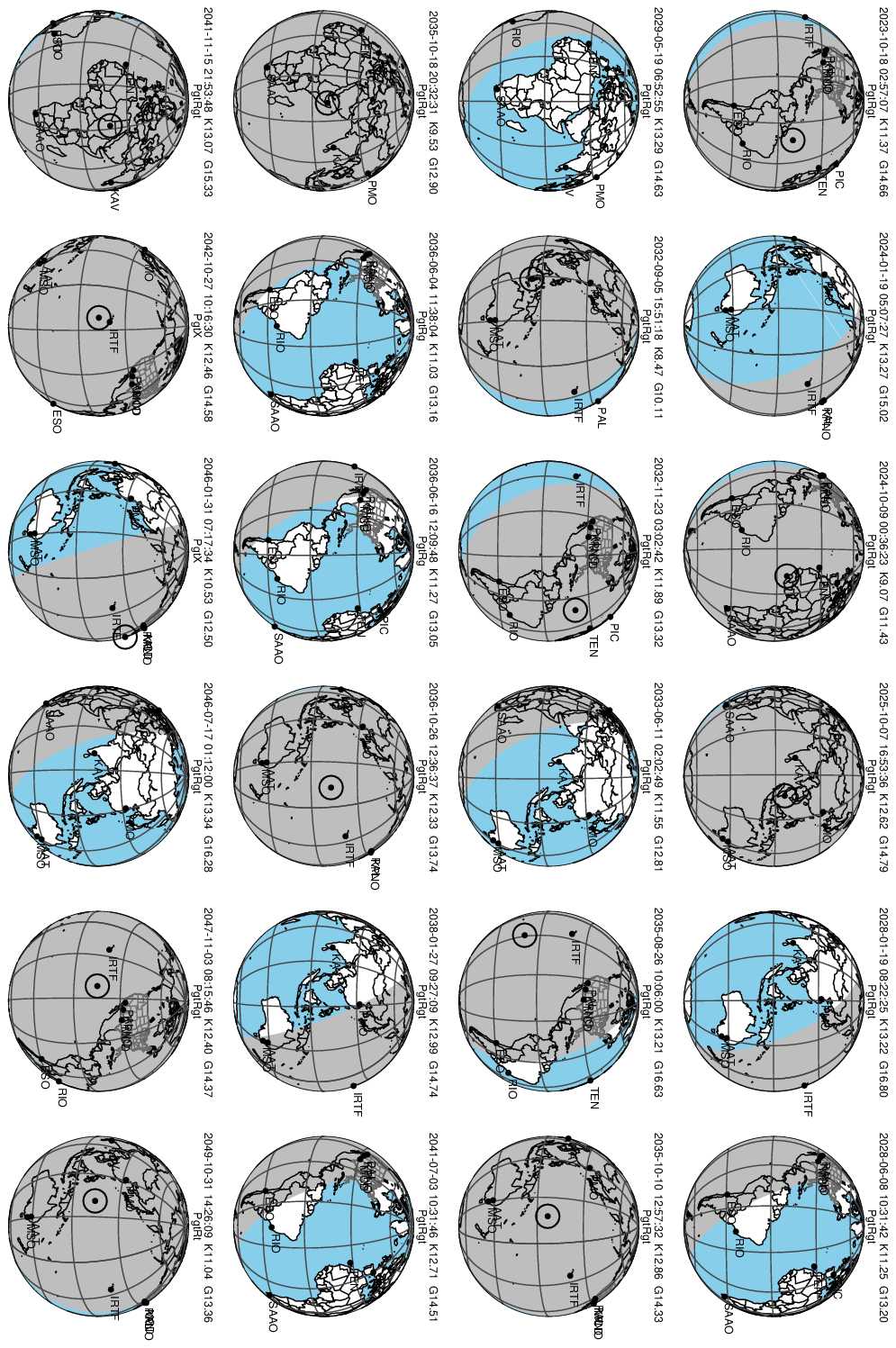}}}
\caption{\red Gallery of Earth views from Neptune at mid-occultation for for the 24 brightest predicted events in the K band with planet event type {\it Pgt} for 2023-2050.}
\label{fig:NeptuneGlobes000}
\end{figure}

\begin{figure}
\centerline{\resizebox{6in}{!}{\includegraphics[angle=0]{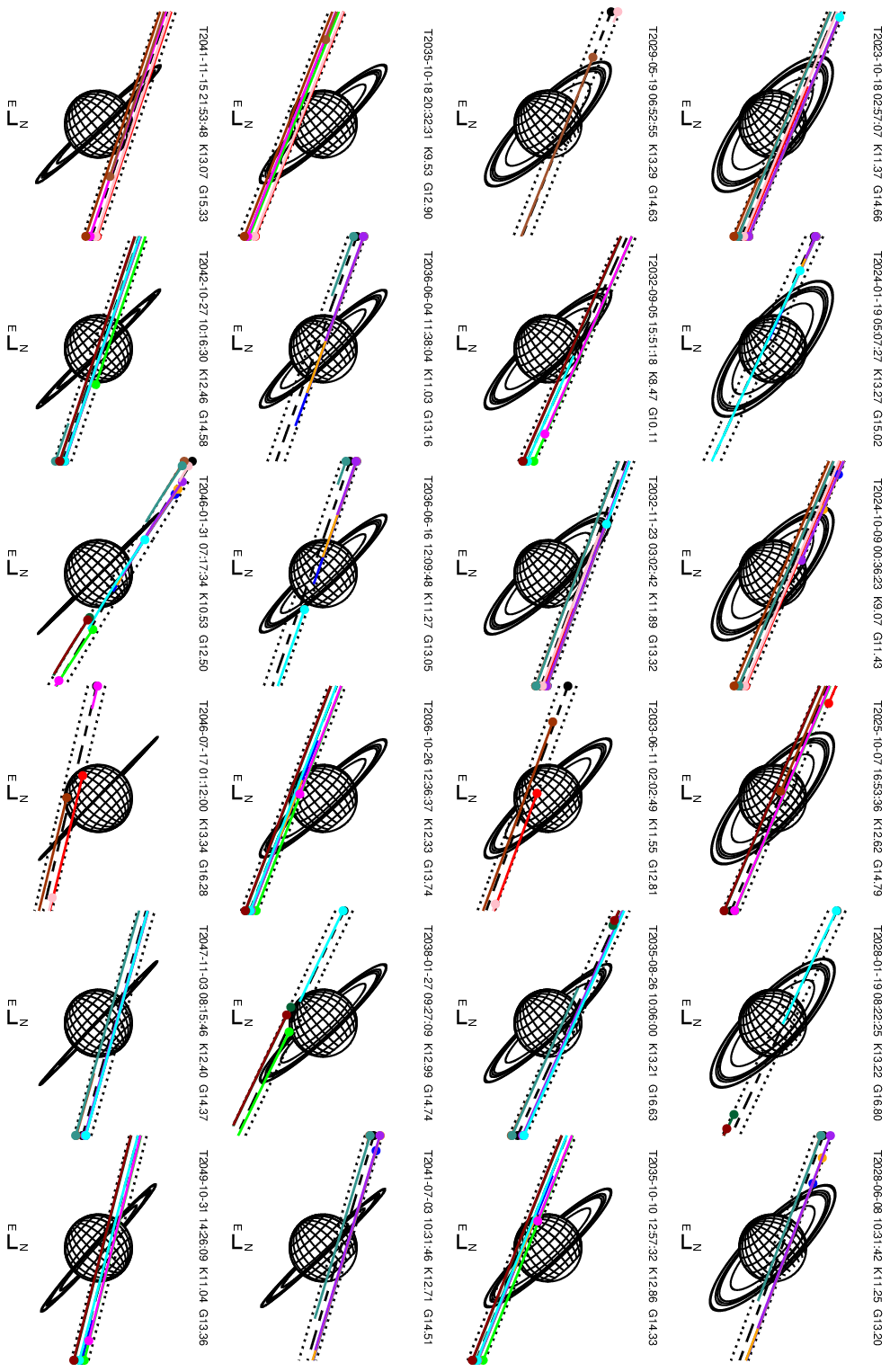}}}
\caption{\red Gallery of geocentric sky plane views of Neptune and its rings at mid-occultation for for the 24 brightest predicted events in the K band with planet event type {\it Pgt} for 2023-2050.}
\label{fig:NeptuneSky000}
\end{figure}

{\red To illustrate the range of event geometries and the varying orientation of Neptune and its rings, {\bf Fig.~\ref{fig:NeptuneGlobes000}} shows galleries of views of Earth at mid-occultation and {\bf Fig.~\ref{fig:NeptuneSky000}} shows the corresponding sky-plane views of Neptune, for the 24 brightest predicted events in the K band with planet event type {\it Pgt} for 2023-2050. (Note that some of these events do not have ring occultations.)
{\bf Table \ref{tbl:nepoccpredK6}} provides detailed information for this set of events.}

The full set of prediction details for K$\leq$15 is included in the SOM. \cite{Saunders2022} identified 14 near-IR Neptune occultations between 2025-01-01 and 2035-12-31 visible from low-Earth orbit, with a solar exclusion angle of $30^\circ$, of which six are visible from the ground. Our computed geometry matches the subset of these events that satisfy our Sun-Geocenter-Target limit of $\geq 45^\circ$, more appropriate for ground-based observations.

\begin{longrotatetable}
\movetabledown=6mm
\begin{deluxetable}{l l c D D D D D D D D D D D D}
\tablecolumns{14}
\tabletypesize{\tiny}
\tablecaption{Geocentric Neptune Occultation Predictions 2023--2050}
\label{tbl:nepoccpredK6}
\tablehead{
\\[-2.75em]
\colhead{Event ID} & 
\colhead{C/A Epoch} &
\colhead{ICRS Star Coord at Epoch} & 
\multicolumn2c{$\sigma(\alpha_*)$ (km)} &
\multicolumn2c{C/A  (${'}{'}$)} & 
\multicolumn2c{PA (deg)} & 
\multicolumn2c{$v_{\rm sky}$ (km/s)} &
\multicolumn2c{Dist (au)} &
\multicolumn2c{Lat I (deg)} &
\multicolumn2c{K}& 
\multicolumn2c{G} & 
\multicolumn2c{RP} & 
\multicolumn2c{E $\lambda$ (deg)} & 
\multicolumn2c{S-G-T} \\[-.95em]
\colhead{Event type} & 
\colhead{Source ID} &
\colhead{Geocentric Object Position} &
\multicolumn2c{$\sigma(\delta_*)$ (km)} &
\multicolumn2c{C/A ($10^3$ km)} &  
\multicolumn2c{P (deg)} & 
\multicolumn2c{B (deg)} &
\multicolumn2c{$\dot r$  (km/s)} &
\multicolumn2c{Lat E (deg)} &
\multicolumn2c{$D_*$ (km)} &
\multicolumn2c{G${}_*$} & 
\multicolumn2c{DUP} & 
\multicolumn2c{$\phi$ (deg)} &
\multicolumn2c{M-G-T} 
}
\decimals
\startdata
N23002 & 2023-10-18 02:57:07.97 & 23 44 21.02634  $-$03 04 57.66860  & 5.1 & 0.514 & 157.20 & -20.44 & 29.030 & 7.4 & 11.366 & 14.660 & 13.616 & 285.8 & 151.0 \\ 
\ \ \ PgtRgt & 2639715216842421504 & 23 44 21.03963  $-$03 04 58.14225  & 3.4 & 10.818 & -41.03 & -21.45 & 24.177 & 41.4 &  .  & 14.684 & 0 & -3.0 & 111.4 \\ 
N24001* & 2024-01-19 05:07:27.08 & 23 43 56.90963  $-$03 04 31.98483  & 121.0 & 0.288 & 335.52 & 21.94 & 30.430 & 1.6 & 13.274 & 15.018 & 14.343 & 161.3 & 56.8 \\ 
\ \ \ PgtRgt & 2639725730922360960 & 23 43 56.90166  $-$03 04 31.72277  & 80.8 & 6.355 & -41.00 & -21.52 & 26.254 & -29.8 & 0.199 & 15.118 & 0 & -2.9 & 46.5 \\ 
N24004* & 2024-10-09 00:36:23.19 & 23 53 34.80647  $-$02 08 27.13561  & 6.1 & 0.502 & 156.84 & -22.69 & 28.946 & 6.1 & 9.068 & 11.430 & 10.707 & 331.5 & 161.8 \\ 
\ \ \ PgtRgt & 2449274030476179456 & 23 53 34.81963  $-$02 08 27.59619  & 5.6 & 10.529 & -42.04 & -20.63 & 27.771 & 41.6 &  .  & 11.567 & 0 & -2.0 & 92.7 \\ 
N25002*** & 2025-10-07 16:53:36.65 & 00 02 12.02238  $-$01 15 34.15602  & 22.4 & 0.214 & 156.75 & -23.27 & 28.918 & -9.0 & 12.624 & 14.794 & 14.096 & 90.9 & 165.6 \\ 
\ \ \ PgtRgt & 2449616081671057024 & 00 02 12.02800  $-$01 15 34.35244  & 10.3 & 4.484 & -42.90 & -19.82 & 31.329 & 29.0 & 0.296 & 14.958 & 0 & -1.1 & 22.3 \\ 
N28001 & 2028-01-19 08:22:25.33 & 00 16 12.87501  +00 13 14.47638  & 31.8 & 0.344 & 154.78 & 17.92 & 30.263 & 35.5 & 13.225 & 16.799 & 15.617 & 120.5 & 65.4 \\ 
\ \ \ PgtRgt & 2545537441718802304 & 00 16 12.88479  +00 13 14.16520  & 18.6 & 7.550 & -44.16 & -18.49 & 23.923 & -1.2 & 0.333 & 16.680 & 0 & 0.4 & 148.0 \\ 
N28002 & 2028-06-08 10:31:42.71 & 00 32 23.72894  +01 56 34.34487  & 5.2 & 0.278 & 159.59 & 15.92 & 30.204 & 38.2 & 11.254 & 13.196 & 12.575 & 313.2 & 69.6 \\ 
\ \ \ PgtRgt & 2544493455427430656 & 00 32 23.73541  +01 56 34.08442  & 4.1 & 6.088 & -45.39 & -16.87 & 26.715 & -10.6 & 0.596 & 12.949 & 0 & 2.1 & 96.1 \\ 
N29001 & 2029-05-19 06:52:55.15 & 00 38 43.65120  +02 35 48.97429  & 8.5 & 0.577 & 158.36 & 25.32 & 30.523 & 53.0 & 13.292 & 14.632 & 14.147 & 29.6 & 48.3 \\ 
\ \ \ PgtRgt & 2550603540326304000 & 00 38 43.66541  +02 35 48.43754  & 6.5 & 12.783 & -45.79 & -16.20 & 39.473 & 6.9 & 0.199 & 14.888 & 0 & 2.8 & 114.4 \\ 
N32002 & 2032-09-05 15:51:18.68 & 01 04 39.16217  +05 06 35.93430  & 10.9 & 0.936 & 155.97 & -19.36 & 29.002 & 29.6 & 8.475 & 10.109 & 9.558 & 153.4 & 146.3 \\ 
\ \ \ PgtRgt & 2552125131275826816 & 01 04 39.18769  +05 06 35.08023  & 6.4 & 19.681 & -47.00 & -13.26 & 26.551 & 71.2 & 1.898 & 10.074 & 0 & 5.3 & 154.5 \\ 
N32003 & 2032-11-23 03:02:42.04 & 00 57 23.03121  +04 21 33.53876  & 9.2 & 1.148 & 159.39 & -15.23 & 29.152 & 51.4 & 11.890 & 13.316 & 12.803 & 266.4 & 134.0 \\ 
\ \ \ PgtRgt & 2551779163070218880 & 00 57 23.05822  +04 21 32.46412  & 6.1 & 24.274 & -46.72 & -14.07 &  .  & 70.2 & 0.376 & 13.020 & 0 & 4.5 & 115.8 \\ 
N33002 & 2033-06-11 02:02:49.21 & 01 13 12.62605  +06 02 16.99800  & 7.2 & 0.656 & 340.02 & 19.95 & 30.325 & -8.4 & 11.555 & 12.815 & 12.370 & 88.3 & 60.9 \\ 
\ \ \ PgtRgt & 2576622765738672384 & 01 13 12.61104  +06 02 17.61501  & 5.7 & 14.418 & -47.28 & -12.33 & 41.034 & -61.3 & 0.451 & 12.812 & 0 & 6.2 & 141.6 \\ 
N35003 & 2035-08-26 10:06:00.86 & 01 31 02.70022  +07 41 16.79568  & 27.5 & 0.672 & 155.31 & -13.21 & 29.194 & 12.4 & 13.205 & 16.625 & 15.510 & 257.2 & 128.6 \\ 
\ \ \ PgtRgt & 2566530554665743232 & 01 31 02.71911  +07 41 16.18487  & 18.0 & 14.234 & -47.58 & -10.20 & 26.455 & 57.5 &  .  & 16.175 & 0 & 7.9 & 39.2 \\ 
N35005 & 2035-10-10 12:57:32.31 & 01 27 12.82180  +07 17 09.94514  & 11.3 & 0.854 & 337.99 & -24.20 & 28.842 & -69.6 & 12.862 & 14.332 & 13.809 & 168.9 & 173.5 \\ 
\ \ \ PgtRgt & 2566173934941174400 & 01 27 12.80029  +07 17 10.73695  & 5.7 & 17.865 & -47.53 & -10.63 & 47.847 & -20.9 & 0.232 & 14.539 & 0 & 7.5 & 84.1 \\ 
N35006* & 2035-10-18 20:32:31.84 & 01 26 20.70963  +07 11 58.53716  & 11.2 & 1.003 & 338.27 & -24.35 & 28.839 & -78.0 & 9.534 & 12.901 & 11.820 & 46.7 & 177.2 \\ 
\ \ \ PgtRgt & 2566151184502925312 & 01 26 20.68469  +07 11 59.46989  & 6.4 & 20.979 & -47.52 & -10.73 &  .  & -32.4 &  .  & 13.114 & 0 & 7.4 & 25.3 \\ 
N36001 & 2036-06-04 11:38:04.68 & 01 37 13.19027  +08 20 50.47105  & 8.1 & 0.210 & 340.27 & 25.17 & 30.495 & 17.4 & 11.032 & 13.163 & 12.496 & 316.8 & 48.4 \\ 
\ \ \ PgtRg & 2571920768686360320 & 01 37 13.18549  +08 20 50.66903  & 4.9 & 4.652 & -47.63 & -9.52 & 73.564 & -39.1 & 0.675 & 13.412 & 0 & 8.5 & 175.3 \\ 
N36003 & 2036-06-16 12:09:48.36 & 01 38 21.67204  +08 26 48.47893  & 9.0 & 0.376 & 340.96 & 20.49 & 30.332 & 9.2 & 11.274 & 13.051 & 12.453 & 297.3 & 59.6 \\ 
\ \ \ PgtRg & 2571943656567597312 & 01 38 21.66377  +08 26 48.83464  & 5.3 & 8.278 & -47.63 & -9.37 & 60.471 & -48.3 &  .  & 13.077 & 0 & 8.6 & 36.5 \\ 
N36004 & 2036-10-26 12:36:37.97 & 01 34 07.81260  +07 56 19.45741  & 10.9 & 0.980 & 338.84 & -24.08 & 28.846 & -78.0 & 12.326 & 13.738 & 13.247 & 159.4 & 171.4 \\ 
\ \ \ PgtRgt & 2565887099845330048 & 01 34 07.78880  +07 56 20.37112  & 5.6 & 20.498 & -47.60 & -9.81 &  .  & -29.8 & 0.297 & 13.939 & 0 & 8.1 & 87.3 \\ 
N38001 & 2038-01-27 09:27:09.21 & 01 38 37.16971  +08 25 12.23641  & 16.9 & 0.997 & 334.47 & 11.11 & 29.995 & -39.4 & 12.995 & 14.737 & 14.162 & 116.6 & 78.9 \\ 
\ \ \ PgtRgt & 2571895656013097984 & 01 38 37.14076  +08 25 13.13604  & 8.6 & 21.688 & -47.62 & -9.32 &  .  & -78.9 & 0.239 & 14.099 & 0 & 8.6 & 174.6 \\ 
N41001 & 2041-07-03 10:31:46.47 & 02 21 31.58020  +12 15 31.13356  & 21.5 & 0.562 & 163.61 & 18.26 & 30.236 & 60.9 & 12.709 & 14.512 & 13.907 & 316.1 & 64.2 \\ 
\ \ \ PgtRgt & 73052518917687552 & 02 21 31.59101  +12 15 30.59482  & 13.4 & 12.315 & -46.87 & -4.17 & 119.164 & -0.7 & 0.280 & 14.413 & 0 & 12.4 & 119.6 \\ 
N41007 & 2041-11-15 21:53:48.53 & 02 15 47.20450  +11 40 57.73039  & 19.5 & 1.029 & 161.43 & -23.11 & 28.864 & 32.7 & 13.074 & 15.326 & 14.592 & 10.7 & 162.2 \\ 
\ \ \ PgtRgt & 72596049793554048 & 02 15 47.22681  +11 40 56.75483  & 17.3 & 21.544 & -47.06 & -4.80 &  .  & 85.2 & 0.244 & 15.483 & 0 & 11.9 & 110.4 \\ 
N42004 & 2042-10-27 10:16:30.91 & 02 26 47.28244  +12 35 47.96408  & 18.5 & 0.081 & 161.22 & -24.41 & 28.817 & -24.6 & 12.459 & 14.581 & 13.889 & 207.2 & 174.7 \\ 
\ \ \ PgtX & 74364167570243072 & 02 26 47.28423  +12 35 47.88708  & 15.2 & 1.700 & -46.66 & -3.47 &  .  & 32.6 & 0.326 & 14.797 & 0 & 12.8 & 24.8 \\ 
N46001 & 2046-01-31 07:17:34.03 & 02 47 01.60013  +14 15 06.48169  & 15.6 & 0.733 & 148.87 & 3.91 & 29.749 & 55.2 & 10.532 & 12.497 & 11.862 & 162.3 & 92.7 \\ 
\ \ \ PgtX & 32828466563598336 & 02 47 01.62622  +14 15 05.85536  & 13.1 & 15.815 & -45.67 & -1.08 &  .  & 26.0 & 0.818 & 10.725 & 0 & 14.4 & 155.8 \\ 
N46003 & 2046-07-17 01:12:00.03 & 03 05 35.37395  +15 37 10.20050  & 38.8 & 0.930 & 346.66 & 17.48 & 30.218 & -25.0 & 13.336 & 16.283 & 15.437 & 94.0 & 65.7 \\ 
\ \ \ PgtRgt & 30960533746904704 & 03 05 35.35908  +15 37 11.10584  & 33.6 & 20.392 & -44.47 & 1.13 &  .  & -86.5 & 0.283 & 16.137 & 0 & 15.8 & 125.9 \\ 
N47006 & 2047-11-03 08:15:46.33 & 03 11 23.56472  +15 56 12.47354  & 16.6 & 0.537 & 164.23 & -24.16 & 28.838 & -1.3 & 12.402 & 14.373 & 13.732 & 242.0 & 170.2 \\ 
\ \ \ PgtRgt & 31391714103977088 & 03 11 23.57484  +15 56 11.95685  & 13.4 & 11.230 & -44.04 & 1.87 & 325.106 & 56.1 & 0.335 & 14.579 & 0 & 16.1 & 6.3 \\ 
N49005 & 2049-10-31 14:26:09.08 & 03 30 10.79228  +17 08 58.17776  & 12.7 & 0.357 & 165.48 & -23.41 & 28.864 & -10.7 & 11.036 & 13.363 & 12.646 & 156.4 & 163.4 \\ 
\ \ \ PgtRt & 54870341808035968 & 03 30 10.79852  +17 08 57.83247  & 9.2 & 7.467 & -42.49 & 4.07 &  .  & 46.3 & 0.671 & 13.534 & 0 & 17.3 & 133.9 \\ 
\enddata
\end{deluxetable}
\end{longrotatetable}

The most promising near-term Neptune occultations are the {\red 2023-10-18 and 2024-10-09} events, both of which intersect the planet and the unblocked orbits of the ring arcs. Uncertainties in the mean motions of the arcs make it challenging to predict the longitudes of the arcs, and thus the prospects for detecting them during any given occultation from a specific observing site are difficult to quantify. Improved ephemerides for the arcs may eventually be derived from {\it JWST} images, at which time it will be warranted to revisit our predictions to estimate the prospects for detecting arcs. In the meantime, the NASA/PDS Ring-Moon Systems node provides online tools that can be used to visualize the predicted locations of the arcs for a variety of assumed mean motions.\footnote{\url https://pds-rings.seti.org/tools/} Next in line are the {\red 2032-09-05 (K=8.5) and 2035-10-18 (K=9.5)} events, both with excellent planet occultations and unblocked sampling of the ring regions.

\subsection{Titan}\label{sec:Titan}
The first extensive occultation observations of Titan's atmosphere were obtained during the {\red 1989-07-03} occultation of 28 Sgr that successfully probed Saturn's rings and atmosphere as well, as described above. 
\cite{Sicardy1999} analyzed a dozen atmospheric lightcurves and derived profiles of density and temperature between altitude levels $z$ = 290 -- 500 km (pressures $P$ from 110 to 1.4 $\mu$bar). The horizontal stratification of the atmosphere was determined from comparison of multiple adjacent chords, with typical horizontal-to-vertical aspect ratios of 15 to 45.
Subsequently, \cite{Sicardy2006} reported on two Titan stellar occultations that occurred on {\red 2003-11-14}. The lightcurves revealed a sharp inversion layer near 515 $\pm$ 6 km altitude {\red ($P\simeq1.5 \ \mu$bar)}. Central flashes observed during the first occultation provided constraints on the zonal wind regime at $z=250$ km. Simultaneous observations of the flashes at various wavelengths enabled the measurement of the wavelength dependence of atmospheric hazes
These results demonstrate the value of continued Earth-based occultation surveillance of Titan's atmosphere even after the highly successful {\it Cassini} and {\it Huygens} remote and in situ observations \citep{Coustenis2010}. 

{\red Our search identified 1769 Titan occultations from 2023-2050, for a limiting magnitude K$\leq$15.} Just as for Saturn, the frequency of predicted occultation opportunities for Titan varies dramatically over the coming decades. {\bf Table \ref{tbl:titanoccpredstats}} shows the number of events by year and K magnitude. {\red Since scientifically useful results can be obtained from Titan occultations using moderate size telescopes at visual wavelengths, we include the statistics for events with magnitude G$_*\leq$19 in} {\bf Table \ref{tbl:titanoccpredstatsG}}.  Each entry is of the form $P/g/t$, where $P$ indicates the number of predicted occultations for which Titan's shadow crosses the Earth but which are not predicted to be observable from any of our selected sites, $g$ indicates the number of predicted occultations in which the geocentric chord intersects Titan, but not from our altitude-limited observing sites, and $t$ indicates the number of events observable from one of the 13 topocentric sites under our usual altitude constraints. Qualitatively, $P$ events can in many instances be challenging to observe, $g$ events may be observable from large telescopes that are not in our selected set, and $t$ events are predicted to be observable from at least one of our observing sites while satisfying the usual Sun and target altitude constraints. The typical cadence of bright (K$\leq$9, G$_*\leq$12) events is about one every other year until 2046--2048, when Saturn and Titan cross the densest regions of the Milky Way. {\red For the years 2047 and beyond (when Saturn crosses the Milky Way), the number of Titan events is quite large. The decrease in the number of predicted events in 2047 for K=13--15 compared to K=12--13 (and similarly for G$_*\geq$18 compared to G$_*$=17--18) reflects the challenge of matching {\it Gaia} stars with their counterparts in the  2MASS catalog in dense star regions.}

\begin{deluxetable}{c c c c c c c  c c c c c}
\tabletypesize{\scriptsize} 
\tablecaption{Titan Occultations 2023--2050 (by K magnitude)}
\label{tbl:titanoccpredstats}
\tablehead{
\colhead{Year}& 
\colhead{K $\leq$5} &
\colhead{K 5--6} &
\colhead{K 6--7} &
\colhead{K 7--8} &
\colhead{K 8--9} &
\colhead{K 9--10} &
\colhead{K 10--11} &
\colhead{K 11--12} &
\colhead{K 12--13} &
\colhead{K 13--14} &
\colhead{K 14--15} \\[-.75em]
\colhead{ }& 
\colhead{P/g/t} &
\colhead{P/g/t} &
\colhead{P/g/t} &
\colhead{P/g/t} &
\colhead{P/g/t} &
\colhead{P/g/t} &
\colhead{P/g/t} &
\colhead{P/g/t } &
\colhead{P/g/t} &
\colhead{P/g/t} &
\colhead{P/g/t}
}
\startdata
2023 & -- & -- & -- & -- & -- & -- & -- & 0/0/1 & -- & 1/0/0 & 5/0/2 \\
2024 & -- & -- & -- & -- & -- & -- & 1/0/0 & -- & 0/1/2 & 1/2/1 & 4/2/2 \\
2025 & -- & -- & -- & -- & -- & -- & -- & 0/0/2 & 1/0/0 & 0/0/1 & 2/2/2 \\
2026 & -- & -- & -- & -- & -- & 0/0/1 & -- & -- & 1/0/1 & 1/0/1 & 1/2/1 \\
2027 & -- & -- & -- & -- & -- & -- & -- & 2/0/0 & 3/2/3 & -- & 0/1/4 \\
2028 & -- & -- & -- & -- & -- & -- & -- & -- & 0/0/2 & 3/0/0 & 1/2/1 \\
2029 & -- & -- & 0/0/1 & -- & -- & -- & 0/0/2 & 1/1/0 & -- & 1/1/1 & 2/1/2 \\
2030 & -- & -- & -- & -- & -- & -- & -- & 0/3/4 & 2/0/1 & 0/3/5 & 5/2/4 \\
2031 & 1/0/0 & -- & -- & -- & 0/1/1 & -- & 1/1/0 & 1/2/2 & 1/2/3 & 7/1/3 & 15/10/11 \\
2032 & -- & -- & -- & -- & -- & 0/1/1 & 0/0/1 & 3/0/1 & 2/2/5 & 10/6/10 & 16/12/22 \\
2033 & -- & -- & -- & -- & -- & -- & 1/0/0 & 0/3/6 & 3/2/3 & 9/9/11 & 14/13/20 \\
2034 & -- & -- & -- & -- & -- & 0/1/1 & 1/0/0 & 1/0/1 & 2/0/0 & 2/0/1 & 3/5/8 \\
2035 & -- & -- & -- & -- & -- & -- & -- & 3/2/4 & 2/0/1 & 2/2/3 & 5/2/4 \\
2036 & -- & -- & -- & -- & 0/1/0 & -- & -- & -- & 0/1/0 & 1/1/3 & 1/2/4 \\
2037 & -- & -- & -- & -- & -- & -- & -- & 2/0/0 & 1/0/0 & 0/3/3 & 2/3/3 \\
2038 & -- & -- & -- & -- & -- & -- & -- & -- & -- & -- & 2/0/1 \\
2039 & -- & -- & -- & -- & -- & -- & -- & 1/0/0 & -- & 0/1/1 & 1/0/1 \\
2040 & -- & -- & -- & -- & -- & -- & -- & 2/0/1 & 1/1/1 & 2/1/1 & 2/0/1 \\
2041 & -- & -- & -- & -- & -- & -- & -- & -- & 1/0/1 & 2/0/0 & 0/3/4 \\
2042 & -- & -- & -- & -- & -- & -- & -- & -- & 0/0/1 & 1/3/3 & 1/4/6 \\
2043 & -- & -- & -- & -- & -- & -- & 0/0/1 & 0/0/2 & 1/0/1 & 1/0/3 & 2/2/2 \\
2044 & -- & -- & -- & -- & -- & 0/0/1 & -- & 1/1/2 & 1/0/2 & 1/1/2 & 4/0/5 \\
2045 & -- & -- & -- & -- & -- & 1/0/0 & 1/0/0 & 1/0/0 & 1/1/5 & 6/2/6 & 10/5/11 \\
2046 & -- & -- & -- & 1/0/1 & 2/0/0 & 0/1/2 & 4/4/8 & 4/2/8 & 14/13/23 & 33/18/41 & 49/46/80 \\
2047 & -- & -- & -- & 1/2/3 & 3/3/6 & 8/4/7 & 22/18/32 & 55/44/81 & 93/71/124 & 89/61/112 & 11/11/18 \\
2048 & -- & -- & -- & 2/0/0 & 0/0/1 & -- & 3/1/3 & 9/1/2 & 6/14/21 & 21/15/36 & 51/46/68 \\
2049 & -- & -- & -- & -- & -- & 1/0/0 & 1/1/2 & 2/1/2 & 2/4/6 & 7/2/5 & 16/11/11 \\[0.5em]
Totals & 1/0/0 & 0/0/0 & 0/0/1 & 4/2/4 & 5/5/8 & 10/7/13 & 35/25/49 & 88/60/119 & 138/114/206 & 201/132/253 & 225/187/298 \\

\enddata
\end{deluxetable}

\begin{deluxetable}{c c c c c c c  c c c c c}
\tabletypesize{\scriptsize} 
\tablecaption{Titan Occultations 2023--2050 (by G$_*$ magnitude)}
\label{tbl:titanoccpredstatsG}
\tablehead{
\colhead{Year}& 
\colhead{G$_*$ $<$ 9} &
\colhead{G$_*$ 9--10} &
\colhead{G$_*$ 10--11} &
\colhead{G$_*$ 11--12} &
\colhead{G$_*$ 12--13} &
\colhead{G$_*$ 13--14} &
\colhead{G$_*$ 14--15} &
\colhead{G$_*$ 15--16} &
\colhead{G$_*$ 16--17} &
\colhead{G$_*$ 17--18} &
\colhead{G$_*$ 18--19} \\[-.75em]
\colhead{ }& 
\colhead{P/g/t} &
\colhead{P/g/t} &
\colhead{P/g/t} &
\colhead{P/g/t} &
\colhead{P/g/t} &
\colhead{P/g/t} &
\colhead{P/g/t} &
\colhead{P/g/t } &
\colhead{P/g/t} &
\colhead{P/g/t} &
\colhead{P/g/t}
}
\startdata
2022 & -- & -- & -- & -- & -- & -- & -- & -- & -- & -- & -- \\
2023 & -- & -- & -- & -- & -- & 0/0/1 & 0/0/1 & -- & 3/0/0 & 2/0/1 & 1/0/0 \\
2024 & -- & -- & -- & -- & 1/0/0 & -- & 1/0/1 & 1/3/2 & 2/2/2 & 1/0/0 & -- \\
2025 & -- & -- & -- & -- & 0/0/1 & -- & 1/0/1 & 0/0/1 & 0/1/1 & 1/0/0 & 1/1/1 \\
2026 & -- & -- & 0/0/1 & -- & -- & -- & 0/0/1 & 2/0/0 & 1/1/1 & 0/1/1 & -- \\
2027 & -- & -- & -- & -- & -- & 3/2/3 & 1/0/0 & 1/0/0 & 0/0/2 & 0/1/2 & -- \\
2028 & -- & -- & -- & -- & 0/0/1 & -- & -- & 2/0/1 & 2/0/0 & -- & 0/2/1 \\
2029 & 0/0/1 & -- & -- & 0/0/1 & 1/0/1 & -- & 0/1/0 & 0/1/1 & -- & 3/0/0 & 0/1/2 \\
2030 & -- & -- & -- & -- & -- & 0/1/1 & 1/2/3 & 1/0/1 & 2/4/8 & 2/1/1 & 1/0/0 \\
2031 & -- & 1/0/0 & 0/1/1 & 1/0/0 & -- & 0/0/1 & 2/4/2 & 3/1/4 & 11/4/8 & 7/7/4 & 1/0/0 \\
2032 & -- & -- & -- & -- & 1/1/1 & 1/0/1 & 3/2/5 & 5/6/9 & 7/4/11 & 11/4/7 & 3/4/6 \\
2033 & -- & -- & -- & -- & 0/0/2 & 1/0/0 & 6/5/4 & 7/5/9 & 6/6/12 & 6/8/10 & 1/3/3 \\
2034 & -- & -- & 0/1/1 & -- & 2/0/1 & 2/0/0 & 1/1/1 & 1/1/2 & 0/2/3 & 2/1/2 & 1/0/1 \\
2035 & -- & -- & -- & -- & 2/2/3 & 3/0/2 & 1/0/0 & 1/3/4 & 1/1/1 & 1/0/1 & 3/0/1 \\
2036 & -- & -- & 0/1/0 & -- & -- & 0/0/1 & 0/1/1 & 1/1/1 & 1/1/2 & 0/1/2 & -- \\
2037 & -- & -- & -- & -- & -- & 2/0/0 & 1/1/1 & -- & 1/2/2 & 1/2/2 & 0/1/1 \\
2038 & -- & -- & -- & -- & -- & -- & -- & -- & -- & -- & 2/0/1 \\
2039 & -- & -- & 1/0/0 & -- & -- & -- & 0/1/1 & -- & 1/0/0 & -- & 0/0/1 \\
2040 & -- & -- & -- & -- & 1/0/0 & 0/0/1 & 1/0/0 & 2/1/2 & 3/1/1 & -- & -- \\
2041 & -- & -- & -- & -- & -- & -- & 1/0/1 & 1/0/0 & 0/1/2 & 1/1/1 & 0/1/1 \\
2042 & -- & -- & -- & -- & -- & -- & 0/0/1 & 1/2/3 & 1/3/3 & 0/2/2 & 0/0/1 \\
2043 & -- & -- & -- & -- & 0/0/2 & 0/0/2 & 1/0/0 & 1/0/2 & 2/1/2 & -- & 0/1/1 \\
2044 & -- & -- & -- & 0/0/2 & 1/1/1 & 0/0/1 & 1/0/1 & 0/1/2 & 4/0/3 & 1/0/2 & -- \\
2045 & -- & -- & -- & -- & -- & -- & 5/0/1 & 2/1/5 & 4/2/7 & 4/3/5 & 5/2/4 \\
2046 & -- & -- & -- & -- & -- & 3/4/7 & 3/2/9 & 10/10/21 & 31/22/45 & 40/31/49 & 20/15/32 \\
2047 & -- & -- & -- & 1/0/1 & 0/2/3 & 8/7/12 & 20/9/20 & 34/38/67 & 65/46/84 & 96/64/120 & 58/48/76 \\
2048 & 1/0/0 & -- & -- & -- & 1/0/1 & 8/2/5 & 10/8/14 & 20/22/40 & 31/18/35 & 16/16/20 & 5/11/16 \\
2049 & -- & -- & -- & -- & 1/1/1 & 3/2/4 & 2/3/6 & 11/2/5 & 6/8/6 & 5/2/4 & 1/1/0 \\[0.5em]
Totals & 1/0/1 & 1/0/0 & 1/3/3 & 2/0/4 & 11/7/18 & 34/18/42 & 62/40/75 & 107/98/182 & 185/130/241 & 200/145/236 & 103/91/149 \\

\enddata
\end{deluxetable}

\begin{figure}
\centerline{\resizebox{6in}{!}{\includegraphics[angle=0]{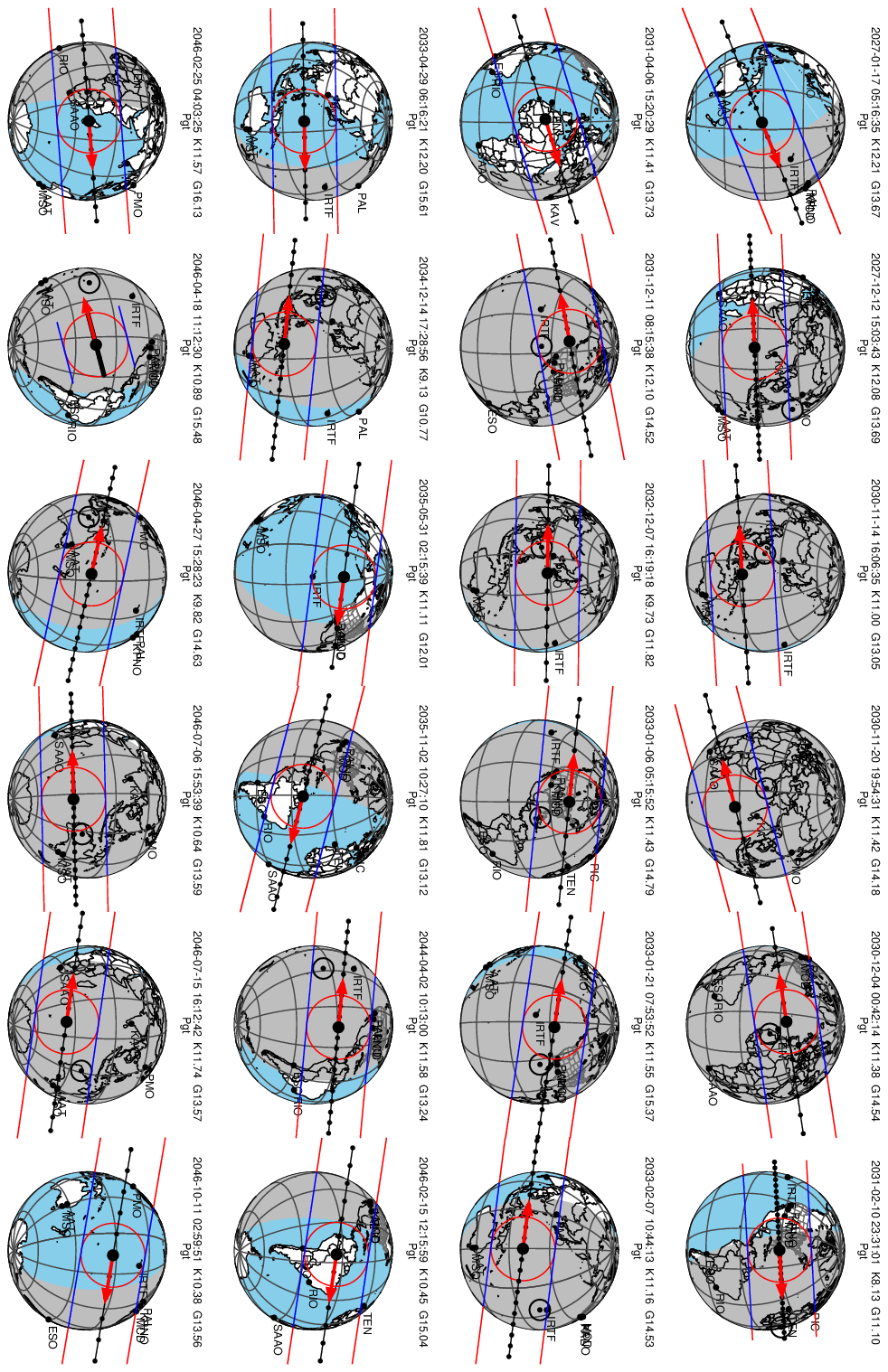}}}
\caption{\red Gallery of the 24 brightest (K band) predicted Titan occultations between 2023 and 2047, restricted to events with at least one topocentric or geocentric sky plane chord that intersects Titan's shadow (event types $Pg, Pgt,$ or $Pt$) Each occultation is labeled by the closest approach epoch, the K and G stellar magnitudes, and the event type.}
\label{fig:TitanGlobes000}
\end{figure}

{\red {\bf Figure \ref{fig:TitanGlobes000}} shows the 24 brightest} (K band) predicted Titan occultations between 2023 and 2047, restricted to events with at least one topocentric or geocentric sky plane chord that intersects Titan's shadow (event types $Pg, Pgt,$ or $Pt$). The solid black lines show the shadow midline, marked by dots at 60 sec time intervals. The direction of the shadow motion is marked by the red arrow. The outline of the target shadow is shown as a red circle, and the projected shadow boundaries of the northern/southern limbs are shown as blue lines when they intersect the Earth and as red lines elsewhere. The antisolar point is marked by an open circle.
{\bf Table \ref{tbl:titanoccpredK10}} includes the geometric information for these 24 events. In the prediction tables for Titan and Triton, we include $C/A_p$, the sky plane separation between the target satellite and the central planet at the epoch of closest approach, to enable observers to estimate the potential contribution of scattered light to the photometric aperture for the target satellite. Details of these and all other predicted Titan occultations are included in machine-readable form on the SOM, as documented in the Appendix.

\begin{longrotatetable}
\movetabledown=6mm
\begin{deluxetable}{c c c D D D D D D D D D D}
\tablecolumns{12}
\tabletypesize{\tiny}
\tablecaption{Geocentric Titan Occultation Predictions 2023--2047}
\label{tbl:titanoccpredK10}
\tablehead{
\\[-2.5em]
\colhead{Event ID} & 
\colhead{C/A Epoch} &
\colhead{ICRS Star Coord at Epoch} & 
\multicolumn2c{$\sigma(\alpha_*)$ (km)} &
\multicolumn2c{C/A  (mas)} & 
\multicolumn2c{C/A$_p$ ({$'$}{$'$})} & 
\multicolumn2c{PA (deg)} & 
\multicolumn2c{K}& 
\multicolumn2c{G} & 
\multicolumn2c{RP} & 
\multicolumn2c{E $\lambda$ (deg)} & 
\multicolumn2c{S-G-T} \\[-.95em]
\colhead{Event type} & 
\colhead{Source ID} &
\colhead{Geocentric Object Position} &
\multicolumn2c{$\sigma(\delta_*)$ (km)} &
\multicolumn2c{C/A (km)} & 
\multicolumn2c{C/A$_p$ (10$^3$ km)} & 
\multicolumn2c{$v_{\rm sky}$ (km/s)} &
\multicolumn2c{Dist (au)} &
\multicolumn2c{G${}_*$} & 
\multicolumn2c{DUP} & 
\multicolumn2c{$\phi$ (deg)} &
\multicolumn2c{M-G-T}
}
\decimals
\startdata
Ti270002 & 2027-01-17 05:16:35.32 & 00 35 52.75277  +01 17 02.67516  & 1.6 & 43.4 & 155.2 & 158.41 & 12.206 & 13.665 & 13.161 & 173.8 & 72.3 \\ 
\ \ \ Pgt & 2544199022533680640 & 00 35 52.75383  +01 17 02.63485  & 1.2 & 303.6 & 1086.6 & 20.76 & 9.657 & 13.706 & 0 & 1.4 & 35.6 \\ 
Ti270012 & 2027-12-12 15:03:43.31 & 01 20 42.16092  +05 39 45.85592  & 1.9 & 138.5 & 70.6 & 177.41 & 12.083 & 13.685 & 13.132 & 73.5 & 120.8 \\ 
\ \ \ Pgt & 2564400353966267904 & 01 20 42.16133  +05 39 45.71753  & 1.5 & 879.7 & 448.6 & -10.74 & 8.756 & 13.010 & 0 & 5.8 & 45.8 \\ 
Ti300013 & 2030-11-14 16:06:35.26 & 04 18 27.83854  +19 15 28.26051  & 1.5 & 320.5 & 202.9 & 176.69 & 11.001 & 13.051 & 12.348 & 129.6 & 165.6 \\ 
\ \ \ Pgt & 47846386650613888 & 04 18 27.83984  +19 15 27.94049  & 1.1 & 1884.3 & 1193.0 & -20.91 & 8.105 & 13.100 & 0 & 19.3 & 34.8 \\ 
Ti300014 & 2030-11-20 19:54:31.10 & 04 15 57.28372  +19 10 10.91518  & 2.6 & 433.0 & 191.6 & 165.26 & 11.419 & 14.184 & 13.326 & 65.9 & 172.3 \\ 
\ \ \ Pgt & 48580826058239104 & 04 15 57.29149  +19 10 10.49641  & 2.0 & 2539.4 & 1123.8 & -22.17 & 8.086 & 14.296 & 0 & 19.2 & 112.1 \\ 
Ti300019 & 2030-12-04 00:42:14.43 & 04 11 38.12678  +19 00 40.41809  & 2.8 & 291.6 & 89.6 & 351.42 & 11.384 & 14.536 & 13.567 & 339.9 & 172.7 \\ 
\ \ \ Pgt & 48515645634144384 & 04 11 38.12371  +19 00 40.70638  & 1.8 & 1708.7 & 525.7 & -25.34 & 8.081 & 14.793 & 0 & 19.1 & 58.7 \\ 
Ti310004 & 2031-02-10 23:31:01.56 & 03 59 27.72426  +18 40 43.98225  & 2.7 & 176.4 & 149.8 & 356.77 & 8.126 & 11.096 & 10.216 & 286.7 & 100.0 \\ 
\ \ \ Pgt & 49815371458307328 & 03 59 27.72358  +18 40 44.15856  & 1.5 & 1130.7 & 959.7 & 7.91 & 8.838 & 10.089 & 0 & 18.8 & 120.2 \\ 
Ti310013 & 2031-04-06 15:20:29.51 & 04 13 49.58603  +19 33 34.61858  & 2.8 & 73.0 & 176.9 & 343.20 & 11.412 & 13.726 & 12.948 & 359.1 & 48.9 \\ 
\ \ \ Pgt & 48650640750399488 & 04 13 49.58453  +19 33 34.68845  & 1.9 & 512.2 & 1241.9 & 28.86 & 9.677 & 14.124 & 0 & 19.6 & 117.1 \\ 
Ti310045 & 2031-12-11 08:15:38.17 & 05 12 58.64159  +21 23 52.08454  & 2.7 & 423.7 & 181.5 & 349.35 & 12.104 & 14.525 & 13.745 & 235.0 & 178.4 \\ 
\ \ \ Pgt & 3414242371731554304 & 05 12 58.63598  +21 23 52.50090  & 1.7 & 2472.5 & 1058.5 & -17.35 & 8.047 & 14.371 & 0 & 21.4 & 139.2 \\ 
Ti320058 & 2032-12-07 16:19:18.61 & 06 19 56.75434  +22 23 19.77808  & 4.2 & 111.7 & 89.5 & 0.68 & 9.734 & 11.816 & 10.757 & 133.7 & 161.0 \\ 
\ \ \ Pgt & 3376835332590836352 & 06 19 56.75447  +22 23 19.88933  & 4.0 & 653.5 & 524.1 & -24.07 & 8.070 & 12.017 & 0 & 22.4 & 136.7 \\ 
Ti330004 & 2033-01-06 05:15:52.96 & 06 09 48.31132  +22 29 37.30113  & 2.7 & 413.9 & 175.2 & 5.74 & 11.427 & 14.792 & 13.821 & 267.9 & 166.5 \\ 
\ \ \ Pgt & 3425060019962276096 & 06 09 48.31431  +22 29 37.71300  & 1.9 & 2417.2 & 1023.5 & -23.10 & 8.051 & 14.948 & 0 & 22.5 & 101.9 \\ 
Ti330015 & 2033-01-21 07:53:52.71 & 06 05 03.96488  +22 32 17.62869  & 4.2 & 207.3 & 199.0 & 8.51 & 11.551 & 15.369 & 14.310 & 212.3 & 150.0 \\ 
\ \ \ Pgt & 3423753280456077184 & 06 05 03.96710  +22 32 17.83372  & 2.7 & 1225.0 & 1176.4 & -18.68 & 8.147 & 15.295 & 0 & 22.5 & 93.6 \\ 
Ti330019 & 2033-02-07 10:44:13.31 & 06 01 01.01860  +22 35 53.71673  & 3.0 & 221.0 & 151.5 & 187.41 & 11.156 & 14.529 & 13.546 & 151.8 & 131.7 \\ 
\ \ \ Pgt & 3424479542243561600 & 06 01 01.01655  +22 35 53.49761  & 2.2 & 1334.0 & 915.3 & -14.62 & 8.324 & 14.189 & 0 & 22.6 & 30.8 \\ 
Ti330048 & 2033-04-29 06:16:21.57 & 06 11 52.91394  +22 47 57.53864  & 5.0 & 105.5 & 95.1 & 179.09 & 12.198 & 15.609 & 14.637 & 141.9 & 53.9 \\ 
\ \ \ Pgt & 3425048921766933632 & 06 11 52.91407  +22 47 57.43319  & 3.8 & 731.7 & 660.2 & 21.88 & 9.566 & 15.707 & 0 & 22.8 & 51.9 \\ 
Ti340019 & 2034-12-14 17:28:56.38 & 08 26 03.16052  +19 31 13.48742  & 1.9 & 392.2 & 159.4 & 185.92 & 9.131 & 10.768 & 10.205 & 141.3 & 138.2 \\ 
\ \ \ Pgt & 663719098892922368 & 08 26 03.15767  +19 31 13.09648  & 1.5 & 2363.4 & 960.9 & -15.57 & 8.308 & 10.496 & 0 & 19.4 & 176.1 \\ 
Ti350016 & 2035-05-31 02:15:39.82 & 08 14 02.35795  +20 23 58.66834  & 2.0 & 354.8 & 146.4 & 7.53 & 11.114 & 12.014 & 11.681 & 201.7 & 52.1 \\ 
\ \ \ Pgt & 675667727972573056 & 08 14 02.36128  +20 23 59.01890  & 1.7 & 2492.9 & 1028.2 & 30.62 & 9.689 & 12.477 & 0 & 20.3 & 132.9 \\ 
Ti350023 & 2035-11-02 10:27:10.27 & 09 22 41.63863  +16 09 03.71639  & 2.2 & 129.2 & 149.6 & 194.46 & 11.811 & 13.122 & 12.667 & 302.8 & 81.3 \\ 
\ \ \ Pgt & 630955851406340480 & 09 22 41.63640  +16 09 03.59132  & 1.8 & 864.4 & 1000.8 & 17.44 & 9.227 & 12.974 & 1 & 16.0 & 107.7 \\ 
Ti440009 & 2044-04-02 10:13:00.69 & 15 49 49.00538  $-$17 44 24.69766  & 3.0 & 304.9 & 144.7 & 5.35 & 11.582 & 13.235 & 12.681 & 253.4 & 133.5 \\ 
\ \ \ Pgt & 6260639759982275712 & 15 49 49.00737  $-$17 44 24.39410  & 1.9 & 2038.4 & 967.8 & -14.31 & 9.219 & 12.872 & 0 & -17.9 & 169.7 \\ 
Ti460039 & 2046-02-15 12:15:59.16 & 17 21 10.67768  $-$21 39 40.67730  & 8.6 & 250.5 & 148.3 & 8.64 & 10.446 & 15.038 & 13.828 & 291.3 & 65.4 \\ 
\ \ \ Pgt & 4115039163112958208 & 17 21 10.68038  $-$21 39 40.42961  & 5.8 & 1889.6 & 1118.3 & 24.14 & 10.399 & 15.242 & 0 & -21.7 & 169.8 \\ 
Ti460055 & 2046-02-25 04:03:25.25 & 17 23 26.53304  $-$21 41 01.04837  & 12.7 & 19.5 & 163.8 & 355.66 & 11.568 & 16.133 & 14.938 & 45.5 & 74.6 \\ 
\ \ \ Pgt & 4114873927148098944 & 17 23 26.53293  $-$21 41 01.02894  & 8.9 & 144.9 & 1217.0 & 19.26 & 10.247 & 16.093 & 0 & -21.7 & 47.7 \\ 
Ti460107 & 2046-04-18 11:12:30.37 & 17 27 08.16215  $-$21 39 01.01547  & 9.5 & 106.9 & 99.5 & 345.24 & 10.888 & 15.477 & 14.282 & 247.6 & 125.6 \\ 
\ \ \ Pgt & 4120809601964982528 & 17 27 08.16020  $-$21 39 00.91207  & 6.7 & 730.5 & 678.6 & -2.83 & 9.419 & 13.354 & 0 & -21.7 & 86.6 \\ 
Ti460115 & 2046-04-27 15:28:23.29 & 17 25 30.94523  $-$21 38 59.78180  & 6.3 & 52.2 & 160.3 & 12.37 & 9.818 & 14.629 & 13.379 & 174.2 & 134.9 \\ 
\ \ \ Pgt & 4120818260633400832 & 17 25 30.94603  $-$21 38 59.73085  & 4.2 & 351.5 & 1080.5 & -14.67 & 9.292 & 14.293 & 0 & -21.7 & 41.2 \\ 
Ti460217 & 2046-07-06 15:53:39.31 & 17 06 10.12253  $-$21 19 29.86495  & 4.2 & 167.8 & 93.7 & 178.63 & 10.640 & 13.591 & 12.734 & 94.0 & 153.3 \\ 
\ \ \ Pgt & 4127597372598625280 & 17 06 10.12282  $-$21 19 30.03273  & 2.8 & 1110.5 & 619.5 & -11.80 & 9.124 & 13.019 & 0 & -21.4 & 119.1 \\ 
Ti460228 & 2046-07-15 16:12:42.40 & 17 03 42.86466  $-$21 19 56.13922  & 4.3 & 252.8 & 146.7 & 188.32 & 11.736 & 13.573 & 12.958 & 79.7 & 144.2 \\ 
\ \ \ Pgt & 4127655023922897664 & 17 03 42.86204  $-$21 19 56.38937  & 2.8 & 1684.8 & 978.0 & -18.46 & 9.189 & 13.486 & 0 & -21.4 & 9.0 \\ 
Ti460264 & 2046-10-11 02:59:51.99 & 17 08 24.28928  $-$21 40 40.27244  & 5.2 & 276.2 & 106.1 & 9.65 & 10.382 & 13.560 & 12.649 & 192.9 & 60.7 \\ 
\ \ \ Pgt & 4115610148914409344 & 17 08 24.29261  $-$21 40 40.00017  & 3.3 & 2102.7 & 807.7 & 28.84 & 10.497 & 13.957 & 0 & -21.7 & 77.2 \\ 
\enddata
\end{deluxetable}
\end{longrotatetable}

\subsection{Triton}\label{sec:Triton}
Our final occultation target is Triton, whose tenuous atmosphere was first studied quantitatively using the {\it Voyager 2} Radio Science Subsystem (RSS) occultation observations made during the spacecraft's final giant planet flyby in 1989 \citep{Tyler1989}. It has been successfully observed during several subsequent extensive Earth-based occultation campaigns as well. As noted by \cite{Bertrand2022}, both Triton and Pluto exhibit volatile cycles of N$_2$, CH$_4$, and CO, and the consequent changes in surface pressure with time are accessible remotely using stellar occultations. 
\cite{Olkin1997} derived the thermal structure of Triton's atmosphere from the 1993 and 1995 occultations, and \cite{Elliot1998} and \cite{Elliot2000} reported on the 1997-11-18 stellar occultation by Triton and found evidence for distortion and increasing pressure in its atmosphere since the 1989 {\it Voyager 2} flyby.
On 2017-10-05, Triton occulted the 13th magnitude star UCAC4 410-143659 as seen from the Eastern US, North Atlantic, and Europe, including the Stratospheric Observatory for Infrared Astronomy (SOFIA) aircraft \citep{Person2018}. A remarkable set of observations of this event from 90 European stations, including 42 detections of the central flash, enabled the detailed investigation of Triton's atmosphere  and a comprehensive reexamination of the full history of occultation observations \citep{Marques2022},  drawing into question the reliability of the evidence for surface pressure changes in the 1990s. 

\begin{figure}
\centerline{\resizebox{6.in}{!}{\includegraphics[angle=0]{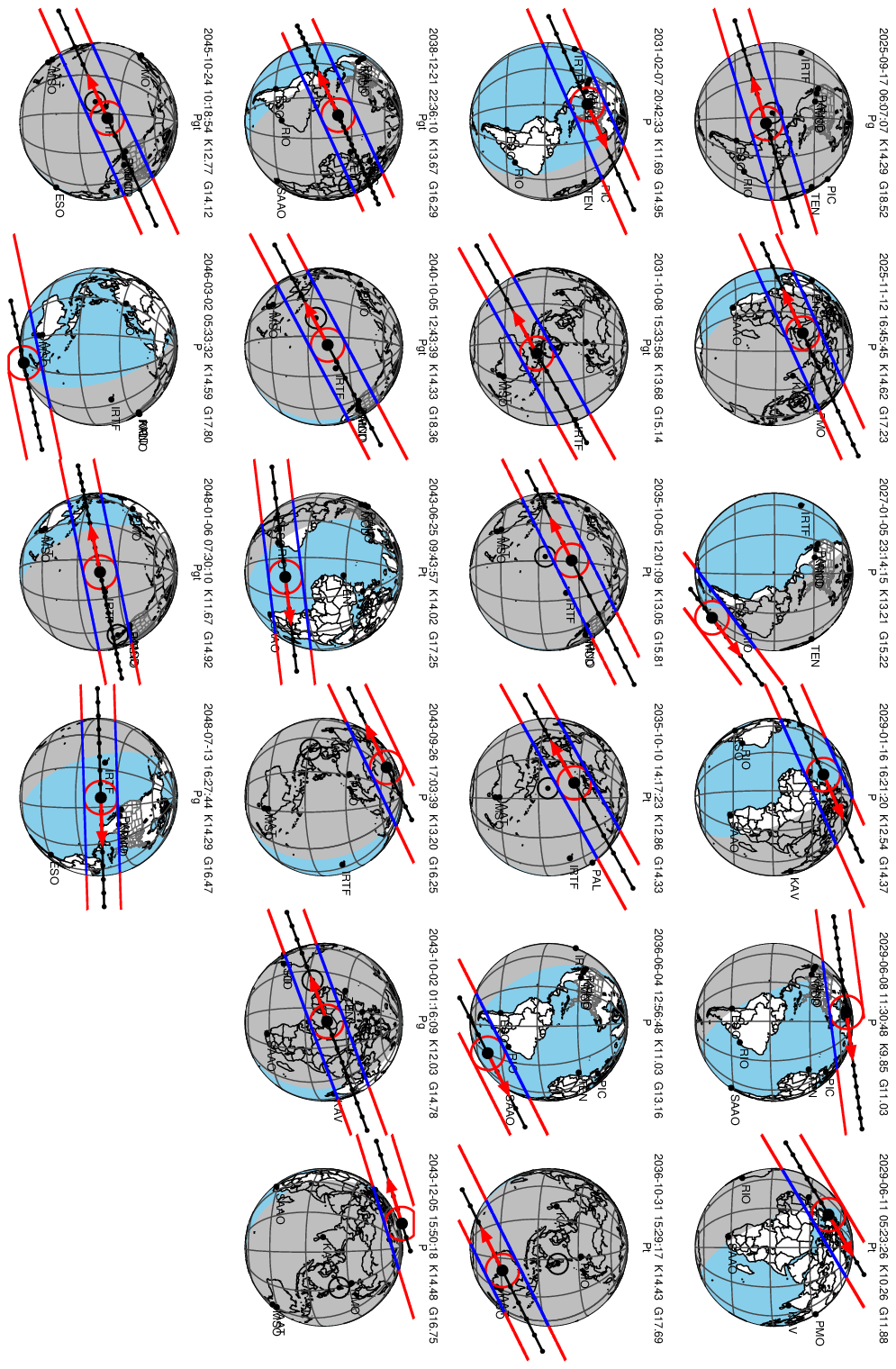}}}
\caption{Gallery of Earth views from Triton at mid-occultation for predicted occultations with K $\leq$15 between 2023 and 2050. Each occultation is labeled by the closest geocentric approach epoch, the K and G stellar magnitudes, and the event type.}
\label{fig:TritonGlobes000}
\end{figure}

In view of their similarities and examples of outer solar system objects with tenuous atmospheres, both Triton and Pluto are appealing targets for future near-term occultation observations. {\red (We excluded Pluto from our list of candidate targets because its long-term ephemeris is much less accurate than Titan's and Triton's.)} Although Pluto continues to be successfully observed on a frequent basis (see \cite{Young2022a} for a recent example), high-SNR Triton occultation opportunities are quite scarce, owing to Neptune's path across fallow star fields (Fig.~\ref{fig:starpaths}). \cite{Marques2022} identified several challenges to accurate Triton occultation predictions, including uncertainties in planetary/satellite ephemeris errors and star positions, {\red although a recent obit and dynamical model for Triton \citep{Wang2023} yielded orbit differences from JPL ephemerides {\tt NEP081} and {\tt NEP097} below 300 km (about 15 mas). This is well below the diameter of both the Earth and Triton and quite adequate to enable secure identification of potential Triton occultation opportunities over the coming decades that are worthy of eventual closer examination.}
{\bf Table \ref{tbl:tritonoccpredstats}} shows the frequency of predicted events by year and K magnitude for K$\leq$15 from 2023--2050 in the same format as for Titan, and  {\bf Table \ref{tbl:tritonoccpredstatsG}} shows 
the corresponding results for G$_*\leq$19, {\red excluding years that have no predicted events}.  Only 22 Triton occultations are predicted over this long period, highlighting the importance of taking advantage of the best opportunities as they arise. 
The complete gallery of predicted Triton events is shown in {\bf Fig.~\ref{fig:TritonGlobes000}}. 
Detailed predictions for the complete set are given in {\bf Table \ref{tbl:tritonoccpredK15}}, and in machine-readable form on the SOM, as documented in the Appendix.
\begin{deluxetable}{c c c c c c c }
\tabletypesize{\scriptsize} 
\tablecaption{Triton Occultations 2023--2050 (by K magnitude)}
\label{tbl:tritonoccpredstats}
\tablehead{
\colhead{Year}& 
\colhead{K 9--10} &
\colhead{K 10--11} &
\colhead{K 11--12} &
\colhead{K 12--13} &
\colhead{K 13--14} &
\colhead{K 14--15} \\[-.75em]
\colhead{ }& 
\colhead{P/g/t} &
\colhead{P/g/t} &
\colhead{P/g/t} &
\colhead{P/g/t} &
\colhead{P/g/t} &
\colhead{P/g/t} \\[-1em]
}
\startdata
2025 & -- & -- & -- & -- & -- & 1/1/0 \\
2027 & -- & -- & -- & -- & 1/0/0 & -- \\
2029 & 1/0/0 & 0/0/1 & -- & 1/0/0 & -- & -- \\
2031 & -- & -- & 1/0/0 & -- & 0/1/1 & -- \\
2035 & -- & -- & -- & 0/0/1 & 0/0/1 & -- \\
2036 & -- & -- & 1/0/0 & -- & -- & 0/0/1 \\
2038 & -- & -- & -- & -- & 0/1/1 & -- \\
2040 & -- & -- & -- & -- & -- & 0/1/1 \\
2043 & -- & -- & -- & 0/1/0 & 1/0/0 & 1/0/1 \\
2045 & -- & -- & -- & 0/1/1 & -- & -- \\
2046 & -- & -- & -- & -- & -- & 1/0/0 \\
2048 & -- & -- & 0/1/1 & -- & -- & 0/1/0 \\
Totals & 1/0/0 & 0/0/1 & 2/1/1 & 1/2/2 & 2/2/3 & 3/3/3 \\

\enddata
\end{deluxetable}

\begin{deluxetable}{c c c c c c c c c c c c}
\tabletypesize{\scriptsize} 
\tablecaption{Triton Occultations 2023--2050 (by G$_*$ magnitude)}
\label{tbl:tritonoccpredstatsG}
\tablehead{
\colhead{Year}& 
\colhead{G$_*$ 10--11} &
\colhead{G$_*$ 11--12} &
\colhead{G$_*$ 12--13} &
\colhead{G$_*$ 13--14} &
\colhead{G$_*$ 14--15} &
\colhead{G$_*$ 15--16} &
\colhead{G$_*$ 16--17} &
\colhead{G$_*$ 17--18} &
\colhead{G$_*$ 18--19} 
\\[-.75em]
\colhead{ }& 
\colhead{P/g/t} &
\colhead{P/g/t} &
\colhead{P/g/t} &
\colhead{P/g/t} &
\colhead{P/g/t} &
\colhead{P/g/t} &
\colhead{P/g/t} &
\colhead{P/g/t} &
\colhead{P/g/t} \\[-1em]
}
\startdata
2025 & -- & -- & -- & -- & -- & -- & -- & 1/0/0 & 0/1/0 \\
2027 & -- & -- & -- & -- & 1/0/0 & -- & -- & -- & -- \\
2029 & 1/0/0 & 0/0/1 & -- & -- & 1/0/0 & -- & -- & -- & -- \\
2031 & -- & -- & -- & -- & -- & 1/1/1 & -- & -- & -- \\
2035 & -- & -- & -- & -- & 0/0/1 & -- & 0/0/1 & -- & -- \\
2036 & -- & -- & -- & 1/0/0 & -- & -- & -- & 0/0/1 & -- \\
2038 & -- & -- & -- & -- & -- & 0/1/1 & -- & -- & -- \\
2040 & -- & -- & -- & -- & -- & -- & -- & -- & 0/1/1 \\
2043 & -- & -- & -- & -- & 0/1/0 & -- & 2/0/0 & 0/0/1 & -- \\
2045 & -- & -- & -- & -- & 0/1/1 & -- & -- & -- & -- \\
2046 & -- & -- & -- & -- & -- & -- & -- & 1/0/0 & -- \\
2048 & -- & -- & -- & -- & 0/1/1 & -- & 0/1/0 & -- & -- \\
Totals & 1/0/0 & 0/0/1 & 0/0/0 & 1/0/0 & 2/3/3 & 1/2/2 & 2/1/1 & 2/0/2 & 0/2/1 \\

\enddata
\end{deluxetable}

\begin{longrotatetable}
\movetabledown=6mm
\begin{deluxetable}{c c c D D D D D D D D D}
\tablecolumns{10}
\tabletypesize{\tiny}
\tablecaption{Geocentric Triton Occultation Predictions 2023 -- 2050 for K $\leq$ 15}
\label{tbl:tritonoccpredK15}
\tablehead{
\\[-2.25em]
\colhead{Event ID} & 
\colhead{C/A Epoch} &
\colhead{ICRS Star Coord at Epoch} & 
\multicolumn2c{$\sigma(\alpha_*)$ (km)} &
\multicolumn2c{C/A  (mas)} & 
\multicolumn2c{C/A$_p$ ({$'$}{$'$})} & 
\multicolumn2c{PA (deg)} & 
\multicolumn2c{K}& 
\multicolumn2c{G} & 
\multicolumn2c{RP} & 
\multicolumn2c{E $\lambda$ (deg)} & 
\multicolumn2c{S-G-T} \\[-.95em]
\colhead{Event type} & 
\colhead{Source ID} &
\colhead{Geocentric Object Position} &
\multicolumn2c{$\sigma(\delta_*)$ (km)} &
\multicolumn2c{C/A (km)} & 
\multicolumn2c{C/A$_p$ (10$^3$ km)} & 
\multicolumn2c{$v_{\rm sky}$ (km/s)} &
\multicolumn2c{Dist (au)} &
\multicolumn2c{G${}_*$} & 
\multicolumn2c{DUP} & 
\multicolumn2c{$\phi$ (deg)} &
\multicolumn2c{M-G-T}
}
\decimals
\startdata
Tr25001 & 2025-09-17 06:07:07.37 & 00 04 15.18297  $-$01 01 59.70659  & 50.4 & 31.6 & 9.0 & 163.16 & 14.288 & 18.519 & 17.207 & 273.2 & 173.6 \\ 
\ \ \ Pg & 2449604188905639040 & 00 04 15.18358  $-$01 01 59.73679  & 32.8 & 661.0 & 188.0 & -20.71 & 28.887 & 18.557 & 0 & -0.9 & 119.5 \\ 
Tr25002 & 2025-11-12 16:45:45.85 & 23 59 18.61005  $-$01 33 51.53200  & 19.0 & 120.8 & 13.6 & 335.68 & 14.615 & 17.226 & 16.431 & 56.7 & 129.0 \\ 
\ \ \ P & 2449519290288023168 & 23 59 18.60673  $-$01 33 51.42195  & 13.6 & 2562.3 & 288.8 & -16.49 & 29.255 & 17.017 & 0 & -1.4 & 146.6 \\ 
Tr27001 & 2027-01-05 23:14:15.27 & 00 07 19.05139  $-$00 42 13.10351  & 11.0 & 286.7 & 8.3 & 143.12 & 13.210 & 15.225 & 14.581 & 268.3 & 76.4 \\ 
\ \ \ P & 2545699104287361664 & 00 07 19.06286  $-$00 42 13.33285  & 6.6 & 6257.3 & 181.7 & 16.01 & 30.090 & 14.983 & 0 & -0.6 & 96.6 \\ 
Tr29001 & 2029-01-16 16:21:20.60 & 00 24 10.27298  +01 02 14.46954  & 7.1 & 200.4 & 15.5 & 335.99 & 12.537 & 14.375 & 13.765 & 4.7 & 69.5 \\ 
\ \ \ P & 2547000135781016320 & 00 24 10.26755  +01 02 14.65264  & 6.0 & 4388.9 & 339.9 & 18.26 & 30.190 & 14.276 & 0 & 1.2 & 47.0 \\ 
Tr29002 & 2029-06-08 11:30:48.14 & 00 40 29.16671  +02 45 54.80459  & 9.1 & 269.8 & 6.6 & 352.41 & 9.849 & 11.026 & 10.601 & 300.7 & 67.2 \\ 
\ \ \ P & 2550622133240584704 & 00 40 29.16433  +02 45 55.07274  & 6.2 & 5916.6 & 145.4 & 14.47 & 30.240 & 10.675 & 0 & 2.9 & 26.4 \\ 
Tr29003* & 2029-06-11 05:23:26.00 & 00 40 39.44754  +02 46 59.57364  & 51.2 & 234.4 & 7.0 & 329.29 & 10.259 & 11.883 & 11.368 & 29.8 & 69.8 \\ 
\ \ \ Pt & 2550622232023925632 & 00 40 39.43955  +02 46 59.77603  & 44.5 & 5133.7 & 154.4 & 18.77 & 30.193 & 11.814 & 0 & 2.9 & 59.1 \\ 
Tr31001 & 2031-02-07 20:42:33.71 & 00 42 01.57927  +02 52 55.87517  & 13.0 & 154.7 & 15.4 & 336.22 & 11.685 & 14.946 & 13.845 & 282.5 & 52.4 \\ 
\ \ \ P & 2550443668759500416 & 00 42 01.57511  +02 52 56.01671  & 12.0 & 3415.0 & 339.6 & 25.85 & 30.443 & 15.225 & 0 & 3.1 & 131.1 \\ 
Tr31002 & 2031-10-08 15:33:58.35 & 00 53 02.38231  +03 55 24.67432  & 14.9 & 53.7 & 6.6 & 150.84 & 13.682 & 15.137 & 14.624 & 123.1 & 178.1 \\ 
\ \ \ Pgt & 2551697971008471808 & 00 53 02.38406  +03 55 24.62743  & 12.9 & 1123.6 & 138.6 & -27.61 & 28.852 & 15.487 & 0 & 4.1 & 93.0 \\ 
Tr35001 & 2035-10-05 12:01:09.75 & 01 27 43.93686  +07 20 07.74648  & 25.8 & 98.8 & 10.9 & 333.31 & 13.049 & 15.808 & 14.941 & 188.1 & 168.5 \\ 
\ \ \ Pt & 2566175343690470144 & 01 27 43.93388  +07 20 07.83479  & 10.9 & 2068.4 & 228.5 & -26.74 & 28.856 & 16.124 & 0 & 7.5 & 144.4 \\ 
Tr35002 & 2035-10-10 14:17:23.75 & 01 27 12.82180  +07 17 09.94514  & 11.3 & 112.7 & 5.6 & 330.87 & 12.862 & 14.332 & 13.809 & 148.9 & 173.5 \\ 
\ \ \ Pt & 2566173934941174400 & 01 27 12.81811  +07 17 10.04359  & 5.7 & 2357.8 & 116.5 & -27.10 & 28.844 & 14.662 & 0 & 7.5 & 83.4 \\ 
Tr36001 & 2036-06-04 12:56:48.45 & 01 37 13.19027  +08 20 50.47105  & 8.1 & 251.0 & 5.3 & 153.15 & 11.032 & 13.163 & 12.496 & 297.0 & 48.4 \\ 
\ \ \ P & 2571920768686360320 & 01 37 13.19791  +08 20 50.24710  & 4.8 & 5551.2 & 118.3 & 27.85 & 30.492 & 13.522 & 0 & 8.5 & 175.8 \\ 
Tr36002 & 2036-10-31 15:29:17.39 & 01 33 36.61897  +07 53 32.49827  & 73.2 & 200.6 & 14.8 & 153.57 & 14.428 & 17.691 & 16.693 & 111.0 & 166.2 \\ 
\ \ \ Pt & 2565881494913346176 & 01 33 36.62498  +07 53 32.31860  & 31.8 & 4200.2 & 310.0 & -24.34 & 28.865 & 17.904 & 0 & 8.1 & 26.1 \\ 
Tr38001 & 2038-12-21 22:36:10.10 & 01 46 59.08374  +09 09 42.25470  & 31.2 & 59.9 & 16.3 & 335.52 & 13.672 & 16.290 & 15.469 & 317.6 & 118.5 \\ 
\ \ \ Pgt & 2572227674164976128 & 01 46 59.08207  +09 09 42.30921  & 30.1 & 1274.5 & 346.5 & -8.80 & 29.336 & 15.399 & 0 & 9.4 & 174.3 \\ 
Tr40001 & 2040-10-05 12:43:39.56 & 02 11 17.56768  +11 18 16.80792  & 109.1 & 16.7 & 2.7 & 332.07 & 14.330 & 18.364 & 17.156 & 187.6 & 157.8 \\ 
\ \ \ Pgt & 73222698406292608 & 02 11 17.56715  +11 18 16.82265  & 118.9 & 349.3 & 55.9 & -25.39 & 28.884 & 18.623 & 0 & 11.5 & 149.6 \\ 
Tr43001 & 2043-06-25 09:43:57.52 & 02 37 43.74095  +13 34 30.89428  & 65.9 & 142.6 & 3.3 & 173.09 & 14.023 & 17.253 & 16.192 & 340.6 & 51.9 \\ 
\ \ \ Pt & 26895600603666432 & 02 37 43.74213  +13 34 30.75276  & 66.3 & 3145.9 & 72.2 & 21.37 & 30.426 & 17.324 & 0 & 13.8 & 95.7 \\ 
Tr43002 & 2043-09-26 17:03:39.54 & 02 38 45.33322  +13 34 09.93323  & 37.9 & 267.6 & 13.8 & 334.38 & 13.201 & 16.245 & 15.317 & 138.9 & 141.3 \\ 
\ \ \ P & 26892856120403968 & 02 38 45.32528  +13 34 10.17456  & 35.0 & 5632.9 & 290.7 & -19.61 & 29.019 & 16.224 & 0 & 13.8 & 63.8 \\ 
Tr43003 & 2043-10-02 01:16:09.36 & 02 38 18.58223  +13 31 54.96788  & 16.9 & 13.3 & 16.9 & 339.82 & 12.025 & 14.776 & 13.924 & 10.4 & 146.7 \\ 
\ \ \ Pg & 26891550450337792 & 02 38 18.58192  +13 31 54.98040  & 14.6 & 280.3 & 354.5 & -19.59 & 28.966 & 14.753 & 0 & 13.7 & 134.2 \\ 
Tr43004 & 2043-12-05 15:50:18.42 & 02 31 40.83793  +13 00 20.93616  & 51.9 & 314.9 & 16.9 & 342.80 & 14.477 & 16.753 & 16.100 & 86.6 & 146.9 \\ 
\ \ \ P & 26451058604260992 & 02 31 40.83156  +13 00 21.23697  & 46.5 & 6618.0 & 354.7 & -19.38 & 28.977 & 16.719 & 0 & 13.2 & 102.5 \\ 
Tr45001 & 2045-10-24 10:18:54.48 & 02 54 06.15989  +14 43 04.64022  & 15.6 & 36.6 & 4.5 & 335.59 & 12.768 & 14.121 & 13.647 & 216.2 & 165.2 \\ 
\ \ \ Pgt & 32518438644356608 & 02 54 06.15885  +14 43 04.67357  & 11.2 & 766.4 & 94.1 & -26.44 & 28.851 & 14.424 & 0 & 14.9 & 32.3 \\ 
Tr46001 & 2046-03-02 05:33:32.53 & 02 48 29.27402  +14 23 42.02432  & 105.5 & 282.4 & 11.7 & 168.18 & 14.585 & 17.799 & 16.743 & 159.2 & 62.9 \\ 
\ \ \ P & 32087322711430784 & 02 48 29.27800  +14 23 41.74795  & 87.2 & 6195.4 & 257.7 & 17.08 & 30.252 & 17.628 & 0 & 14.6 & 123.1 \\ 
Tr48001 & 2048-01-06 07:30:10.46 & 03 05 20.15502  +15 33 34.13969  & 23.0 & 5.1 & 16.7 & 348.41 & 11.666 & 14.919 & 13.891 & 188.9 & 123.3 \\ 
\ \ \ Pgt & 30958747040513152 & 03 05 20.15495  +15 33 34.14473  & 18.4 & 109.2 & 354.8 & -11.25 & 29.265 & 14.294 & 0 & 15.7 & 115.3 \\ 
Tr48002 & 2048-07-13 16:27:44.57 & 03 22 56.20484  +16 46 51.21500  & 55.8 & 7.8 & 2.3 & 358.27 & 14.285 & 16.472 & 15.799 & 232.2 & 58.6 \\ 
\ \ \ Pg & 54599793226129664 & 03 22 56.20482  +16 46 51.22279  & 43.0 & 171.5 & 50.4 & 18.41 & 30.336 & 16.382 & 0 & 17.0 & 86.9 \\ 
\enddata
\end{deluxetable}
\end{longrotatetable}

Several of the predicted Triton occultations have challenging observing geometry, with the nighttime occultation path passing over remote areas of the Earth, such as the 2029-06-08 occultation of a bright star (K=9.85, G=11.03). Three days later, the 2029-11-11 occultation (K=10.26, G=11.88) is observable from Tenerife (TEN) at an elevation angle of 39$^\circ$ when the Sun is 10$^\circ$ below the horizon. {\red Given the rarity of high-SNR Triton occultation opportunities, the possibility of observing any of the upcoming events from {\it CHEOPS} and other space platforms should be explored.}


\section{Discussion and Conclusions}\label{section:discussion}
\subsection{Comparison of {\it Gaia} and 2MASS star positions}
\begin{figure}
\centerline{\resizebox{4.5in}{!}{\includegraphics[angle=0]{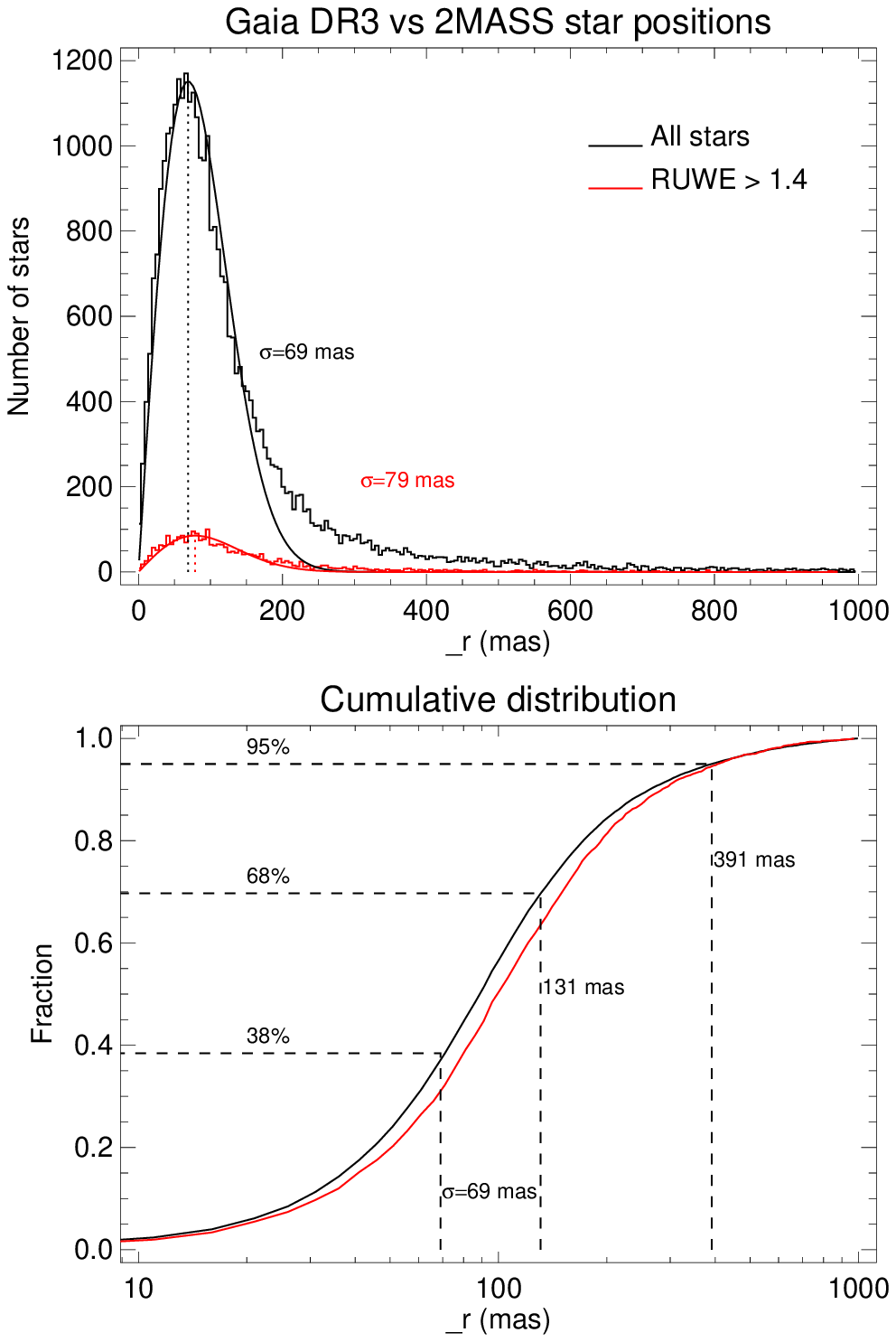}}}
\caption{Comparison of {\it Gaia} DR3 and 2MASS star positions.}
\label{fig:2MASSpos}
\end{figure}

An underlying assumption of our prediction catalog is that we have accurately matched the 2MASS and {\it Gaia} counterparts for each event. Since it is likely that the proper motion-corrected {\it Gaia} positions are much more accurate than the typical positional uncertainty of 100 mas (rms) for K $\leq$14  2MASS stars \citep{Skrutskie2006}, our working hypothesis is that the offsets $_{\_}r$ between these positions are due primarily to scatter in the 2MASS catalog. To test this notion, we show in {\bf Fig.~\ref{fig:2MASSpos}} a histogram of the differences between the 2MASS and {\it Gaia} positions for all stars in our initial lists of occultation candidates for all targets. {\red The upper panel shows a fit to the histogram assuming a Rayleigh distribution appropriate for the one-sided variable $_{\_}r$ with a Gaussian random distribution of standard deviation $\sigma$}:
\beq
f(_{\_}r) = \frac{-_{\_}r}{\sigma^2} e^{-_{\_}r^2/2\sigma^2}.
\eeq

Our fitted value of $\sigma=69$ mas is consistent with quoted 2MASS accuracy of 70-80 mas over the magnitude range of 9 $\leq$ K  $\leq$14  \citep{Cutri2003}, although the actual distribution of position differences has a somewhat larger contribution over the range 150 -- 500 mas than implied by the model Rayleigh distribution. To explore this in more detail, we plotted the cumulative distribution of the offsets $_{\_}r$  in the lower panel of Fig.~\ref{fig:2MASSpos}. We find that 38\% of the stars have  computed $_{\_}r<\sigma$, very close to the expected 1-$\sigma$ cumulative probability of  $1-e^{-1/2}=0.393$ for a Rayleigh distribution. For a two-sided Gaussian distribution, it is customary to think of 1-$\sigma$ and 2-$\sigma$ results as having cumulative probabilities of 68\% and 95\%, and we have estimated the position offsets corresponding to these probabilities to be $_{\_}r=131$ and $391$ mas, respectively, as labeled in the figure.

These results suggest that there is a somewhat more pronounced tail to the distribution of 2MASS position errors than expected for a Gaussian process. An alternative possibility that there are uncertainties in the {\it Gaia} positions or proper motions at the level of a few hundred mas seems to be less likely because of the many internal tests of the accuracy of the {\it Gaia} catalog astrometry \citep{Gaia2021,Gaia2022}. However, \cite{Dunham2021} point out that {\it Gaia} stars with high RUWE (Renormalized Unit Weight Error), a measure of the quality of the astrometric solution, have been shown to have significant astrometric offsets that can be important for occultations of small targets such as asteroids. Also of concern is the ``Duplicated Source Flag" indicating that the star's astrometric information might be degraded by unresolved close duplicity. To assess whether stars with high RUWE have significantly larger {\it Gaia}-2MASS offsets, we performed a separate Rayleigh distribution fit to the approximately 8\% of stars with RUWE$>$1.4, as shown by the red curve in the upper panel of Fig.~\ref{fig:2MASSpos}. The fitted value of $\sigma=79$ mas is modestly greater than $\sigma=69$ mas for the full set of stars, and this high-RUWE population is not responsible for the excess of stars with $_{\_}r$ between 150 -- 500 mas relative to the model distribution. The lower panel shows the corresponding cumulative distribution for the high-RUWE stars (plotted in red), and the differences are slight relative to the full set of stars. We conclude that the larger astrometric error for the high-RUWE stars has only a slight effect on the differences in the {\it Gaia} and 2MASS catalog positions computed at the 2MASS epoch. 

To summarize, the offsets between the {\it Gaia} and 2MASS positions are broadly consistent with the estimated astrometric uncertainties of the 2MASS catalog positions, but there is a tail to the distribution that either reflects a non-Gaussian distribution of 2MASS astrometric errors or is the result of occasional misidentification of the 2MASS star that corresponds to a given {\it Gaia} star. We have reduced the chances for misidentification by eliminating unrealistically reddened stars from our candidate pool. We do this by requiring that the {\it Gaia} G magnitude be no more than 5 magnitudes fainter than the K magnitude for the Jupiter and Saturn searches, and no more than 4 magnitudes fainter for the other searches. Nevertheless, it would be prudent for observers to make an independent assessment of the correspondence between the {\it Gaia} star positions quoted in our predictions and the coordinates of the 2MASS star we have claimed as a match, in cases where $_{\_}r$ approaches our cutoff value of 1 arcsec. 
We include $_{\_}r$ and RUWE in the machine-readable tables, and set {\tt 2MASS DUPFLAG} to 1 in instances where there are two or more predicted events within a day of each other that have separate {\it Gaia} DR3 stars but have been matched to the same 2MASS star, so that observers can take appropriate cautions. 

\subsection{{\it JWST} and Earth-orbiting spacecraft occultation opportunities}
The recent successes of {\it CHEOPS} in observing the 2020-06-11 Quaoar occultation \citep{Morgado2022} and of {\it JWST} in observing the 2022-10-18 stellar occultation by Chariklo's rings \citep{Santos2022b} are reminders of the possibility that current and future spacecraft may be in a position to observe occultations by the targets we have considered in this work. Given the substantial parallax associated with {\it JWST}'s orbit compared to our geocentric positions, the prediction tables presented here cannot be used to identify events observable by {\it JWST}, but there may well be opportunities for {\it CHEOPS} or {\it HST} to observe some of our predicted events. We encourage those searches, which are beyond the scope of the present work.

\subsection{Practical {\red considerations}}
Our goal in presenting occultation predictions extending to 2050 is to provide planetary scientists with an accurate sense of prospects for high-SNR Earth-based stellar occultations by Jupiter, Saturn, Uranus, Neptune, Titan, and Triton in the coming decades, with sufficient detail to support planning for observing campaigns. Given this long duration, however, it is inevitable that future improvements in astrometry, stellar proper motion estimates, photometry, and planetary and satellite ephemerides will affect the details of some of our predicted events. For the giant planets and ring systems, the resulting geometrical changes are less likely to be consequential than for the smaller targets, Titan and Triton. As a cautionary example, {\it Gaia} proper motion estimates change with each new catalog release. For the 2020-10-13 occultation of Plutino (28978) Ixion, \cite{Levine2021} found that the proper motion solution for the star changed by roughly 0.46
mas~yr$^{-1}$ between GDR2 and GEDR3. Over the span of roughly 20 years, comparable changes in the estimated proper motion of our candidate stars 
 could account for a change in star position at epoch of 10 mas, {\red corresponding to potential north-south shifts in the predicted paths of Titan and Triton of $\sim65$ and $\sim290$  km, respectively. These are substantially less than the target diameters and do not call into question our identification of potentially observable occultations by these small objects. Nevertheless, for optimal positioning of portable telescopes to sample multiple occultation chords for Titan and Triton events, detailed updated predictions using the latest stellar catalog positions and planetary ephemerides should be made well in advance of any observing campaign.}

\subsection{Scientific value of long-term occultation observations}

With the rapid increase in the discovery and characterization of extra-solar planets, the four giant planets in our own solar system provide valuable case studies of ice giant and Jupiter-scale worlds. In the era of JWST and future Earth-based and orbiting observing platforms, occultations can provide complementary information about planetary atmosphere and ring systems over the coming decades, such as documenting latitudinal and seasonal variations in planetary stratospheres and extending the long time baseline of high-resolution observations of the ring systems of Saturn, Uranus, and possibly Neptune. Continued reconnaissance of Titan's atmosphere will be especially important in advance of the {\it Dragonfly} mission, scheduled to reach Saturn in 2034. Finally, provocative similarities between Triton's and Pluto's surface and atmospheric interactions and seasonal variability invite concerted efforts to take advantage of the rare opportunities to observe Triton's atmosphere through occultations in the coming decades.

%
\section*{Acknowledgements}
{\red We are grateful to two anonymous referees for their detailed helpful suggestions.} We thank the creators of SORA \citep{Gomes2022} for producing open-source software for the prediction and characterization of stellar occultations, which was developed with the support of ERC Lucky Star and LIneA/Brazil. We used SORA version 0.2.1 to determine the projected diameters and positional uncertainties of our candidate occultation stars.\footnote{\url https://github.com/riogroup/SORA}  We also used SORA to confirm the completeness of our catalog search and to provide an independent check for the geometric quantities computed using our occultation prediction algorithms. 
This publication makes use of data products from the Two Micron All Sky Survey, which is a joint project of the University of Massachusetts and the Infrared Processing and Analysis Center/California Institute of Technology, funded by the National Aeronautics and Space Administration and the National Science Foundation. This research has also made use of the VizieR catalogue access tool, CDS,
 Strasbourg, France (DOI: 10.26093/cds/vizier). The original description of the VizieR service was published in \cite{Ochsenbein2000}. This work has made use of data from the European Space Agency (ESA) mission
{\it Gaia} (\url{https://www.cosmos.esa.int/gaia}), processed by the {\it Gaia}
Data Processing and Analysis Consortium (DPAC,
\url{https://www.cosmos.esa.int/web/gaia/dpac/consortium}). Funding for the DPAC
has been provided by national institutions, in particular the institutions
participating in the {\it Gaia} Multilateral Agreement.
We made extensive use of NASA's NAIF SPICE toolkit and ephemerides for this project \citep{Acton1996}.

\section*{Data availability}
Details of all of the occultation predictions are included in the Supplementary Online Material SOM described in the Appendix.
\vfill
\eject
\restartappendixnumbering
\appendix
\section{Description of the Supplementary Online Material (SOM)}

Detailed occultation prediction tables and figures are available as Supplementary Online Material (SOM) in NASA's Planetary Data System Ring-Moon Systems node at \dataset[DOI: 10.17189/m9sk-g963]{https://doi.org/10.17189/m9sk-g963}. Users should download the entire repository to a local storage device, using the command {\tt wget}, freely available from \url{https://www.gnu.org/software/wget/}. To download the entire SOM contents, enter the following commands from the command line of a terminal:

{\tt cd destdir } (where {\tt destdir} is the local directory within which the SOM directory will reside)

{\tt wget -c -r -nH --cut-dirs=2 https://pds-rings.seti.org/rms-annex/french23\_occult\_pred/SOM/}

The top directory name is {\tt SOM/}. The approximate data volume is 28 GB.

The abbreviated directory structure of the {\tt SOM/} directory is shown below.
Each directory has its own {\tt aareadme.txt}, with its filename including its home directory. The entire SOM can be navigated by opening the {\tt index.html} file in a web browser. The {\tt SOM/docs/} directory contains two sample python programs to illustrate simple occultation searches of the SOM. They can be easily modified for more complex searches.

\input{SOM_directorystruct.txt.vb}
\subsection{\tt SOM/}
Contents of the {\tt SOM/aareadme.txt} file:
\begin{verbatim}
SOM/ is the top-level directory for Supplementary Online Material (SOM)
for the paper "Earth-based Stellar Occultation Predictions for Jupiter, Saturn, 
Uranus, Neptune, Titan, and Triton: 2023-2050" 

Authors: Richard G. French and Damya Souami
Publication: Planetary Science Journal (accepted July 2023)

SOM: Supplementary Online Material

The directory structure is shown below, with aareadme.txt files in each subdirectory.

SOM/
    aareadme.txt           Brief description of SOM/ contents
    index.html             Open this file in a browser to navigate SOM/ directory
    doc/
        aareadme.txt                  Description of SOM/doc/ contents
        EventSearchExample*.*         Example occultation search codes and output files
                           To perform the example searches:
                             python3 EventSearchExample1.py
                             python3 EventSearchExample2.py

        Occpred-manuscript.pdf        Typeset version of occultation prediction paper
        Occpred-SOM.pdf               Typeset version of SOM documentation
    tables/
        aareadme.txt                  Description of SOM/tables/ contents
        Jupiter/
        Saturn/
        Uranus/
        Neptune/
        Titan/
        Triton/
    events/
        aareadme.txt                  Description of SOM/events/ contents
        Jupiter/
        Saturn/
        Uranus/
        Neptune/
        Titan/
        Triton/
        Jupiter/
\end{verbatim}

\subsection{\tt SOM/doc/}
Contents of the {\tt SOM/doc/aareadme.txt} file:
\input{README_doc.txt.vb}
\subsection{\tt SOM/events/}
Contents of the {\tt SOM/events/aareadme.txt} file:
\begin{verbatim}
The SOM/events/ directory contains both summary and detailed occultation predictions
with a subdirectory for each of six targets:
      Jupiter Saturn Uranus Neptune Titan Triton

For example, the SOM/events/Neptune/ directory has the following structure and
abbreviated contents:

SOM/events/Neptune:

-------------gallery of views of Earth from Neptune at mid-occultation time
-------------for 25 events per file
    Neptune_globes_2023-05-03_to_2032-02-16.pdf
    Neptune_globes_2032-09-05_to_2040-07-12.pdf
    Neptune_globes_2041-07-03_to_2047-03-16.pdf
    Neptune_globes_2047-07-01_to_2049-11-05.pdf

-------------gallery of views of skyplane from Earth at mid-occultation time
-------------for 25 events per file
    Neptune_skyplanes_2023-05-03_to_2032-02-16.pdf
    Neptune_skyplanes_2032-09-05_to_2040-07-12.pdf
    Neptune_skyplanes_2041-07-03_to_2047-03-16.pdf
    Neptune_skyplanes_2047-07-01_to_2049-11-05.pdf

-------------detailed information for each event is in subfolders, by year:
Neptune/2023:
-------------Overview summary PDF file for the event 
-------------with C/A time 2023-05-03T13_00_52.170
-------------from prediction series YYYYMMDD 20230528 
-------------version a :
    Neptune_2023-05-03T13_00_52.170_20230528a.pdf

-------------Text file containing summary of this event:
    Neptune_2023-05-03T13_00_52.170_20230528a.txt

-------------full-size images of :
-------------    altitude of target as a function of time
-------------    globe (view of Earth from target)
-------------    skyplane (view of target and occultation chords)
    Neptune_2023-05-03T13_00_52.170_20230528a.alt.jpg
    Neptune_2023-05-03T13_00_52.170_20230528a.globe.jpg
    Neptune_2023-05-03T13_00_52.170_20230528a.skyplane.jpg

-------------Text file containing detailed predictions
-------------for all observatories from which this event was
-------------visible, subject to altitude constraints
-------------    sun more then 5 degrees below horizon 
-------------    target more then 5 degrees above horizon
-------------The observatory code in this case is PAL
-------------for Mt. Palomar 5m telescope.
-------------See full list of observatory codes in the main
-------------body of the paper:
    Neptune_2023-05-03T13_00_52.170_PAL_20230528a.txt
\end{verbatim}

\subsection{\tt SOM/tables/}
Contents of the {\tt SOM/tables/aareadme.txt} file:
\normalsize
\begin{verbatim}
The SOM/tables/ directory contains tables of occultation predictions in both typeset 
form and in machine-readable form, with with a sub-directory for each target:
        Jupiter, Saturn, Uranus, Neptune, Titan, and Triton. 

The table contents differ slightly for ringed/non-ringed planets and for the satellite targets. 

For example, the Neptune directory contains the following files:

./Neptune:
Neptune_tables.pdf               Complete set of typeset tables, produced
                                 from the following LaTeX source files:

Neptune_tables.tex               LaTeX file to produce Neptune_tables.pdf
Neptune_predictions_01_of_01.tex Data tables referred to in Neptune_tables.tex
aastex631.cls                    LaTeX style sheet used for Neptune_tables.tex

Note that to get proper column alignment in the typeset output file, the
input Neptune_tables.tex file must be typeset three times in succession.

Neptune_predictions_MR.txt       Machine-readable version of complete
                                 set of predictions for Neptune.
The format of the machine-readable file is listed at the beginning of the file. 

The header, file format, and first line of data (line-wrapped) for the file
Neptune/Neptune_predictions_MR.txt are as follows:

head -59 Neptune/Neptune_predictions_MR.txt | fold -124
\end{verbatim}
\scriptsize
\begin{verbatim}
Title: Earth-based Stellar Occultation Predictions for Jupiter, Saturn, Uranus, Neptune, Titan, and Triton: 2023-2050
Authors: Richard G. French &  Damya Souami
Table: Occultation predictions for Neptune 2023-2050 for K<15
============================================================
Byte-by-byte Description of file: Neptune_predictions_MR.txt
---------------------------------------------------------
   Bytes      Format    Units    Label     Explanations
---------------------------------------------------------
     1-  8       A8       ---               TARGET Occultation target
    10- 80      A71       ---          SUMMARY_PDF SOM pathname to event summary PDF
    82- 92      A11       ---              EVENTID Event identification (target letter, YYnnn)
    94-112    F19.7         d                   JD Julian date of closest approach (CA)
   114-140      A27       ---               UTC_CA UTC of closest approach
   142-151      A10       ---            EVENTTYPE Event type P: planet, g: geocentric t: topocentric
   153-174      I22       ---               STARID Gaia star ID
   176-211      A36       ---              STARPOS J2000 star position at epoch hh mm ss dd mm ss
   213-219     F7.1        km            STARERR_F E-W error in star position in skyplane (E positive)
   221-227     F7.1        km            STARERR_G N-S error in star position (N positive)
   229-264      A36       ---              TARGPOS J2000 target position at epoch hh mm ss dd mm ss
   266-274     F9.3    arcsec                   CA closest approach distance of target and star in skyplane
   276-285    F10.2       deg                   PA position angle of CA
   287-295     F9.3   1000 km               RKM_CA sky plane separation at CA
   297-305     F9.2      km/s                 VSKY sky plane velocity of target relative to star
   307-316    F10.3        au               DISTAU Observer-target distance
   318-327    F10.3      km/s                 RDOT ring plane radial velocity
   329-338    F10.2       deg                    P position angle of target pole
   340-349    F10.2       deg                 BDEG ring opening angle
   351-357     F7.1       deg                 LATI Ingress geodetic latitude
   359-367     F9.1       deg                 LATE Egress geodetic latitude
   369-377     F9.3       mag                 KMAG apparent K magnitude of occultation star
   379-388    F10.3       mag                 GMAG apparent G magnitude of occultation star
   390-399    F10.3       mag                GSTAR apparent G magnitude of occultation star, corrected for vsky
   401-410    F10.3       mag                RPMAG apparent RP magnitude of occultation star
   412-412       I1       ---                  DUP Source with multiple source identifiers (Gaia catalog entry)
   414-422     F9.3        km                SDIAM projected diameter of occultation star at target
   424-432     F9.1       deg               LONDEG Sub-target Earth longitude (East) at CA
   434-441     F8.1       deg               LATDEG Sub-target Earth latitude at CA
   443-450     F8.1       deg                  SGT Sun-Earth-Target separation at CA
   452-460     F9.1       deg                  MGT Moon-Earth-Target separation at CA
   462-469     F8.2       ---                 RUWE Renormalized unit weight error (<1.4 is good astronometric solution)
   471-491      A21       ---            _2MASS_ID 2MASS catalog ID
   493-493       I1       ---       _2MASS_DUPFLAG another nearby event with a different STARID shares this 2MASS catalog ID
   495-503     F9.3    arcsec                   _R angular separation of Gaia and 2MASS positions
   505-505       I1       ---     EAS_N_TARGETOCCS number of target occultations by candidate observatories in EAS region
   507-507       I1       ---       EAS_N_RINGOCCS number of ring occultations by candidate observatories in EAS region
   509-509       I1       ---     ENA_N_TARGETOCCS number of target occultations by candidate observatories in ENA region
   511-511       I1       ---       ENA_N_RINGOCCS number of ring occultations by candidate observatories in ENA region
   513-513       I1       ---     GEO_N_TARGETOCCS number of target occultations by candidate observatories in GEO region
   515-515       I1       ---       GEO_N_RINGOCCS number of ring occultations by candidate observatories in GEO region
   517-517       I1       ---     NAM_N_TARGETOCCS number of target occultations by candidate observatories in NAM region
   519-519       I1       ---       NAM_N_RINGOCCS number of ring occultations by candidate observatories in NAM region
   521-521       I1       ---     OCN_N_TARGETOCCS number of target occultations by candidate observatories in OCN region
   523-523       I1       ---       OCN_N_RINGOCCS number of ring occultations by candidate observatories in OCN region
   525-525       I1       ---     SAF_N_TARGETOCCS number of target occultations by candidate observatories in SAF region
   527-527       I1       ---       SAF_N_RINGOCCS number of ring occultations by candidate observatories in SAF region
   529-529       I1       ---     SAM_N_TARGETOCCS number of target occultations by candidate observatories in SAM region
   531-531       I1       ---       SAM_N_RINGOCCS number of ring occultations by candidate observatories in SAM region
---------------------------------------------------------
 Neptune   SOM/events/Neptune/2023/Neptune_2023-05-03T13_00_52.170_20230528a.pdf      N23001     2460068.0422705      2023-0
5-03 13:00:52.17      PgRgt    2447704782568325504      23 48 59.15652  -02 29 12.98136     4.4     3.0      23 48 59.14834 
 -02 29 12.68060     0.324     337.79     7.199     26.29     30.600     33.510     -41.58     -21.13     2.4     -34.4     
9.623     11.116     11.413     10.610 0     1.181     301.2     -2.4     46.1     160.6     0.96      23485925-0229122 0   
  0.324 0 0 0 0 1 1 0 1 0 0 0 0 0 0
\end{verbatim}
\normalsize

\section{Examples of SOM files}

For each predicted occultation, the SOM includes visual  and tabular overviews of the observing circumstances. We illustrate these using the 2028-12-19 Uranus occultation.

\begin{figure}
\includegraphics[angle=0,width=6.5in]{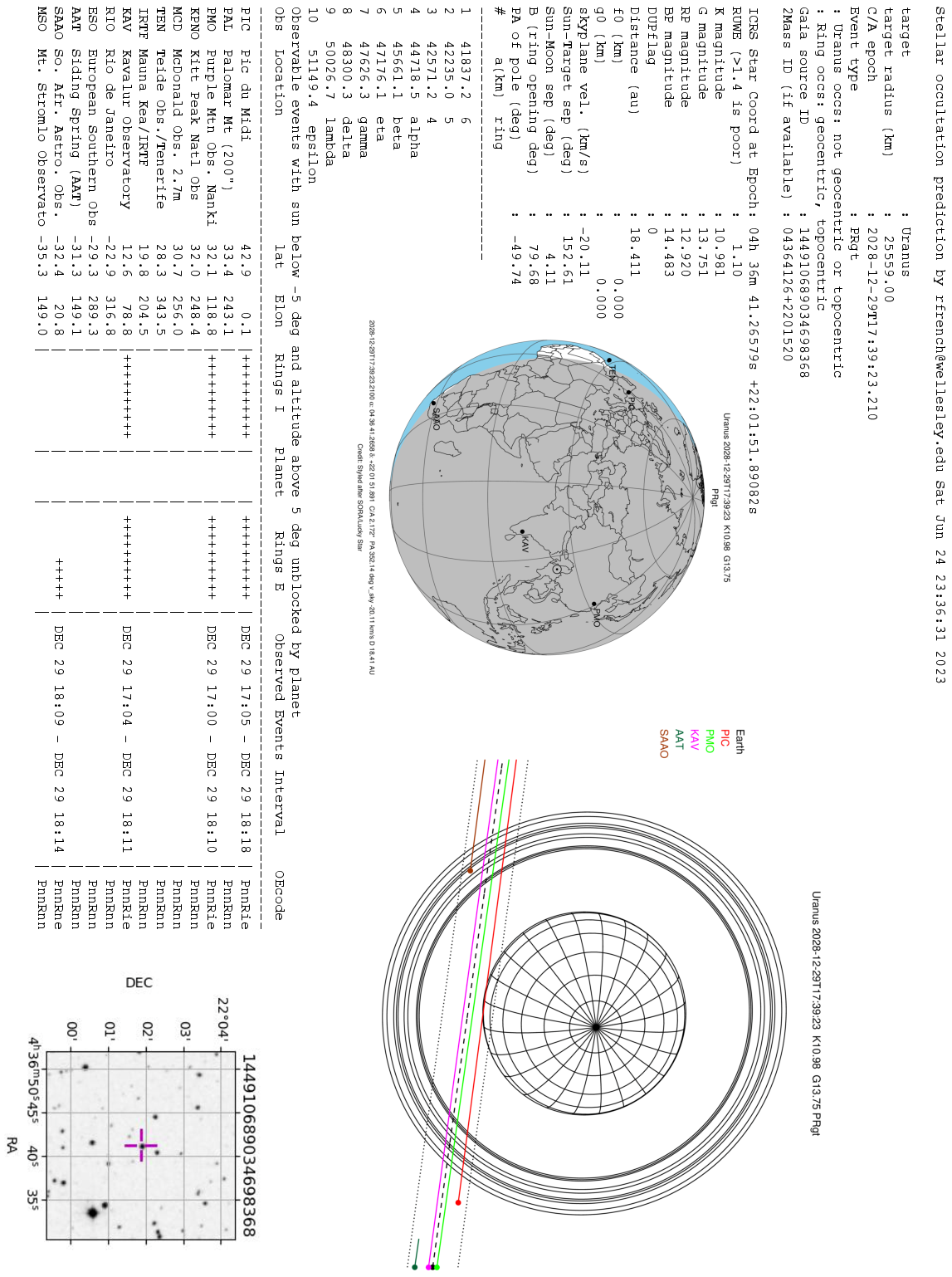}
\caption{Summary page of the PDF file contained in the Supplementary Online Material (SOM) for the 2028-12-29 Uranus occultation.}
\label{fig:pdfsummary}
\end{figure}

\begin{figure}
\centerline{\resizebox{\textwidth}{!}{\includegraphics[angle=0]{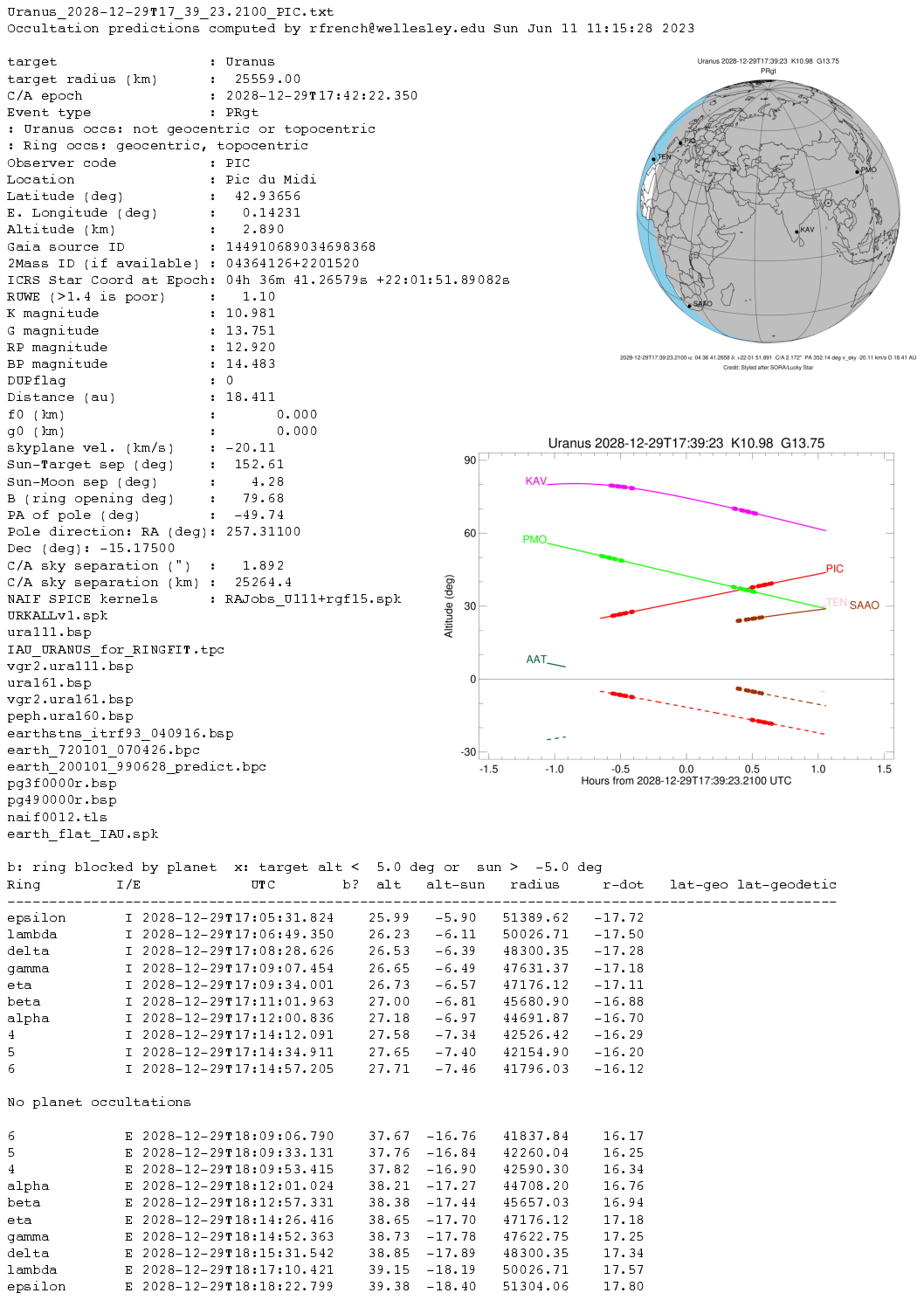}}}
\caption{Site-specific page for PIC (Pic du Midi) observations from the PDF file contained in the Supplementary Online Material (SOM) for the 2028-12-29 Uranus occultation.}
\label{fig:pdfsummaryPIC}
\end{figure}

A single PDF file for each predicted occultation  includes key observational data and plotted figures showing the event geometry. Figure \ref{fig:pdfsummary} shows the summary page for this event.\footnote{\tt SOM/events/Uranus/2028/Uranus\_2028-12-29T17\_39\_23.210\_20230528a.pdf}
Both the view of Earth and the sky plane plots are included in the same SOM subdirectory. 
The summary page includes information about the target object and occultation star, along with other geometrical information defined in more detail below. This text is included in a separate plain text file on the SOM.\footnote{\tt SOM/events/Uranus/2028/Uranus\_2028-12-29T17\_39\_23.210\_20230528a.txt} At lower right, a finder chart image (created using the {\tt plot\_finder\_image} from the Python {\tt astronplan.plots} package) is shown, with the event star marked by crosshairs. 
At the bottom of the page, a convenient summary of the observability of the occultation from all 13 observing sites is included:
\begin{itemize}
\item The observing site code, name, and topocentric Earth location are shown. 
\item The observability of each individual ring event and the planet limb occultations is summarized. In time order, each of the ten Uranus rings is marked during ingress with a $+$ if the ring was observable, given the usual altitude constraints. 
\item The ingress and egress planet occultations are marked, followed in time order by the egress rings.\footnote{For some Saturn and Neptune ring occultations, the ingress and egress ring events all precede or follow the planet occultation, rather than being interrupted by the planet event, but for simplicity we retain the same format for all ringed planets.} For example, from PIC (Pic du Midi), all ingress and egress ring events were observable, but the grazing occultation chord missed the atmosphere. 
\item The next column lists the complete interval over which any marked events were predicted to be observable. 
\item We include a summary observed event code (OEcode) for each site, in the following format: {\tt PXYRxy}, where {\tt X} is set to {\tt i} if the target planet/satellite (denoted by {\tt P}) ingress limb event is observable and  {\tt Y} is set to {\it e}
if the egress limb event is observable. Similarly,  {\tt x} is set to {\tt i} if any ingress ring events (denoted by {\tt R }) are observable from the given site (unblocked by the planet and meeting the standard altitude criteria), and  {\tt y} is set to {\tt e}
if any egress ring events are observable. If the given events are not observable, the appropriate letters are set to {\tt n}. In this example, the OEcodes for the PIC and KAV observations are {\tt PnnRie} and {\tt PnnRne}, respectively; from KAV, only the outer five rings are observable during egress. An OEcode of {\tt PnnRnn} indicates that neither the planet nor any ring occultations were observable for the site in question, such as PAL for this example.
\end{itemize}
For each site that has an OEcode indicating that a planet/target limb and/or ring occultation is observable, we include a separate page in the SOM PDF file that provides additional detailed information about the geometric circumstances of these events. Figure~\ref{fig:pdfsummaryPIC} shows this page for the predicted Pic du Midi observations of the 2028-12-29 Uranus event. Inset figures showing the Earth from Uranus and the altitude of the target and sun over time are included, available at full resolution in the SOM. The text shown includes details of the occultation event and included as a separate text file in the SOM.\footnote{{\tt SOM/events/Uranus/2022/Uranus\_2028-12-29T17\_39\_23.210\_PIC\_20230528a.txt}, where the event time in the filename is the geocentric C/A time.}

At the bottom of the page, we include detailed predictions for each ring event and planet limb that intersects the sky plane chord for the occultation as observed from this site:
\begin{itemize}
\item For each listed ring, we computed the ingress (I) and egress (E) predicted event times. (For Uranus, we use the full eccentric and inclined ring orbital elements for the rings, taken from \cite{French2023b}; for the other planets, we assume circular and equatorial ring orbits). 
\item The UTC time of each predicted event is given, along with the altitudes of the target object and sun at the event time, the ring plane radius probed (taking into account the orientation of the possibly inclined, eccentric ring at the observed time and accounting for general relativistic bending by the oblate planet), and the ring plane radial velocity, labeled as {\tt r-dot}, negative for ingress and positive for egress. 
\item If the planet occultation is observable, we include the predicted occultation time assuming the planetary shape/oblateness as specified in the kernel file {\tt pck00010.tpc}. Our atmospheric event times do not take into account the refractive bending of the atmospheric half-light ray that typically amounts to one atmospheric scale height. 
\item Ring events that are blocked by the planet are marked with a {\tt b} in cases where the rings are viewed nearly edge-on and the occultation sky plane chord intersects the rings only when in the shadow of the planet (not applicable in this instance). 
\item Events for which the target altitude is less than 5$^\circ$  or the sun's altitude is above $-5^\circ$ are marked by an {\tt x} to indicate that they are not observable. In this example, none of the  events are so marked.
\end{itemize}

\section{Machine readable tables, LaTeX source and typeset files}
The complete prediction list for each target is contained in both machine-readable and typeset form. The {\tt SOM/tables/} directory contains subdirectories for each target, within which is a single machine-readable file in the form support by the American Astronomical Society journals and described at \url{https://journals.aas.org/mrt-overview/}.
Also included are the LaTeX source files used to typeset the tables for each target, and a PDF file containing the typeset tables. These make use of the document class {\tt aastex631.cls}, provided in each target subdirectory.

\section{Example occultation searches}
The body of this paper contains only a small subset of the full list of predicted events contained in the SOM. As part of the SOM documentation, we provide two example programs written in Python3 that perform searches of the entire database for occultations that match requested criteria. To run these example codes, users should first download the entire SOM repository to their local machines and then navigate to the {\tt SOM/doc/} directory. Both programs are provided as Python source files {\tt *.py} and as Jupyter notebooks {\tt *.ipynb}, with sample output files {\tt *.out} produced by running the codes in their default configurations. The codes are intended to be illustrative only, and can be modified to conduct more sophisticated searches.

\subsection{Example 1 - find selected occultations by geographical region}
In the first example, the user specifies the following search criteria:
\begin{itemize}
\item{The list of targets to search}
\item{The corresponding upper limits on the K magnitude for each target}
\item{The range of dates for the search}
\item{The geographical regions to search (see Table \ref{tbl:obslocs} in the main body of the paper)}
\end{itemize}
Additional options are included that control the output of the search. In its default mode, the output file produce by the program includes:
\begin{itemize}
\item{A list of the available quantities in the Python table read from the Machine Readable (MR) file for each target}
\item{A summary of the requested search criteria}
\item{For each occultation found, a listing of the summary text file of observing circumstances for all sites}
\item{The SOM pathnames for the summary PDF and text files for each event}
\item{The summary PDF file is opened for user viewing for each identified occultation}
\end{itemize}

In its tersest mode, the default search results in the following output:
\begin{verbatim}
********** Contents of  EventSearchExample1.out  ****************
Results of  EventSearchExample1.py
------------------------------------------------------------------------------
Results of search for occultations by Jupiter between 2024-01-01 and 2030-01-01
K magnitude limit 5
Visible from  N. Am., Eur.&N.Afr., S. Afr., S. Am., Oceania, E. Asia
 
  2027-02-22 11:12:09.02 Pgt     K= 4.926 is visible from N. Am., Oceania, E. Asia
 
------------------------------------------------------------------------------
Results of search for occultations by Saturn between 2024-01-01 and 2030-01-01
K magnitude limit 8
Visible from  N. Am., Eur.&N.Afr., S. Afr., S. Am., Oceania, E. Asia
 
  2024-08-06 17:12:21.22 PgtRgt  K= 7.716 is visible from N. Am., Oceania, E. Asia
  2029-07-19 00:57:19.09 PgtRgt  K= 7.241 is visible from Eur.&N.Afr., S. Afr., E. Asia
 
------------------------------------------------------------------------------
Results of search for occultations by Uranus between 2024-01-01 and 2030-01-01
K magnitude limit 10.9
Visible from  N. Am., Eur.&N.Afr., S. Afr., S. Am., Oceania, E. Asia
 
  2026-07-21 02:13:50.22 PgtRgt  K= 10.802 is visible from Eur.&N.Afr., S. Afr.
 
------------------------------------------------------------------------------
Results of search for occultations by Neptune between 2024-01-01 and 2030-01-01
K magnitude limit 11
Visible from  N. Am., Eur.&N.Afr., S. Afr., S. Am., Oceania, E. Asia
 
  2024-10-09 00:36:23.19 PgtRgt  K= 9.068 is visible from N. Am., Eur.&N.Afr., S. Afr., S. Am.
 
------------------------------------------------------------------------------
Results of search for occultations by Titan between 2024-01-01 and 2030-01-01
K magnitude limit 7
Visible from  N. Am., Eur.&N.Afr., S. Afr., S. Am., Oceania, E. Asia
 
  2029-03-04 02:37:38.25 Pt      K= 6.151 is visible from N. Am.
 
------------------------------------------------------------------------------
Results of search for occultations by Triton between 2024-01-01 and 2030-01-01
K magnitude limit 15
Visible from  N. Am., Eur.&N.Afr., S. Afr., S. Am., Oceania, E. Asia
 
  2029-06-11 05:23:26.00 Pt      K= 10.259 is visible from Eur.&N.Afr.
\end{verbatim}
\subsection{Example 2 - find selected occultations by observing site}
In the second example, the user specifies the following search criteria:
\begin{itemize}
\item{The list of targets to search}
\item{The corresponding upper limits on the K magnitude for each target}
\item{The range of dates for the search}
\item{Specific observing sites for the search (see Table \ref{tbl:obslocs} in the main body of the paper)}
\end{itemize}
Additional options are included that control the output of the search. In its default mode, the output file produce by the program includes:
\begin{itemize}
\item{A list of the available quantities in the Python table read from the Machine Readable (MR) file for each target}
\item{A summary of the requested search criteria}
\item{For each occultation found, a listing of the summary text file of observing circumstances for all sites}
\item{The SOM pathnames for the summary PDF and text files for each event}
\item{The SOM pathnames for the individual event summary text files for each observing site}
\item{The summary PDF file is opened for user viewing for each identified occultation}
\end{itemize}

In its tersest mode,  the default search results in the following output:
\begin{verbatim}
********** Contents of  EventSearchExample2.out  ****************
Results of  EventSearchExample2.py
------------------------------------------------------------------------------
Results of search for occultations by Jupiter between 2024-01-01 and 2030-01-01
K magnitude limit 5
Visible from  IRTF, TEN, KPNO
 
  2027-02-22 11:12:09.02 Pgt     K= 4.926 is visible from IRTF, KPNO
 
------------------------------------------------------------------------------
Results of search for occultations by Saturn between 2024-01-01 and 2030-01-01
K magnitude limit 8
Visible from  IRTF, TEN, KPNO
 
  2024-08-06 17:12:21.22 PgtRgt  K= 7.716 is visible from IRTF
  2029-07-19 00:57:19.09 PgtRgt  K= 7.241 is visible from TEN
 
------------------------------------------------------------------------------
Results of search for occultations by Uranus between 2024-01-01 and 2030-01-01
K magnitude limit 10.9
Visible from  IRTF, TEN, KPNO
 
No events found with requested conditions
 
------------------------------------------------------------------------------
Results of search for occultations by Neptune between 2024-01-01 and 2030-01-01
K magnitude limit 11
Visible from  IRTF, TEN, KPNO
 
  2024-10-09 00:36:23.19 PgtRgt  K= 9.068 is visible from KPNO, TEN
 
------------------------------------------------------------------------------
Results of search for occultations by Titan between 2024-01-01 and 2030-01-01
K magnitude limit 7
Visible from  IRTF, TEN, KPNO
 
  2029-03-04 02:37:38.25 Pt      K= 6.151 is visible from KPNO
 
------------------------------------------------------------------------------
Results of search for occultations by Triton between 2024-01-01 and 2030-01-01
K magnitude limit 15
Visible from  IRTF, TEN, KPNO
 
  2029-06-11 05:23:26.00 Pt      K= 10.259 is visible from TEN
\end{verbatim}

\vfill
\eject   

\end{document}